\documentclass[fleqn,nofootinbib,aps,reprint,amsmath,amssymb,10pt,superscriptaddress,floatfix,a4paper,letterpaper,fleqn]{revtex4-1}
\usepackage{amssymb}
\usepackage{amsfonts}
\usepackage{amsmath}
\usepackage{makeidx}
\usepackage[bookmarksnumbered,colorlinks,bookmarks,breaklinks=true,linkbordercolor={1 1 1},linktocpage,citecolor=blue]{hyperref}
\usepackage{graphicx,color}
\usepackage{subfigure}
\usepackage{multirow}
\usepackage{dcolumn}
\usepackage{bm}
\usepackage{fancyhdr}
\usepackage{array}
\usepackage{longtable}
\usepackage{threeparttable}
\usepackage{floatflt}
\usepackage{color}
\usepackage{verbatim}
\usepackage{nicefrac}
\setlongtables
\usepackage[activate={true,nocompatibility},final,tracking=true,kerning=true,spacing=true,factor=1100,stretch=10,shrink=10]{microtype}
\newcommand\T{\rule{0pt}{2.6ex}}       % Top strut
\newcommand\B{\rule[-1.2ex]{0pt}{0pt}} % Bottom strut
\def\etal{\textit{et al.~}}
\def\etals{\textit{et al.}}
\def\ql{``}
\def\qr{''\hspace{0.5mm}}
\def\qrs{''}

\def\g{$\gamma$}

\def\pos{$\beta^+$}

\newcommand{\halflife}{T$_{\nicefrac{1}{2}}$}

\begin{document}
\setcounter{page}{1}
\title{
% Do not remove the line below, it generates blank space for Elsevier stamp
     \qquad \\ \qquad \\ \qquad \\  \qquad \\  \qquad \\ \qquad \\
     Upgrade of IAEA recommended data of selected nuclear reactions for production of PET and SPECT isotopes }

\author{A.\thinspace Hermanne}
\affiliation{Cyclotron Department - TONA, Vrije Universiteit Brussel, Brussels, Belgium}

\author{F.\thinspace T.\thinspace T\'ark\'anyi}
\affiliation{Institute for Nuclear Research, Debrecen, Hungary}

\author{A.\thinspace V.\thinspace Ignatyuk}
\affiliation{Institute of Physics and Power Engineering (IPPE), Obninsk, Russia}

\author{S.\thinspace Tak\'acs}
\affiliation{Institute for Nuclear Research, Debrecen, Hungary}

\author{R.\thinspace Capote}
\email{r.capotenoy@iaea.org}
\affiliation{NAPC--Nuclear Data Section, International Atomic Energy Agency, Vienna, Austria}

\date{\today}

% Do not touch the line below, it serves editorial purposes
\received{XXX 2017; revised received XXX 2017; accepted XX 2017}
\begin{abstract}
An IAEA research project was dedicated to the compilation, evaluation and recommendation of cross-section data for the accelerator production of $^{11}$C, $^{13}$N, $^{15}$O, $^{18}$F, $^{64}$Cu, and $^{124}$I positron-emitting radionuclides clinically used for PET imaging, and for the  accelerator production of two gamma emitters, $^{81}$Rb and $^{123}$I, used in SPECT imaging. Cross sections for 19 charged-particle induced reactions that can be employed for radionuclide accelerator production were evaluated including uncertainties. The resulting reference cross-section data were obtained from Pad\'e fits to selected and corrected experimental data, and integral thick target yields were subsequently deduced. Uncertainties in the fitted results were estimated via a Pad\'e least-squares method with the addition of a 4\% assessed systematic uncertainty to the estimated experimental uncertainty. Experimental data were also compared with theoretical predictions available from the TENDL library. All of the numerical reference cross-section data with their corresponding uncertainties and deduced integral thick target yields are available on-line at the IAEA-NDS medical portal \href{https://www-nds.iaea.org/medportal/}{\textit{www-nds.iaea.org/medportal/}} and also at the IAEA-NDS web page \href{https://www-nds.iaea.org/medical/}{\textit{www-nds.iaea.org/medical/}}.
\end{abstract}
\maketitle
%\newpage

\lhead{ Upgrade of IAEA recommended data ...}
\chead{NUCLEAR DATA SHEETS}
\rhead{A.~Hermanne \etals} \lfoot{} \rfoot{} %
\renewcommand{\headrulewidth}{0.4pt} \renewcommand{\footrulewidth}{0.4pt}
\tableofcontents
\vfill

%======================================================================
\section{INTRODUCTION} % instead of IAEA mission
%======================================================================
\label{sect-I}

%\subsection{Mission of IAEA: importance of medical and industrial applications of radionuclides}
%Article of IAEA statutes:\textit{\ql The (International Atomic Energy) Agency shall seek to accelerate and enlarge the contribution of atomic energy to peace, health and prosperity throughout the world\qrs.}
%
%$\smallskip$
The IAEA document on \ql Medium Term Strategy 2012--2017\qrs clearly stated:
\textit{The Agency will seek to support the safe and effective use of radiation medicine for the diagnosis and treatment of patients\ldots
In the area of utilization of research reactors and accelerators for radioisotope production and radiation technology, the Agency will support Member States in building capacity for sustainable production and related quality assurance systems, and ensure accessibility to products and techniques that have a unique added value \ldots}

Hence, development and optimization of radionuclide production both for industrial and medical applications are of considerable interest to the IAEA. Particle accelerators are broadly being used for the production of radioisotopes for both diagnostic and therapeutic purposes often with the IAEA support.

The IAEA Nuclear Data Section has sponsored over the last 25 years several actions to set up a database for recommended cross sections and nuclear data for various charged-particle reactions used for medical radionuclides production. In 1995 a first Coordinated Research Project (CRP) entitled \ql Development of Reference Charged Particle Cross-section Database for Medical Radioisotope Production\qr was initiated in order to meet data needs at that time and focused on the radionuclides most commonly used for diagnostic purposes (PET and SPECT imaging) and on the related beam monitor reactions.

%\squeezetable
\begin{table*}[t]
\vspace{-3mm}
\caption{Decay data of studied radionuclides as present in IAEA-TECDOC-1211 \cite{Gul:2001}, the updates undertaken in 2003--2004 \cite{database:2001,Takacs:2003,Takacs:2005} and in IAEA TRS 473~\cite{Betak:2011}, ENSDF \cite{ENSDF} and the DDEP \cite{DDEP} evaluations$^{\mathrm{a}}$.}
\begin{tabular}{c|c|c|c|c|c}
\hline\hline
\T \multirow{2}{*}{ } &  IAEA TECDOC-1211 & updates 2003--2004 & IAEA TRS 473  &ENSDF$^{\mathrm{b}}$ &  DDEP \\ \hline
\T Radionuclide       & \multicolumn{5}{c}{\halflife~of radionuclide} \B \\ \hline
\T
 $^{11}$C    & 20.39 min & 20.36 min  &  --        & 20.364(14) min & 20.361(23) min \\
 $^{13}$N    &  9.96 min & 9.965 min  &  --        & 9.965(4) min   & 9.967(37) min  \\
 $^{15}$O    & 122.4 s   & 122.22 s   &  --        & 122.24(16) s   & 122.46(36) s   \\
 $^{18}$F    & 109.8 min & 109.77 min &  --        & 109.77(5) min  & 1.82890(23) h  \\

 $^{64}$Cu   & --        & 12.701 h   & 12.701 h   & 12.701(2) h    & 12.7004(20) h  \\

 $^{124}$I   & --        & 4.18 d     & 4.1760 d   & 4.1760(3) d    & --             \\

 $^{81}$Rb   & 4.58 h    & 4.572 h    & --         & 4.572(4) h     & --             \\

 $^{123}$I   & 13.2 h    & 13.2235 h  & --         & 13.2235(19) h  & 13.2234(37) h  \\

\hline
\T \multirow{1}{*}{Radionuclide}  &  \multicolumn{5}{c}{E$_{\gamma}$(keV)}  \B    \\
\hline
\T
 $^{11}$C    & 511   & 511    &           & 511       & 511 \\
 $^{13}$N    & 511   & 511    &           & 511       & 511 \\
 $^{15}$O    & 511   & 511    &           & 511       & 511 \\
 $^{18}$F    & 511   & 511    &           & 511       & 511 \\

 $^{64}$Cu   & --     & 511    & 511      & 511       & 511 \\
             & --     & 1345.77 & 1346.   & 1345.77(6) & 1345.75(5) \\

 $^{124}$I   & --     & 602.73  & 602.7   & 602.73(8)  & --   \\
             & --     & 722.78  & 722.78  & 722.28(8)  & --   \\
             & --     & 1690.98 & 1691.   & 1690.96(8) & --   \\
             & --     & --      & 511     & 511       & --   \\

 $^{81}$Rb   & 190.4  & 190.46  & --      & 190.46(16) & --        \\
             & --     & 446.15  & --      & 446.15(3)  & --        \\
             & --     & 511    & --       & 511       & --        \\

 $^{123}$I   & 159.   & 158.97  & --      & 158.97(5) & 158.97(5) \\

\hline
\T \multirow{1}{*}{Radionuclide} & \multicolumn{5}{c}{I$_{\gamma}$ (\%)$^{\mathrm{c}}$} \B \\
\hline
\T
 $^{11}$C    & 199.6  & 200.    &         & 199.534(5)   & 199.500(26)  \\
 $^{13}$N    & 199.6  & 200.    &         & 199.607(4)   & 199.636(26)  \\
 $^{15}$O    & 199.8  & 200.    &         & 199.8006(20) & 199.770(12)  \\
 $^{18}$F    & 197.   & 200.    &         & 193.46(8)    & 193.72(38)   \\

 $^{64}$Cu   & --     & 35.2    & 34.5    & 35.2(4)      & 35.04(30)    \\
             & --     & 0.475   & 0.54    & 0.475(11)    & 0.4749(34)   \\

 $^{124}$I   & --     & 61.     & 62.9    & 62.9(7)      & --   \\
             & --     & 10.36   & 10.36   & 10.35(12)    & --   \\
             & --     & 10.41   & 10.9    & 11.15(17)    & --   \\
             & --     & --      & 44.2    & 45(3)        & --   \\

 $^{81}$Rb   & 64.3   & 64.9    & --      & 64.9(22)     & --        \\
             & --     & 23.5    & --      & 23.5(9)      & --        \\
             & --     & 54.4    & --      & 54.4(20)     & --        \\

 $^{123}$I   & 83.3   & 83.3    & --      & 83.3(4)         & 83.25(21) \\

\hline\hline
\end{tabular}
\label{table1}
\vspace{+2mm}
\begin{tablenotes}
\item{${}^{\mathrm{a}}${\small~Table~2 exactly provides the recommended decay data.}}
\item{${}^{\mathrm{b}}${\small~ENSDF nuclear structure and decay data can be easily extracted, understood and studied in an attractive user-friendly manner by means of LiveChart of Nuclides \cite{Livechart} and NuDat \cite{NUDAT}.}}
\item{${}^{\mathrm{c}}${\small~$\gamma$-ray intensities I$_{\gamma}$ defined in energy order per each product radionuclide as presented in the listing of the E$_\gamma$ data above.}}
\end{tablenotes}
\vspace{-4mm}
\end{table*}

%% Table 2
\begin{table*}[!htb]
\vspace{-3mm}
\caption{Production reactions for studied radionuclides and recommended decay data of the activation products (\halflife~is the product half-life, and E$_{\gamma}$ is the \g-ray energy in~keV of the transition with intensity I$_{\gamma }$ in \%).
Reaction threshold or the Q value if positive is given in the one to last column, the IAEA document with last update is shown in the last column.}
\begin{tabular}{l|c|c|c|c|c|c}
\hline\hline
\T Radionuclide     &             &                   &                   & Production & Threshold/    & Latest    \\
  Decay path (\%)   & ~\halflife~ & E$_{\gamma}$(keV) &  I$_{\gamma}$(\%) & Reaction   & Q-value [MeV] & update \B \\
 \hline
\T
 $^{11}$C           &                 &              &             &                                &        &          \\
 $\epsilon$: 0.2331 &  20.364(14) min &  511         & 199.534(5)  &  $^{14}$N(p,$\alpha$)$^{11}$C  & 3.132  & DB-2003  \\
 $\beta^+$: 99.7669 &                 &              &             &                                &        &          \\
 \hline
\T
 $^{13}$N           &                 &              &             &                                &        &          \\
$\epsilon$: 0.2     &  9.965(4) min   &  511         & 199.607(4)  &  $^{16}$O(p,$\alpha$)$^{13}$N  & 5.547  & DB-2003  \\
$\beta^+$: 99.8036  &                 &              &             &                                &        &          \\
 \hline
\T
 $^{15}$O           &                 &              &             &                                &        &          \\
$\epsilon$: 0.1     &  122.24(16) s   &  511         & 199.8006(20)&  $^{14}$N(d,n)$^{15}$O         & 5.072  & DB-2003  \\
$\beta^+$: 99.9003  &                 &              &             &  $^{15}$N(p,n)$^{15}$O         & 3.774  &          \\
 \hline
\T
 $^{18}$F           &                 &              &             &  $^{18}$O(p,n)$^{18}$F         & 2.574  &          \\
$\epsilon$: 3.27    &  109.77(5) min  &  511         & 193.46(8)   &  ${^\mathrm{nat}}$Ne(d,x)$^{18}$F       & 2.795  & DB-2003  \\
$\beta^+$: 96.73    &                 &              &             &                                &        &          \\
 \hline
\T
 $^{64}$Cu          &                 &              &             &  $^{64}$Ni(p,n)$^{64}$Cu       & 2.495  &          \\
$\epsilon$: 46.9    &  12.701(2) h    &  511         & 35.2(4)     &  $^{64}$Ni(d,2n)$^{64}$Cu      & 4.828  & TRS 473  \\
$\beta^+$: 17.6     &                 &  1345.77(6)  & 0.475(11)   &  $^{68}$Zn(p,x)$^{64}$Cu       & 7.905  &          \\
$\beta^-$: 35.5     &                 &              &             &  ${^\mathrm{nat}}$Zn(d,x)$^{64}$Cu      &  --    &          \\
 \hline
\T
 $^{124}$I          &                 &  602.73(8)   & 62.9(7)     &  $^{124}$Te(p,n)$^{124}$I      & 3.973  &          \\
$\epsilon$: 6.9     &  4.1760(3) d    &  722.78(8)   & 10.36(12)   &  $^{125}$Te(p,2n)$^{124}$I     & 10.595 & TRS 473  \\
$\beta^+$: 22.7     &                 &  1690.96(8)  & 11.15(17)   &  $^{124}$Te(d,2n)$^{124}$I     &  6.266 &          \\
$\beta^-$: 93.1     &                 &  511         & 45(3)       &                                &        &          \\
 \hline
\T
 $^{81}$Rb          &                 &  190.46(16)  & 64.9(22)    & $^{82}$Kr(p,2n)$^{81}$Rb       & 14.16  &          \\
$\epsilon$: 72.8    &  4.572(4) h     &  446.15(3)   & 23.5(9)     & ${^\mathrm{nat}}$Kr(p,x)$^{81}$Rb       &  --    & DB-2004  \\
$\beta^+$: 27.2     &                 &  511         & 54.4(20)    &                                &        &          \\
 \hline
\T
$^{123}$I           &                 &              &             & $^{123}$Te(p,n)$^{123}$I       & 2.027  &          \\
$\epsilon$: 100     & 13.2235(19) h   &  158.97(5)   & 83.3(4)     & $^{124}$Te(p,2n)$^{123}$I      & 11.528 & DB-2004  \\
                    &                 &              &             & $^{127}$I(p,5n)$^{123}$Xe$\rightarrow$$^{123}$I & 37.095 &          \\
                    &                 &              &             & $^{127}$I(p,3n)$^{125}$Xe$\rightarrow$$^{125}$I & 18.864 &          \\
\hline\hline
\end{tabular}
\label{table2}
\vspace{-4mm}
\end{table*}

That project represented the first major international effort dedicated to the standardisation of cross-section data for radionuclide production through light charged-particle nuclear reactions and for monitoring the characteristics of the particle beams used in these productions (protons, deuterons, $^3$He and $\alpha$-particles). That CRP produced the much needed IAEA TECDOC-1211 Handbook that was published in 2001~\cite{Gul:2001}. The on-line version of the database, containing tables of recommended cross sections and yields as well as all references, was made available at the IAEA webpage \href{https://www-nds.iaea.org/medical/}{\textit{www-nds.iaea.org/medical/}}~\cite{database:2001}. This database was partly updated by inclusion of new experimental data, corrections of factual errors and spline fits of cross sections used for production of positron and gamma  emitters in 2003--2005~\cite{Takacs:2003,Takacs:2005} and for monitors in 2007~\cite{Takacs:2008}.

The growing number of emerging radionuclides for therapeutic applications, asked for a new CRP of which the results were published in 2011 as IAEA Technical Report 473 entitled \ql Nuclear Data for Production of Therapeutic Radionuclides\qr~\cite{Betak:2011}. Although the recommended cross sections were believed to be accurate enough to meet the demands of all current applications, further development of procedures and research in nuclear medicine, of the evaluation methodology, and publications of more experimental results necessitated a widening of the existing database while it also became clear that determination of the uncertainties and their correlations was needed.

In June 2011 recommendations from four IAEA consultants' meetings~\cite{Capote:2008,Capote:2011,Nichols:2011,Nichols:2012} were brought together to formulate the scope, work programme and deliverables of a new CRP designed to improve, update and especially broaden the cross-section database given in IAEA-TECDOC-1211~\cite{Gul:2001}. The most important novelty in this CRP is that after compilation and selection of the experimental datasets (corrected for outdated nuclear decay data, see Table 1 and 2) recurrent Pad\'e fitting was performed at IPPE Obninsk and uncertainties on the evaluated cross sections were derived. The result of this new large international joint effort was recently published in four peer reviewed articles, authored by all CRP participants,  that respectively contained recommended cross-section values for 34 monitor reactions~\cite{Hermanne:2018} (lead author A. Hermanne), for 25 reactions for production of  diagnostic SPECT radionuclides~\cite{Tarkanyi:2019} (lead author F. T. T\'ark\'anyi), for 70 reactions for production of diagnostic PET radionuclides~\cite{Tarkanyi:2019b}, (lead author F. T. T\'ark\'anyi), and for 15 reactions for production of therapeutic radionuclides~\cite{Engle:2019} (lead author J.W. Engle).
All numerical recommended values with uncertainties and yields for these 144 reactions were added in 2018--2019 to the on-line medical database of the IAEA (prepared by S. Tak\'acs) and are available at the IAEA-NDS medical portal at \href{https://www-nds.iaea.org/medportal/}{\textit{www-nds.iaea.org/medportal/}} and at the IAEA webpage \href{https://www-nds.iaea.org/medical/}{\textit{www-nds.iaea.org/medical/}}.

An additional publication dealing with an update of evaluated nuclear decay data for selected medical radionuclides is in preparation~\cite{Nichols:2020}.

During the preparation of the final reports it appeared that although the database was now more complete, an important number of reactions that had been previously evaluated in IAEA TECDOC-1211~\cite{Gul:2001} and its subsequent updates and in the IAEA Technical Report Series 473 (from now on IAEA TRS-473)~\cite{Betak:2011} had not been updated and hence information on uncertainties were still missing. To remediate this deficiency a limited working party was given the assignment to update the existing data using the same methodology as for the 2012--2017 CRP. This means:
\begin{itemize}
\item Select production routes for re-evaluation for PET and SPECT radionuclides from IAEA-TECDOC-1211~\cite{Gul:2001} (including the updates) and IAEA TRS-473~\cite{Betak:2011}.
\item Undertake a full survey of new or missing literature for experimental data of selected production routes.
\item Correct (if needed) published datasets for up to date monitor cross sections~\cite{Hermanne:2018} or nuclear decay characteristics (see Table 2 for recommended decay data).
\item Select datasets for fitting from all available corrected datasets.
\item Fit with Pad\'e statistical approach and derive new recommended data with uncertainties..
\end{itemize}
More detailed information on the motivation, possible problems or limitations of the chosen approach for selection of experimental data and full discussion of the Pad\'e fitting, including obtaining uncertainties, can be found in the introductory chapters of IAEA TECDOC-1211~\cite{Gul:2001} and of the different publications containing the results of the last CRP~\cite{Hermanne:2018,Tarkanyi:2019,Tarkanyi:2019b,Engle:2019}.

The present publication presents the results for reactions, often multiple, leading to formation of six positron-emitting radionuclides clinically used for PET imaging and for two gamma emitters used in SPECT imaging. A separate section is devoted to each imaging radionuclide containing first a short reminder of the decay data useful for its identification and quantification (taken from Table 2) and also indications on the clinical use and principal molecules labelled.
For each reaction all bibliographic references of the studies containing  experimental cross sections in the relevant energy domain are then listed, indicating which sets were not included in the updates of the IAEA charged particle database~\cite{database:2001} (recent  or  previously missed publications). An overview of the deselected datasets with reason for rejection, possibly different and more detailed than in the publications by Tak\'acs \etal of 2003 and 2005~\cite{Takacs:2003,Takacs:2005} is given.
The parameters (order L of the Pad\'e function, number of included experimental data points N, and $\chi^2$) of the Pad\'e fit on the selected datasets are given together with the energy behaviour of the uncertainties (including a correction for estimated unrecognized systematic uncertainty~\cite{Capote:2020}). For each reaction two figures are shown, the first (a) includes all the data points of identified literature datasets with uncertainty bars as defined by the compilers and a comparison with the theoretical predicted cross section found in the  TENDL-2017~\cite{TENDL} (one case was compared with TENDL-2019~\cite{TENDL19}) library, calculated with a standard input set for TALYS 1.9~\cite{TENDL,TENDL19,TALYS}; the second figure (b) includes only data points (with uncertainties) retained for the Pad\'e fit, the curve of the fit and a histogram showing evaluated uncertainties.

A comparison of decay data of the radionuclides studied as presented in the IAEA-TECDOC-1211~\cite{Gul:2001}, the on-line database update of 2003/2007~\cite{database:2001}, the IAEA TRS-473~\cite{Betak:2011}, ENSDF~\cite{ENSDF}, NUDAT~\cite{NUDAT} and the evaluation of the last CRP usually released through the DDEP project \cite{DDEP} is given in Table 1.  In Table 2 the decay paths, gamma lines and emission probabilities used for measurement; reactions studied with thresholds or Q-values if positive, and the IAEA document containing the  last updates for the studied radionuclides are shown.

%======================================================================
\section{RADIONUCLIDES USED IN POSITRON EMISSION TOMOGRAPHY (PET)} %-- $\beta^+$--EMITTERS}
%======================================================================
\label{sect-PET}

Apart from the four short-lived, low-Z, \ql pure\qr \pos-emitters with widespread use in clinical PET-imaging ($^{11}$C, $^{13}$N, $^{15}$O and $^{18}$F) we included here two longer-lived emerging radionuclides that have complex decay schemes: $^{64}$Cu and $^{124}$I.  These two isotopes were categorized as therapeutic in the TRS-473~\cite{Betak:2011}, but because of their referenced clinical use and their theranostic capabilities (pairs $^{64}$Cu--$^{67}$Cu, $^{124}$I--$^{125,131}$I) we preferred to consider them together with the \pos-emitters.
A particularity for the four low Z,  short-lived isotopes, is that since the last update of the database in 2003 (see also Ref.~\cite{Takacs:2003}) a majority of  the datasets that were not available in tabulated forms in the original publications and were digitized by the compilers for the IAEA TECDOC-1211, were re-digitized and are now included in the EXFOR database. The EXFOR data, often only marginally different from those used in the earlier compilations, are used in this study. It appears however that not always sufficient attention has been paid to the labelling of the reactions mentioned in the original publications:  most of the experimental studies have been done on targets with natural composition from which reaction cross sections on isotopic targets can be deduced, but the adequate conversion was not done or mentioned in the EXFOR data. The needed corrections were made in this study and are explained in the individual discussions. Only a few new sets of experimental data sets are available since 2003.
It has also to be remarked that as the cross sections are often obtained on gas targets or in the frame of physics experiments the measuring procedures are quite different from what is used in stacked foils activation analysis and rely on neutron or charged particle detection. The low Z-value also results in quasi resonances for energies below 20 MeV, especially for proton induced reactions. Special selection of data was done to represent well the resonances (only datasets with high energy-resolution in the concerned energy region). The codes for fitting can analyze simultaneously only 5--6 resonances, because each resonance requires 5-6 parameters and for a total number of parameters larger than 35---40 any fitting becomes unstable (the covariance uncertainty matrices are too large!). Thus, the analysis can only be performed with a piece by a piece approach over limited energy intervals. A detailed example is given in the remark of Sect.~\ref{sssect:N15pnO15} (see footnote in that Section) for fitting the $^{15}$N(p,n)$^{15}$O reaction.

%-----------
\subsection{Production of $^{11}$C \textit{via} $^{14}$N(p,$\alpha$)$^{11}$C reaction}\label{ssect:N14paC11}

\noindent\textbf{Decay data:} \halflife=20.364(14)~min; \newline decay branches: $\epsilon$: 0.2331\%, $\beta^+$: 99.7669\%.\newline
\noindent\textbf{Most abundant gammas:} \newline $E_{\gamma}=511$~keV, $I_{\gamma}=199.534$(5)\%.\newline
\T\noindent\textbf{Applications:} \textit{PET imaging,} [$^{\mathrm{11}}$C]\textit{-methionine for oncology in central nervous system, many other molecules labelled. Addition to the nitrogen gas of up to 2\% of} O$_2$ \textit{results in } [$^{\mathrm{11}}$C]O$_2$; \textit{5-–10\% of }H$_2$ \textit{results in } [$^{\mathrm{11}}$C]H$_4$.\newline

Evaluation has been made of the $^{14}$N(p,$\alpha$)$^{11}$C reaction.

\begin{figure}[!thb]
%\vspace{-1mm}
\centering
\subfigure[~All experimental data are plotted with uncertainties and compared to TENDL-2017 evaluation \cite{TENDL}.]
{\includegraphics[width=\columnwidth]{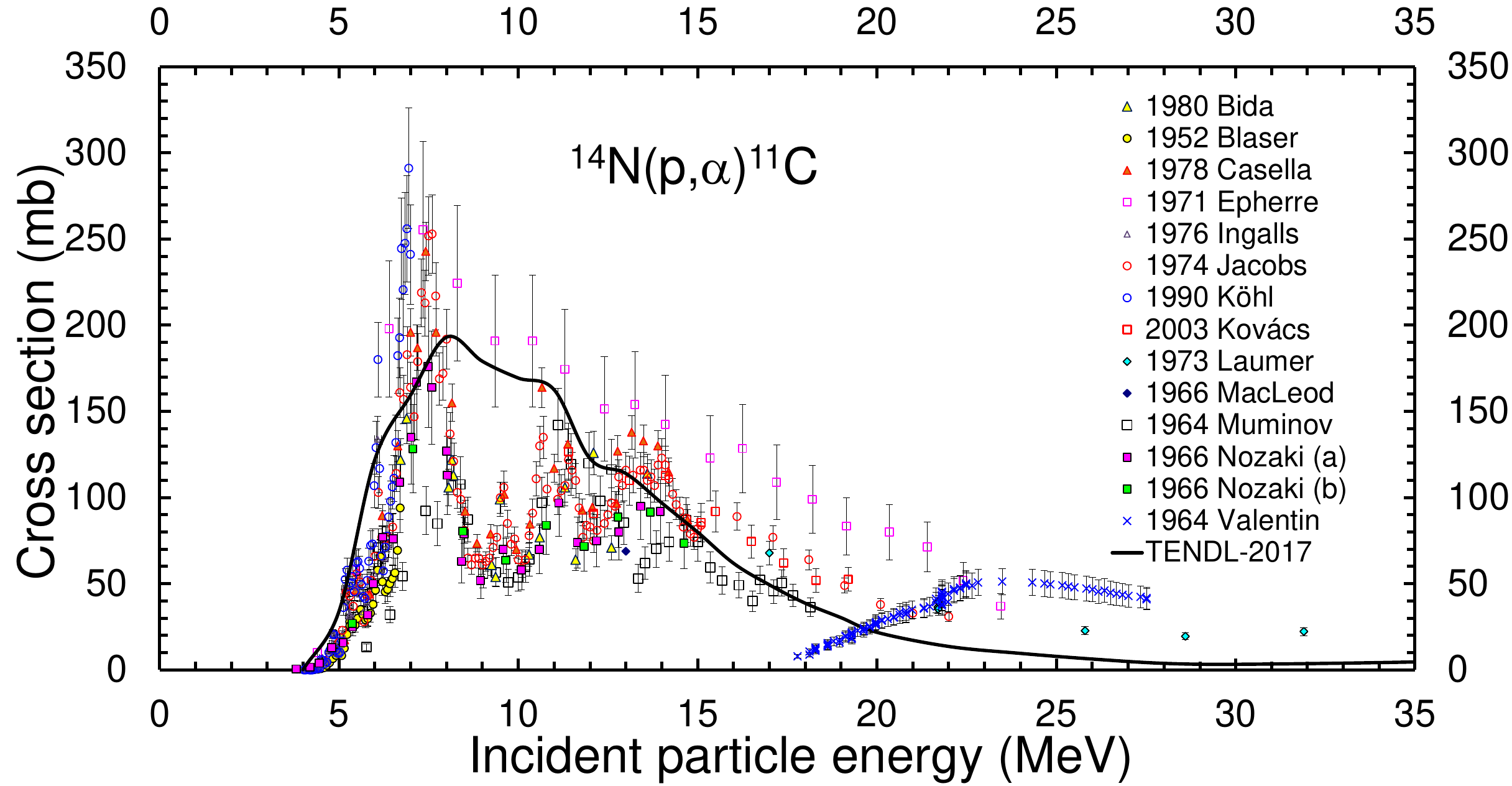}}
\subfigure[~Selected data compared with evaluated Pad\'e fit (L = 75, N = 306, $\chi^2$=3.61, solid line) and estimated total uncertainty in percentage including
a 4\% systematic uncertainty (dashed line, right-hand scale).]
{\includegraphics[width=\columnwidth]{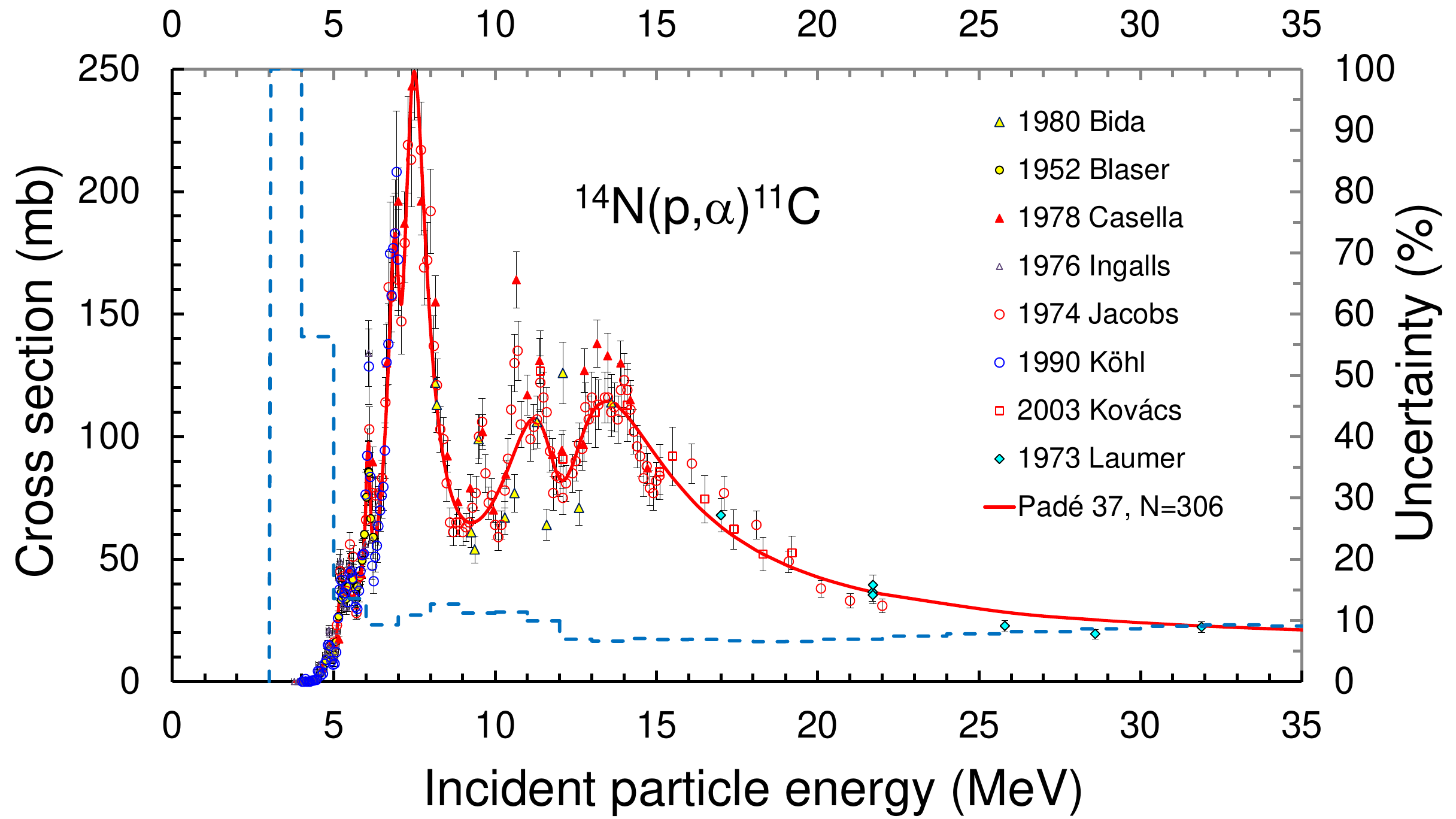}}
\centering
%\vspace{-2mm}
\caption{(Color online) Evaluated Pad\'e fit and experimental data from
Refs.~\cite{Bida:1980, Blaser:1952, Casella:1978, Epherre:1971, Ingalls:1976, Jacobs:1974, Kohl:1990, Kovacs:2003, Laumer:1973, MacLeod:1966, Muminov:1974, Nozaki:1966, Valentin:1964}
for the $^{14}$N(p,$\alpha$)$^{11}$C reaction.}
\label{fig1:n14pa-c11}
%\vspace{-2mm}
\end{figure}

For the formation of short-lived $^{11}$C, a grand total of thirteen publications~\cite{Bida:1980,Blaser:1952,Casella:1978,Epherre:1971,Ingalls:1976,Jacobs:1974,Kohl:1990,Kovacs:2003,Laumer:1973,MacLeod:1966,Muminov:1974,Nozaki:1966,Valentin:1964} with experimental cross-section data were identified in the literature in the considered energy range and are represented with uncertainties in Fig.~\ref{fig1:n14pa-c11}(a). An additional publication by Vdovin \etal(1979)~\cite{Vdovin:1979} contains only a single data point at 50 MeV and is not represented in the figure. For the publication by Nozaki \etal(1966)~\cite{Nozaki:1966} two datasets are identified in EXFOR and are indicated as (a) and (b) in the figure. No new references were added after the previous update of the IAEA on-line charged particle database~\cite{database:2001} in 2003. For the publication by Laumer \etal\cite{Laumer:1973} only 1 data point is mentioned in EXFOR but we conserved the 10 data points shown in TECDOC-1211 for this set obtained by its compiler. In order to have better agreement with the other datasets the values of K\"ohl \etal(1990)~\cite{Kohl:1990} were arbitrarily divided by a factor of 1.4 and the values of Blaser \etal(1952)~\cite{Blaser:1952} uniformly shifted by -0.7 MeV and multiplied by a factor of 1.3. The results of 5 studies were deselected for the further analysis; the reason for deselecting is indicated in parentheses: Epherre and Seide (1971)(thick targets, averaged cross sections)~\cite{Epherre:1971}, MacLeod and Reid (1966) (1 data point, low, detection through cloud chamber)~\cite{MacLeod:1966}, Muminov \etal(1964) (values too low and deviant shape, old monitor values)~\cite{Muminov:1974}, Nozaki et al.(1966) series (a) and (b) (values do not represent well the resonances)~\cite{Nozaki:1966}, Valentin et al.(1994) (discrepant values, only higher energy data)~\cite{Valentin:1964}. The datasets of the remaining 8 papers from Refs.~\cite{Bida:1980,Blaser:1952,Casella:1978,Ingalls:1976,Jacobs:1974,Kohl:1990,Kovacs:2003,Laumer:1973} were used as input for a least-squares Pad\'e fit.  To allow better fitting of the resonances at low energy several data points of Bida \etal(1980)~\cite{Bida:1980} were removed (4 low energy points and a point at 12.2 MeV removed).		

The Pad\'e functions with 75 parameters were fitted to 306 selected data points with a $\chi^2=3.61$ and covering the energy range up to 35 MeV as shown in Fig.~\ref{fig1:n14pa-c11}(b). The uncertainties (including 4\% systematic uncertainty) range between 60\% around 5 MeV, decrease to below 8\% between 13 and 28 MeV and rise again to 10\% at the highest energy.

%-----------
\subsection{Production of $^{13}$N \textit{via} $^{16}$O(p,$\alpha$)$^{13}$N reaction}\label{ssect:O16paN13}

\noindent\textbf{Decay data:} \halflife=9.965(4)~min; \newline decay branches: $\epsilon$: 0.2\%, $\beta^+$: 99.8036\%.\newline
\noindent\textbf{Most abundant gammas:} \newline $E_{\gamma}=511$~keV, $I_{\gamma}=199.607$(4)\%.\newline
\T\noindent\textbf{Applications:} $^{\mathrm{13}}$N \textit{is used to tag ammonia molecules for PET imaging of the myocardium under stress or rest conditions to assess myocardial blood flow.}\newline

Evaluation has been made of the $^{16}$O(p,$\alpha$)$^{13}$N reaction.

\begin{figure}[!thb]
\vspace{-1mm}
\centering
\subfigure[~All experimental data are plotted with uncertainties and compared to TENDL-2017 evaluation \cite{TENDL}.]
{\includegraphics[width=0.95\columnwidth]{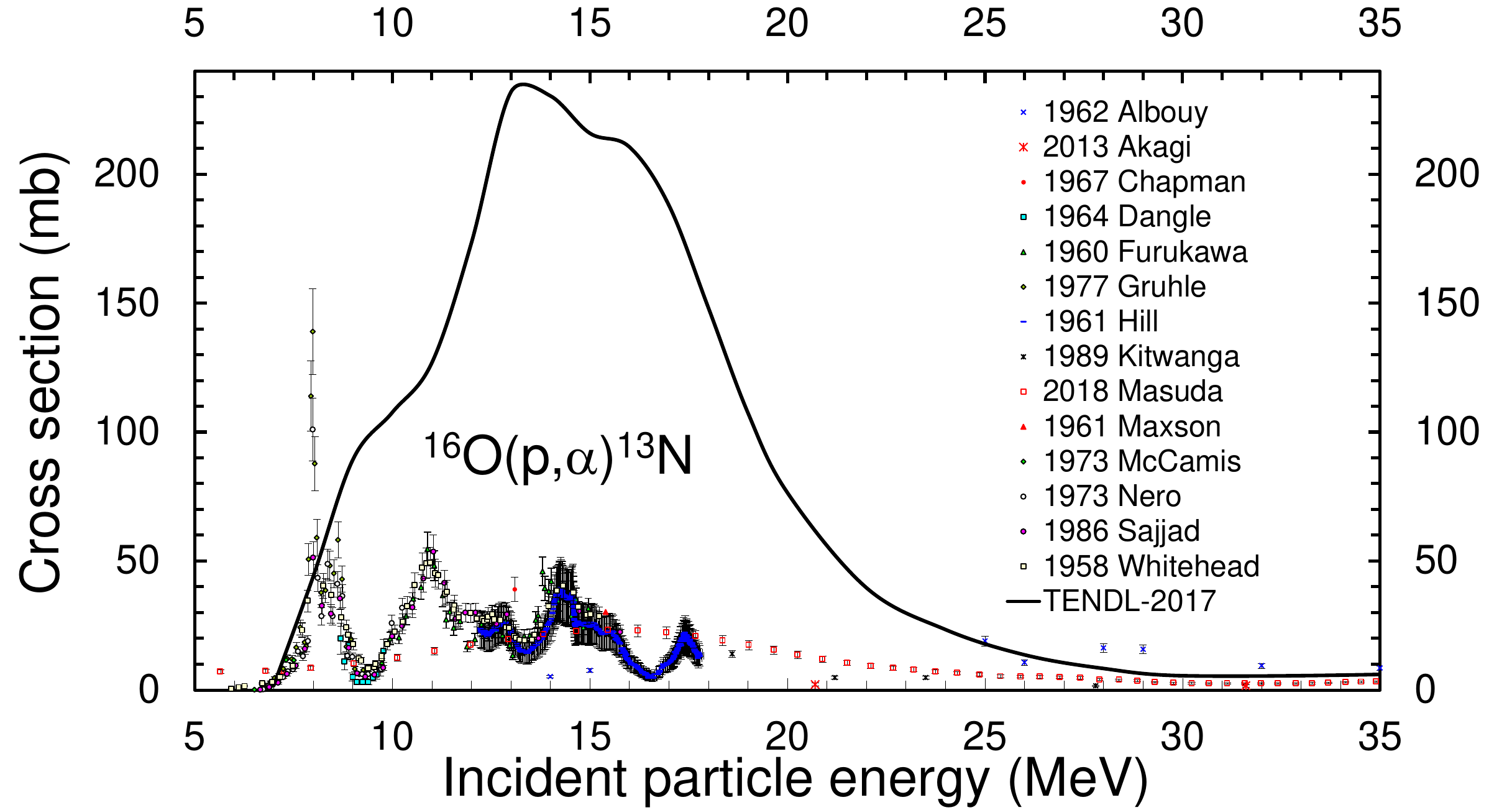}}
\subfigure[~Selected data compared with evaluated Pad\'e fit (L = 40, N = 607, $\chi^2$=1.96, solid line) and estimated total uncertainty in percentage including
a 4\% systematic uncertainty (dashed line, right-hand scale).]
{\includegraphics[width=0.95\columnwidth]{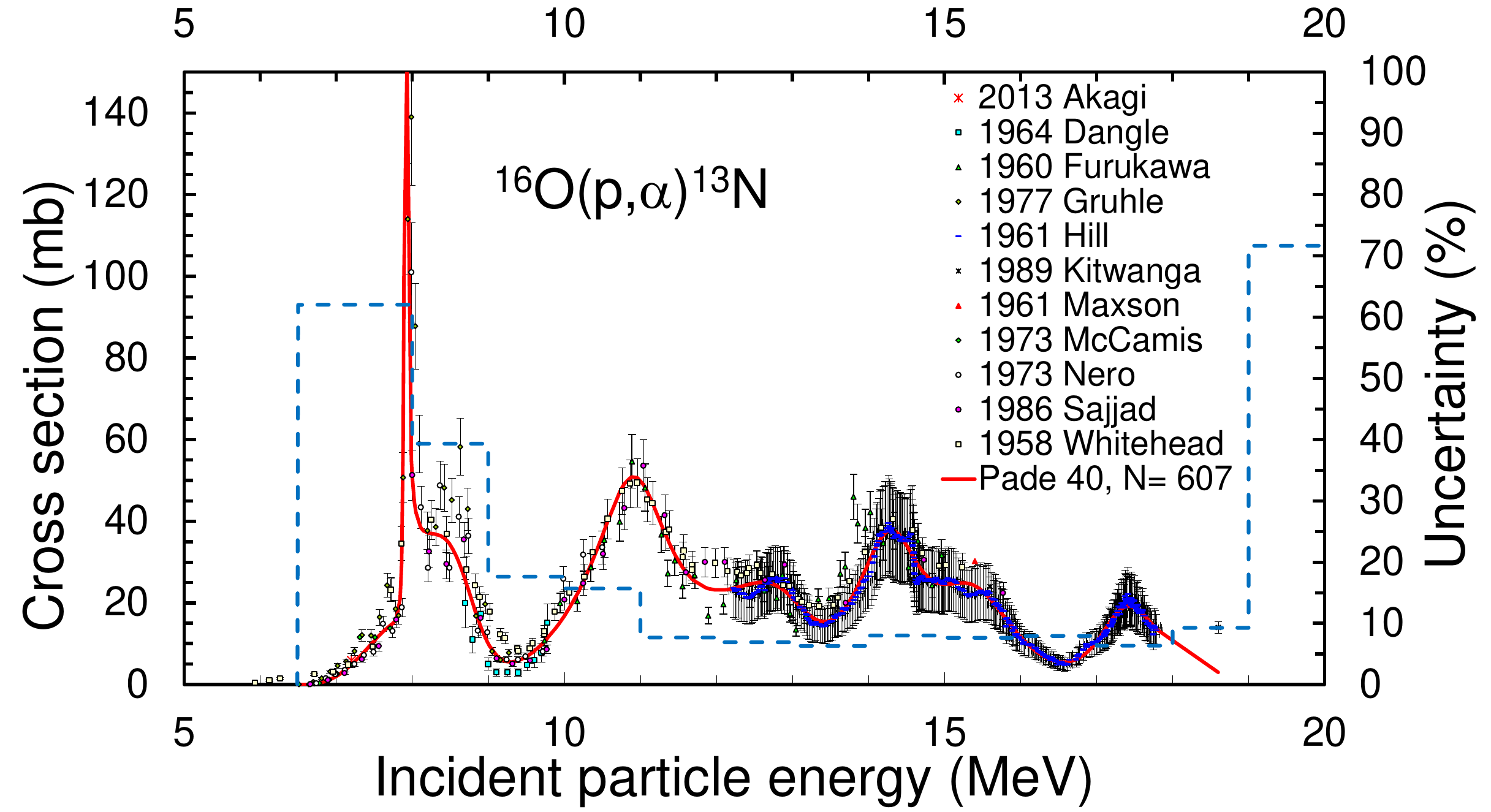}}
\centering
\vspace{-2mm}
\caption{(Color online) Evaluated Pad\'e fit and experimental data from
Refs.~\cite{Akagi:2013, Albouy:1962, Chapman:1967, Dangle:1964, Furukawa:1960, Gruhle:1977, Hill:1961, Kitwanga:1989, Masuda:2018, Maxson:1961, McCamis:1973, Nero:1973, Sajjad:1986, Whitehead:1958} for the $^{16}$O(p,$\alpha$)$^{13}$N reaction.}
\label{fig2:o16pa-n13}
\vspace{-6mm}
\end{figure}

For formation of short-lived $^{13}$N, a grand total of fourteen publications~\cite{Akagi:2013, Albouy:1962, Chapman:1967, Dangle:1964, Furukawa:1960, Gruhle:1977, Hill:1961, Kitwanga:1989, Masuda:2018, Maxson:1961, McCamis:1973, Nero:1973, Sajjad:1986, Whitehead:1958} with experimental cross-section data were identified in the literature for incident particle energies up to 35 MeV  and are represented with uncertainties in Fig.~\ref{fig2:o16pa-n13}(a). Three additional datasets containing each only one data point at higher energy are not shown: Gambarini \etal(1969)~\cite{Gambarini:1969}, Valentin \etal(1963)~\cite{Valentin:1963}, and  Vdovin \etal(1979)\cite{Vdovin:1979} as in the TECDOC-1211 evaluation. Two new datasets were added after the previous update of the IAEA on-line charged particle database~\cite{database:2001}: Akagi \etal(2013)~\cite{Akagi:2013} and Masuda \etal(2018)~\cite{Masuda:2018}. The data taken from the original publication of Whitehead and Foster (1958)~\cite{Whitehead:1958} were shifted uniformly to lower energy by 0.3 MeV and the original data of Furukawa et al.~(1960)~\cite{Furukawa:1960} multiplied by a factor of 0.7 to have better agreement with other datasets. The EXFOR data are used for Refs.~\cite{Akagi:2013, Albouy:1962, Chapman:1967, Gruhle:1977, Hill:1961, Kitwanga:1989, Masuda:2018, Maxson:1961, McCamis:1973, Sajjad:1986}. No EXFOR data exist for Dangle \etal(1964)~\cite{Dangle:1964} and Nero and Howard~(1973)~\cite{Nero:1973}.
The results of 3 studies were rejected and not considered for further analysis, and the reasons for their removal are indicated: Albouy \etal(1962) (too low values in the 14--15 MeV region, other data points above 25 MeV)~\cite{Albouy:1962}, Masuda \etal(2018) (no structure, too low values)~\cite{Masuda:2018}, Chapman and MacLeod; (1967) (single point, too high)~\cite{Chapman:1967}. The remaining eleven datasets from Refs.~\cite{Akagi:2013, Dangle:1964, Furukawa:1960, Gruhle:1977, Hill:1961, Kitwanga:1989, Maxson:1961, McCamis:1973, Nero:1973, Sajjad:1986, Whitehead:1958} were used as input for a least-squares Pad\'e fit.

The Pad\'e functions with 40 parameters were fitted to 607 selected data points with a $\chi^2$=1.96 and covering the energy range up to 20 MeV as shown in Fig.~\ref{fig2:o16pa-n13}(b). The uncertainties (including 4\% systematic uncertainty) range between 65\% near the reaction threshold, decrease to below 7\% between 11 and 18 MeV and then monotonically increase to reach 12\% at the highest energy.

%-----------
\subsection{Production of $^{15}$O}\label{ssect-O15}

\noindent\textbf{Decay data:} \halflife=122.24(16)~s; \newline decay branches: $\epsilon$: 0.1\%, $\beta^+$: 99.9003\%.\newline
\noindent\textbf{Most abundant gammas:} \newline $E_{\gamma}=511$~keV, $I_{\gamma}=199.8006$(20)\%.\newline
\T\noindent\textbf{Applications:} $^{15}$O \textit{has been used for cerebral and myocardial blood flow studies (labelling of the oxygen in water:} H$_2$[$^{\mathrm{15}}$O]) \textit{or to study the oxygen metabolism (labelled }[$^{\mathrm{15}}$O]$_2$ \textit{or }C[$^{\mathrm{15}}$O]\textit{)}.\newline

Evaluation has been made of the $^{15}$N(p,n)$^{15}$O and  $^{14}$N(d,n)$^{15}$O reactions.

\subsubsection{~~$^{15}\mathrm{N(p,n)}^{15}\mathrm{O}$ reaction}\label{sssect:N15pnO15}
\begin{figure}[t]
\vspace{-3mm}
\centering
\subfigure[~All experimental data are plotted with uncertainties and compared to TENDL-2017 evaluation \cite{TENDL}.]
{\includegraphics[width=0.95\columnwidth]{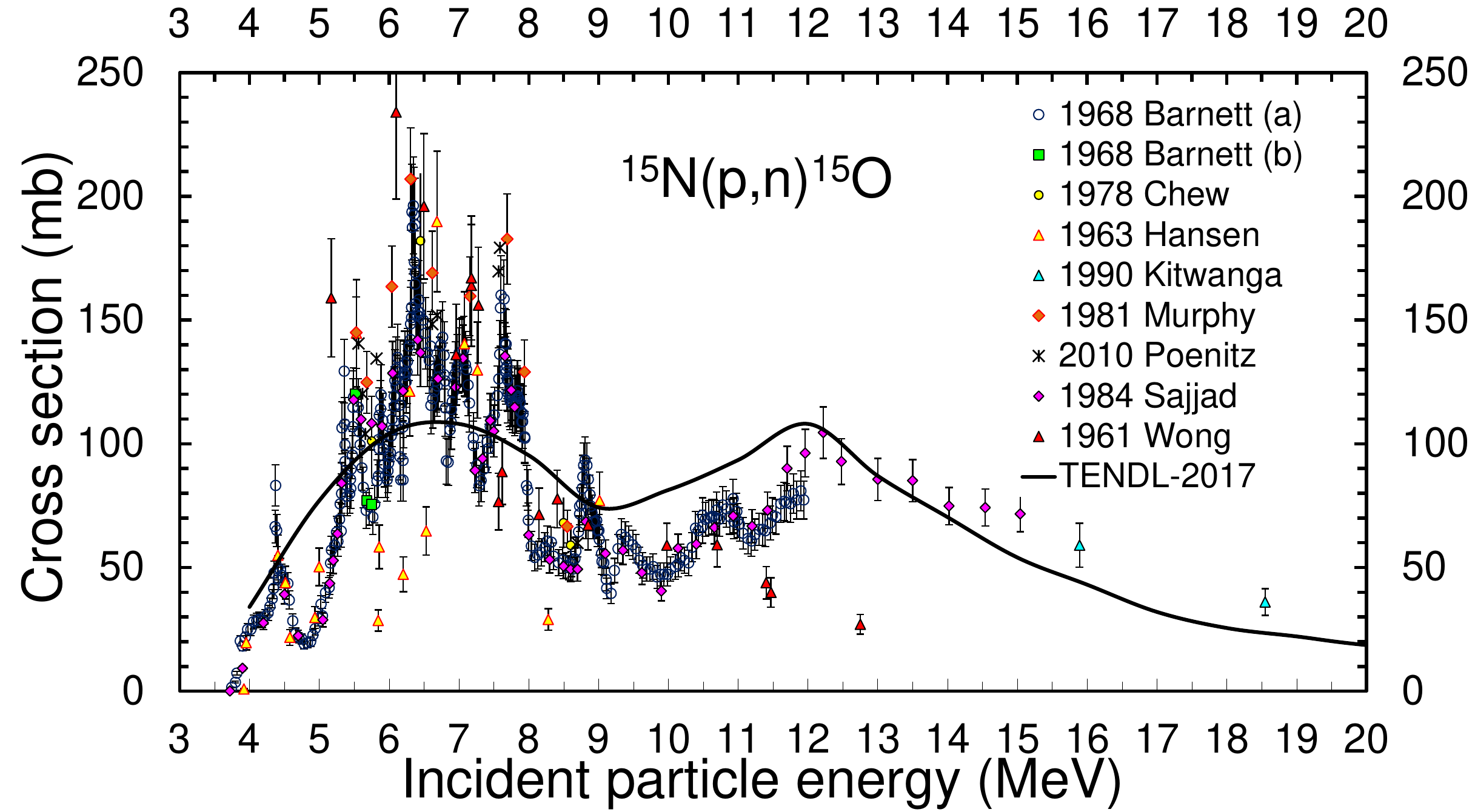}}
\subfigure[~Selected data compared with evaluated Pad\'e fit (L = 93, N = 389, $\chi^2$=2.4, solid line) and estimated total uncertainty in percentage including
a 4\% systematic uncertainty (dashed line, right-hand scale).]
{\includegraphics[width=0.95\columnwidth]{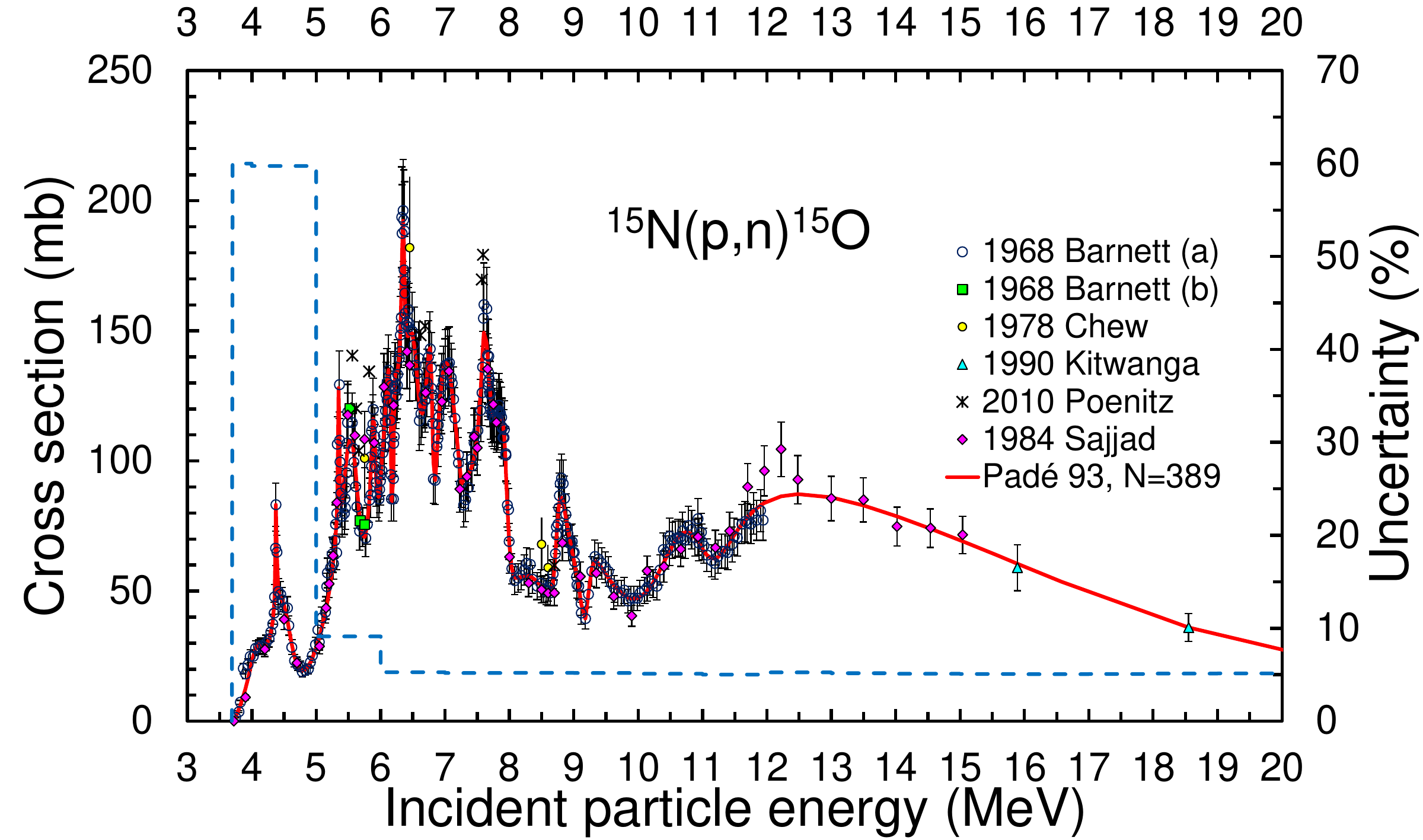}}
\centering
\vspace{-2mm}
\caption{(Color online) Evaluated Pad\'e fit and experimental data from
Refs.~\cite{Barnett:1968, Chew:1978, Hansen:1963, Kitwanga:1990, Murphy:1981, Poenitz:2010, Sajjad:1984, Wong:1961} for the $^{15}$N(p,n)$^{15}$O reaction.}
\label{fig3:n15pn-o15}
\vspace{-4mm}
\end{figure}

For formation of short-lived $^{15}$O  through a (p,n) reaction on low abundance $^{15}$N (0.366\%), a grand total of eight publications~\cite{Barnett:1968, Chew:1978, Hansen:1963, Kitwanga:1990, Murphy:1981, Poenitz:2010, Sajjad:1984,  Wong:1961} with experimental cross-section data were identified in the literature for incident particle energies up to 20 MeV  and are represented with uncertainties in Fig.~\ref{fig3:n15pn-o15}(a). One new dataset was added after the previous update of the IAEA on-line charged particle database ~\cite{database:2001} in 2003: Poenitz (2010) \cite{Poenitz:2010}. The original publication of Barnett (1968) ~\cite{Barnett:1968} contains  a large number of relative measurements using $\beta^+$ counting (marked as (a)) and 3 absolute measurements around 5.5 MeV (b). The data set Barnett (a) was normalised to these 3 data points and are as such in EXFOR. As was already remarked in ~\cite{Gul:2001} all values were divided by two (photon multiplicity from positron annihilation) for use in the evaluation.
No EXFOR data exist for Hansen and Stelts (1963) \cite{Hansen:1963} and for Chew \etal(1978)~\cite{Chew:1978}.
The results of 3 studies were rejected and not considered for further analysis, and the reasons for their removal are indicated: Hansen and Stelts ~\cite{Hansen:1963} (too low near maximum), Murphy \etal(1981) (disagreement in energy)~\cite{Murphy:1981}, Wong \etal(1961) (discrepant values)~\cite{Wong:1961}.

The remaining six datasets of Refs.~\cite{Barnett:1968, Chew:1978, Kitwanga:1990, Poenitz:2010, Sajjad:1984} were used as input for a least-squares Pad\'e fit. 			
The Pad\'e functions with 93 parameters were fitted to 389 selected data points with a $\chi^2$=2.40 and covering the energy range up to 20 MeV as shown in Fig.~\ref{fig3:n15pn-o15}(b)\footnote{The fit over all points (N= 389) and the whole energy region with L=93 and $\chi^2$=2.40 is in fact an assembly of 3 fits over limited energy regions, made continuous at the respective end-points. The partial fits are: from 2.518 MeV to 3.563 MeV Pad\'e functions with L=27, N=106, $\chi^2$=1.25; from 3.563 MeV to 5.836 MeV Pad\'e functions with L=36, N=189, $\chi^2$=0.56; from 5.836 MeV up to 30 MeV Pad\'e functions with L=30, N=189, $\chi^2$=0.59.}. The uncertainties (including a 4\% systematic uncertainty) are very high near the reaction threshold, reach 5.5\% around 6 MeV and remain at that figure over the whole studied energy range.

\subsubsection{~~$^{\mathrm{14}}\mathrm{N(d,n)}^{\mathrm{15}}\mathrm{O}$ reaction}\label{sssect:N14dnO15}
\begin{figure}[!thb]
%\vspace{-2mm}
\centering
\subfigure[~All experimental data are plotted with uncertainties and compared to TENDL-2017 evaluation \cite{TENDL}.]
{\includegraphics[width=0.99\columnwidth]{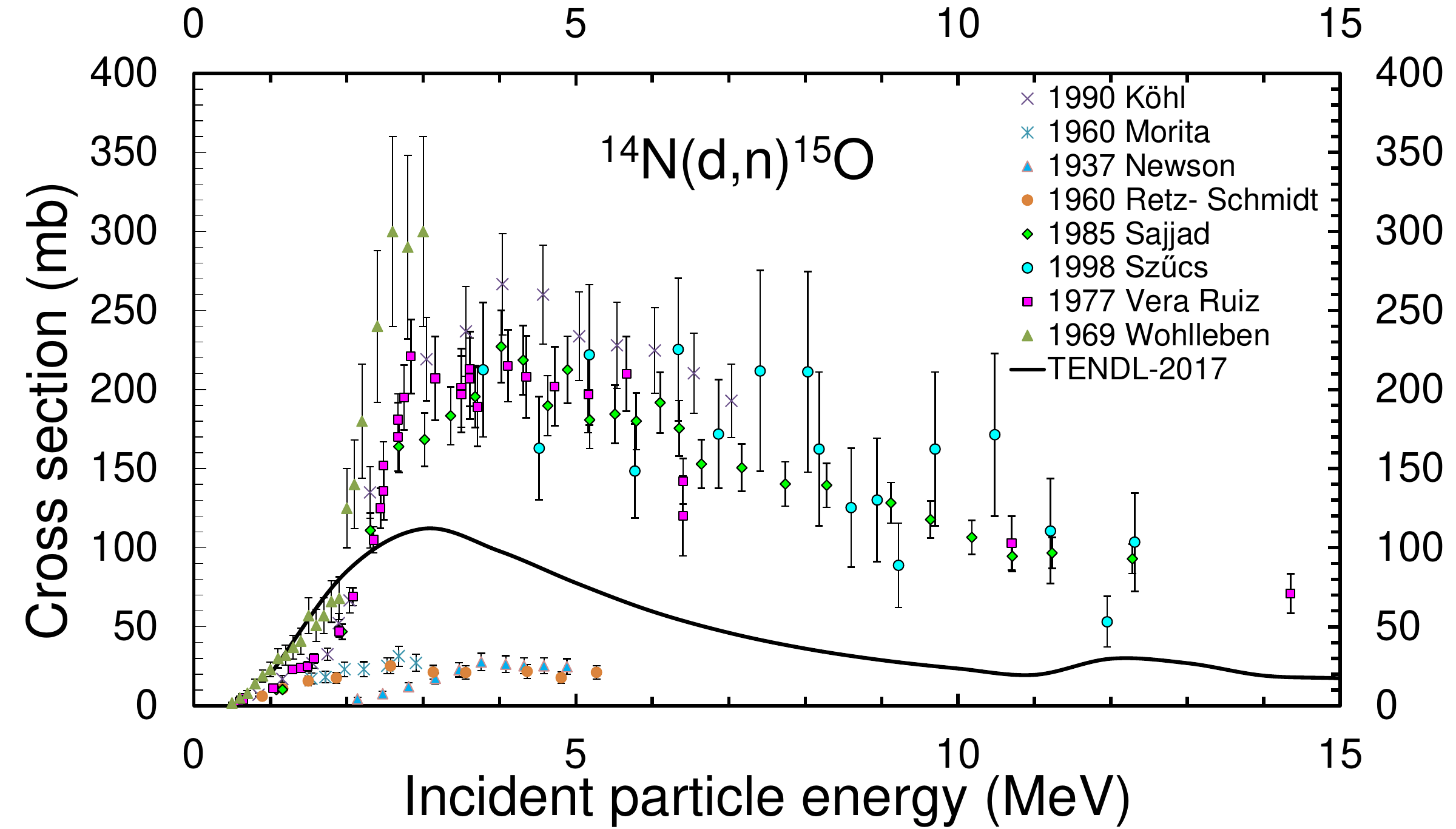}}
\subfigure[~Selected data compared with evaluated Pad\'e fit (L = 8, N = 93, $\chi^2$=1.22, solid line) and estimated total uncertainty in percentage including
a 4\% systematic uncertainty (dashed line, right-hand scale).]
{\includegraphics[width=0.99\columnwidth]{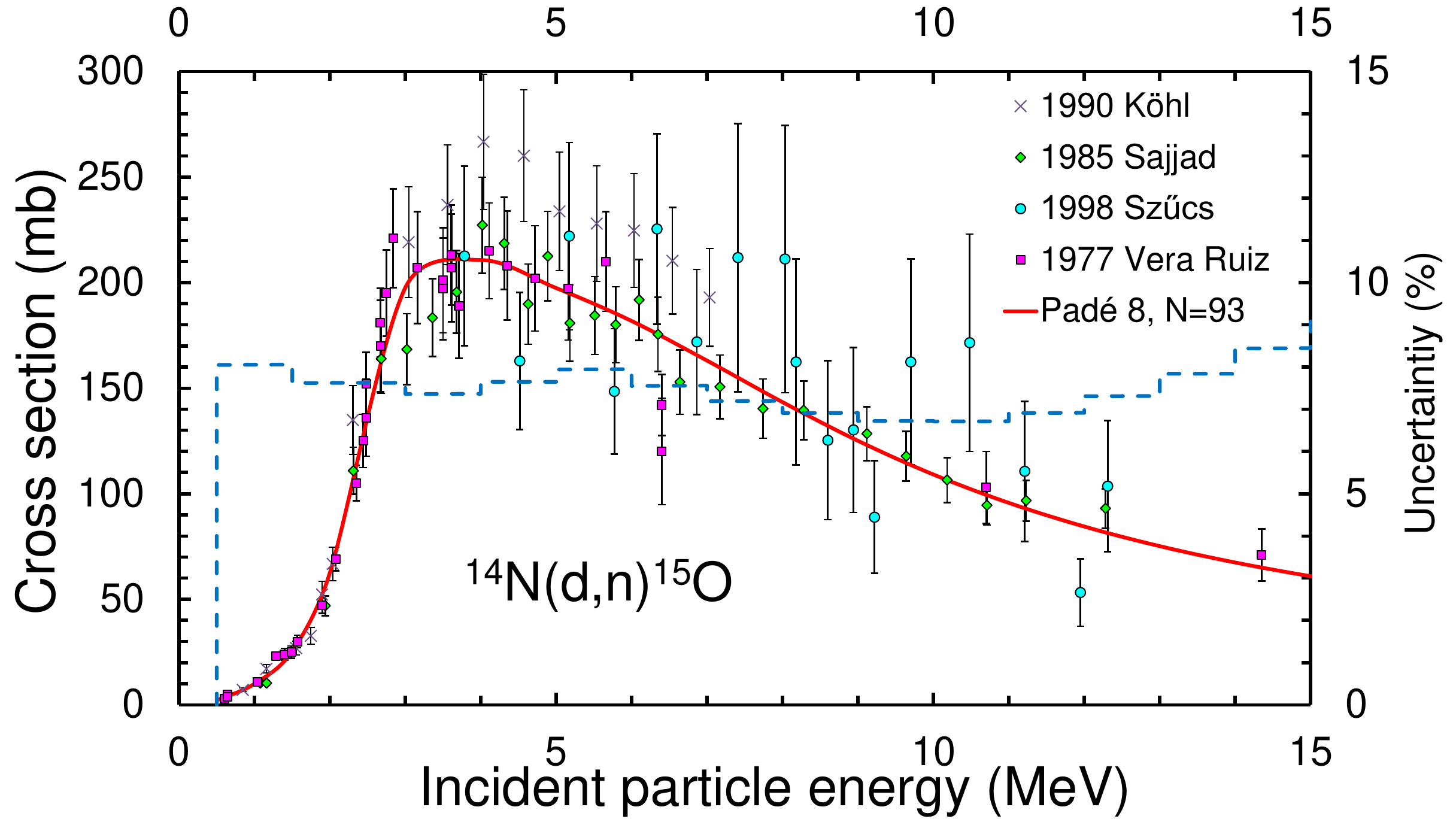}}
\centering
\vspace{-2mm}
\caption{(Color online) Evaluated Pad\'e fit and experimental data from
Refs.~\cite{Kohl:1990, Morita:1960, Newson:1937, Retz:1960, Sajjad:1985, Szucs:1998, Vera Ruiz:1977, Wohlleben:1969} for the $^{14}$N(d,n)$^{15}$O reaction.}
\label{fig4:n14dn-o15}
%\vspace{-2mm}
\end{figure}

For formation of short-lived $^{15}$O through the $^{14}$N(d,n)$^{15}$O reaction a grand total of eight publications from Refs.~\cite{Kohl:1990, Morita:1960, Newson:1937, Retz:1960, Sajjad:1985, Szucs:1998, Vera Ruiz:1977,  Wohlleben:1969} with experimental cross-section data were identified in the literature for incident particle energies up to 15 MeV  and are represented with uncertainties in Fig.~\ref{fig4:n14dn-o15}(a).

Two additional references are not represented in the figure: Nonaka \etal(1957)~\cite{Nonaka:1957} contains only one data point for a partial cross section at 1.92 MeV while in Marion \etal(1955)~\cite{Marion:1955} only neutrons emitted in forward direction were measured and the derived cross sections (by normalization of lower energy values to Vera Ruiz \etal(1977)~\cite{Vera Ruiz:1977}) are extremely high.
No new data were made available after the previous update of the IAEA on-line charged particle database ~\cite{database:2001} in 2003.
The results of 4 studies were rejected and not considered for further analysis, and the reasons for their removal are indicated:  Morita \etal(1960) (measurement of differential cross sections, discrepant integrated data, five times too low)~\cite{Morita:1960}, Newson (1937) (measurement of differential cross sections, discrepant data, orders of magnitude too low)~\cite{Newson:1937},  Retz-Schmidt  and Weil (1960) (measurement of differential cross sections discrepant data, five times too low)~\cite{Retz:1960}, Wohlleben and Schuster (1969) (values near maximum are too high, uncertainties over 20\%)~\cite{Wohlleben:1969}.

The remaining four datasets from Refs. ~\cite{Kohl:1990, Sajjad:1985, Szucs:1998, Vera Ruiz:1977} were used as input for a least-squares Pad\'e fit. 			
The Pad\'e functions with 8 parameters were fitted to 93 selected data points with a $\chi^2$=1.22 and covering the energy range up to 15 MeV as shown in Fig.~\ref{fig4:n14dn-o15}(b). The uncertainties (including 4\% systematic uncertainty) vary between 6 and 8\% over the whole energy region.

%-----------
\subsection{Production of $^{18}$F}\label{ssect-F18}

\noindent\textbf{Decay data:} \halflife=109.77(5)~min; \newline decay branches: $\epsilon$: 3.27\%, $\beta^+$: 96.73\%.\newline
\noindent\textbf{Most abundant gammas:} \newline $E_{\gamma}=511$~keV, $I_{\gamma}=193.46$(8)\%.\newline
\T\noindent\textbf{Applications:} $^{\mathrm{18}}$F \textit{labelled fluorodeoxyglucose }([$^{\mathrm{18}}$F]--FDG) \textit{is the most commonly used PET imaging compound. Increased FDG uptake occurs in cancer cells, but also with infection and inflammation due to the activation of granulocytes and macrophages. Therefore, }[$^{\mathrm{18}}$F]--FDG \textit{is used most commonly for tumor, cardiac, and brain imaging, and is increasingly being used to detect infection.}\newline

Evaluation has been made of the $^{18}$O(p,n)$^{18}$F and ${^\mathrm{nat}}$Ne(d,x)$^{18}$F reactions. The reactions on liquid targets are used for nucleophilic labelling, on gas targets for electrophilic substitution.

\subsubsection{~~$^{\mathrm{18}}\mathrm{O(p,n)}^{\mathrm{18}}\mathrm{F}$ reaction}\label{sssect:O18pnF18}
\begin{figure}[!thb]
%\vspace{-2mm}
\centering
\subfigure[~All experimental data are plotted with uncertainties and compared to TENDL-2017 evaluation \cite{TENDL}.]
{\includegraphics[width=\columnwidth]{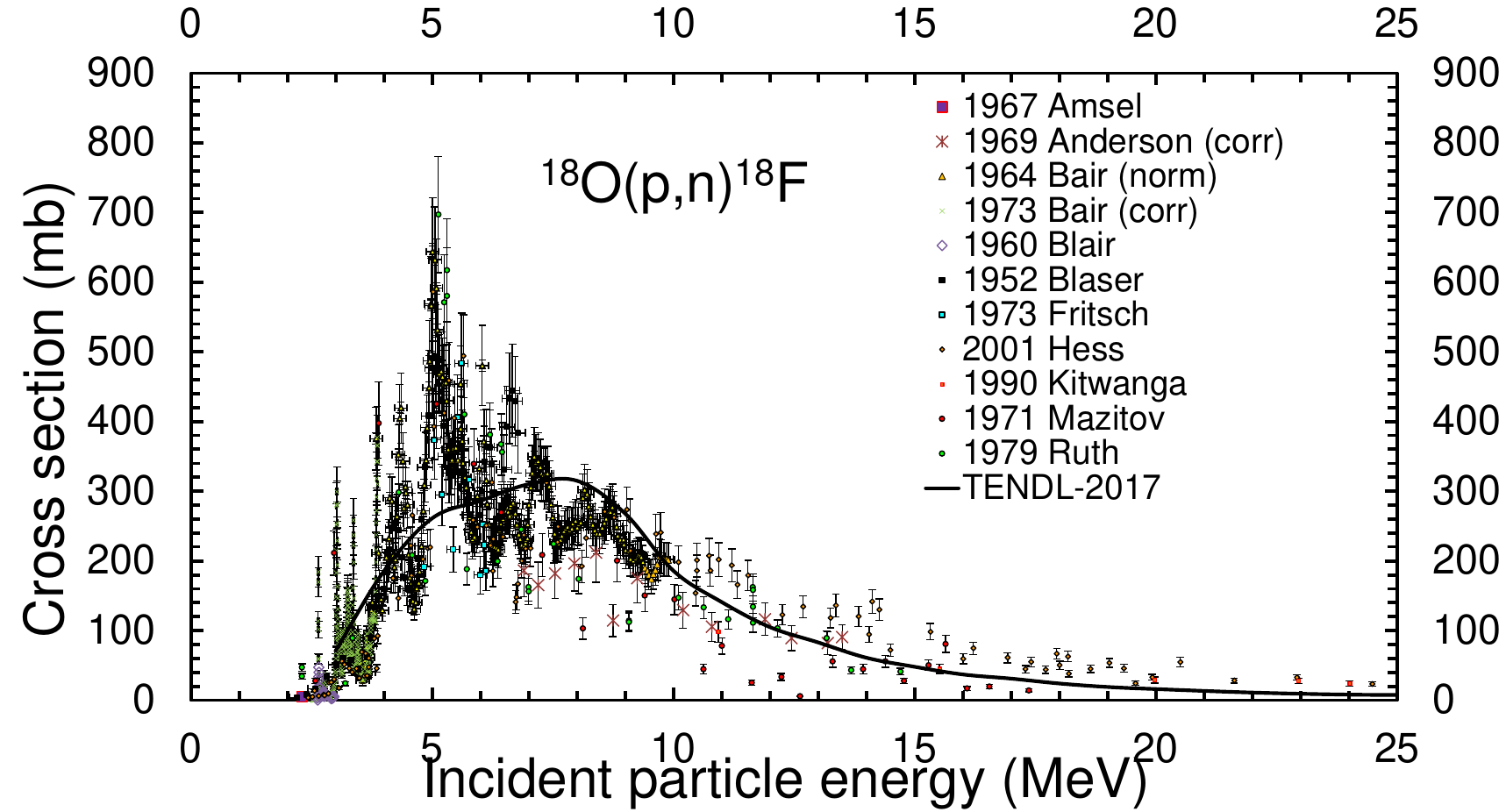}}
\subfigure[~Selected data compared with evaluated Pad\'e fit (L = 98, N = 534, $\chi^2$=4.6, solid line) and estimated total uncertainty in percentage including
a 4\% systematic uncertainty (dashed line, right-hand scale).]
{\includegraphics[width=\columnwidth]{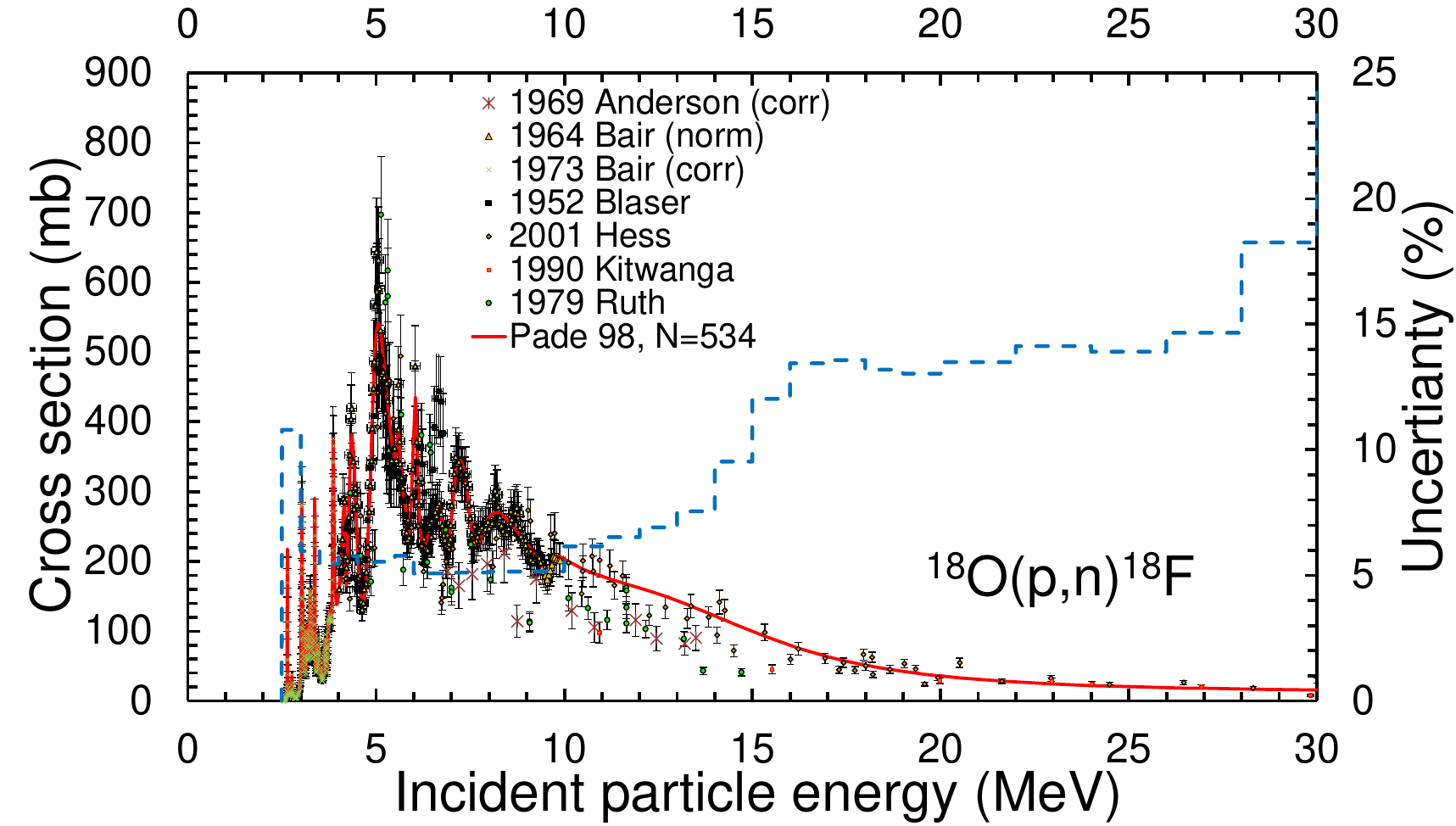}}
\centering
%\vspace{-2mm}
\caption{(Color online) Evaluated Pad\'e fit and experimental data from
Refs.~\cite{Amsel:1967, Anderson:1969, Bair:1964, Bair:1973, Bair:1981, Blair:1960, Blaser:1952a, Fritsch:1973, Hess:2001, Kitwanga:1990, Mazitov:1971, Ruth:1979} for the $^{18}$O(p,n)$^{18}$F reaction.}
\label{fig5:o18pn-f18}
%\vspace{-2mm}
\end{figure}

For formation of $^{18}$F through the commercially used $^{18}$O(p,n)$^{18}$F reaction, a grand total of eleven publications~\cite{Amsel:1967, Anderson:1969, Bair:1964, Bair:1973, Bair:1981, Blair:1960, Blaser:1952a, Fritsch:1973, Hess:2001, Kitwanga:1990, Mazitov:1971, Ruth:1979} with experimental cross-section data were identified in the literature for incident particle energies up to 25 MeV  and are represented with uncertainties in Fig.~\ref{fig5:o18pn-f18}(a). No new datasets were added after the previous update of the IAEA on-line charged particle database~\cite{database:2001} in 2003. The large number of data presented by Bair \etal(1964) are relative measurements~\cite{Bair:1964} and were normalised to the absolute values published in Bair (1973)~\cite{Bair:1973} and corrected according to recommendations in another article of Bair \etal(1981) (multiplication by 1.35 and uncertainties of 7\%)~\cite{Bair:1981}. The values of Anderson \etal(1969)~\cite{Anderson:1969} are summed partial cross sections (from different levels) and are multiplied by 1.2 as proposed by Bair \etal(1981)~\cite{Bair:1981}.

The EXFOR data are used for Blaser \etal(1952),  Hess \etal(2001), Kitwanga \etal(1990); Ruth and Wolf (1979) from Refs.~\cite{Blaser:1952a, Hess:2001, Kitwanga:1990, Ruth:1979}, respectively. No EXFOR entries exist for Mazitov \etal(1971)~\cite{Mazitov:1971} and Fritsch \etal(1973)~\cite{Fritsch:1973}.
The results of 3 studies were rejected and not considered for further analysis, and the reasons for their removal are indicated: Blair  and Leigh (1960) (only data points lower than 3~MeV, have resonance structure, are lower than Bair \etal 1973~\cite{Bair:1973}, has energy gaps, seem energy shifted)~\cite{Blair:1960}, Fritsch \etal(1973) (energy shifted)~\cite{Fritsch:1973}, Mazitov \etal(1971) (all data points are too low)~\cite{Mazitov:1971}.

The remaining eight datasets~\cite{Amsel:1967, Anderson:1969, Bair:1964, Bair:1973, Bair:1981, Blaser:1952a, Hess:2001, Kitwanga:1990, Ruth:1979} were considered  as possible input for a least-squares Pad\'e fit.  In order to make a reasonable fit of the many pronounced resonances possible a further selection of datasets was made. For energy below 4 MeV only the data points of Bair (1973) ~\cite{Bair:1973} were retained. The single point of Amsel (1967)~\cite{Amsel:1967} cannot be checked and the data of Hess \etal(2001)~\cite{Hess:2001}, Bair \etal(1964)~\cite{Bair:1964},  Ruth and Wolf (1979)~\cite{Ruth:1979} have lower energy resolution in this domain and hinder a good fitting. On the other hand data from Bair (1973)~\cite{Bair:1973} and of Blaser \etal(1952)~\cite{Blaser:1952a} (problem with resonance around 6.6 MeV, data not selected in 2003) were not considered for fitting above 4 MeV.			

The Pad\'e functions with 98 parameters were fitted to 534 selected data points with a $\chi^2$=4.6 and covering the energy range up to 30~MeV as shown in Fig.~\ref{fig5:o18pn-f18}(b).
The uncertainties (including a 4\% systematic uncertainty) and averaged over the resonances, range between 25\% near the reaction threshold, decrease to below 6\% between 3.5 and 11 MeV and increase slowly to about 20\% at the highest energy.

\subsubsection{~~$^{\mathrm{nat}}\mathrm{Ne(d,x)}^{\mathrm{18}}\mathrm{F}$ reaction}\label{sssect:NedxF18}
\begin{figure}[!thb]
\vspace{-2mm}
\centering
\subfigure[~All experimental data are plotted with uncertainties and compared to TENDL-2019 evaluation~\cite{TENDL19}.]
{\includegraphics[width=0.97\columnwidth]{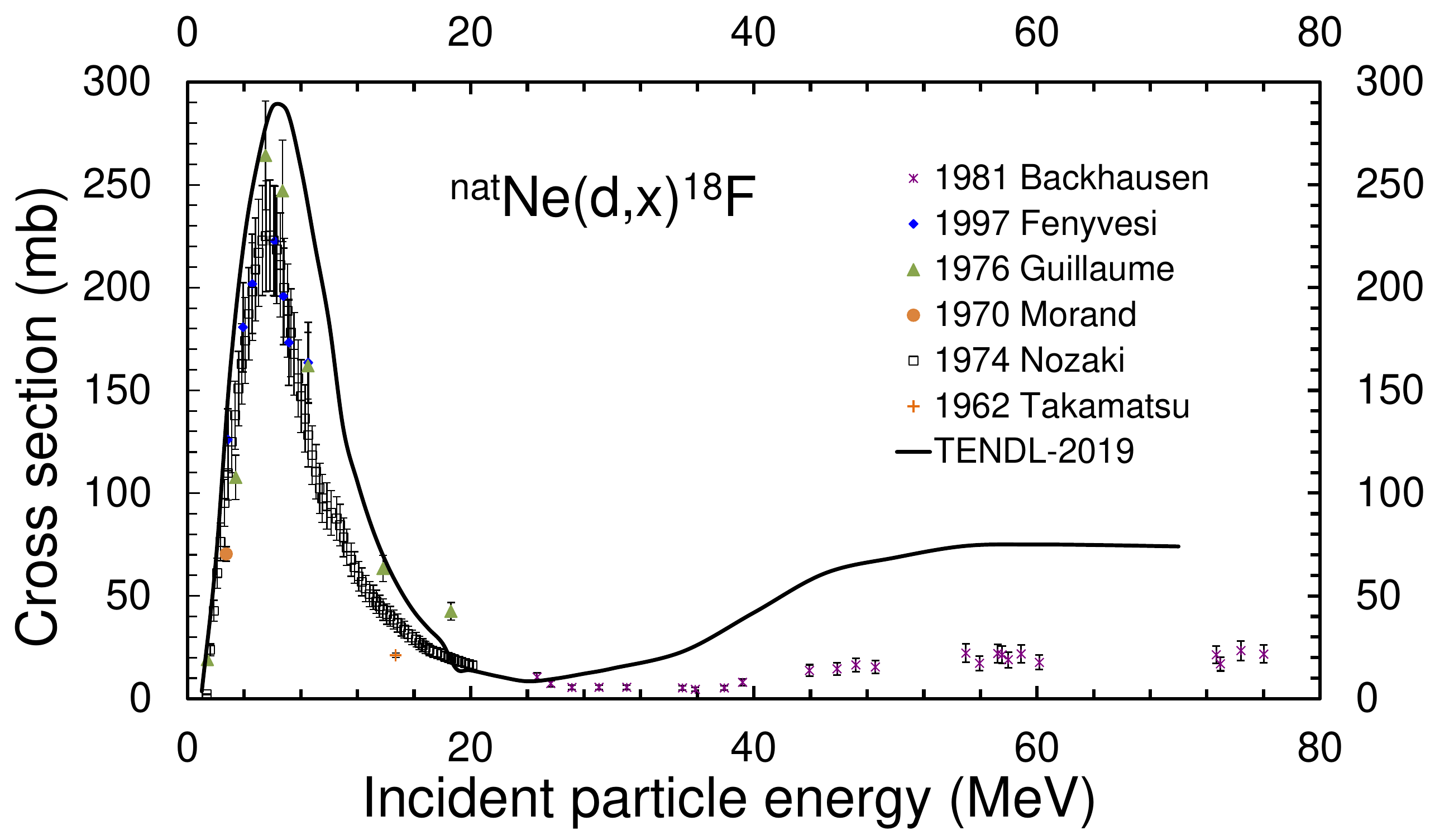}}
\subfigure[~Selected data compared with evaluated Pad\'e fit (L = 15, N = 91, $\chi^2$=0.75, solid line) and estimated total uncertainty in percentage including a 4\% systematic uncertainty (dashed line, right-hand scale).]
{\includegraphics[width=0.97\columnwidth]{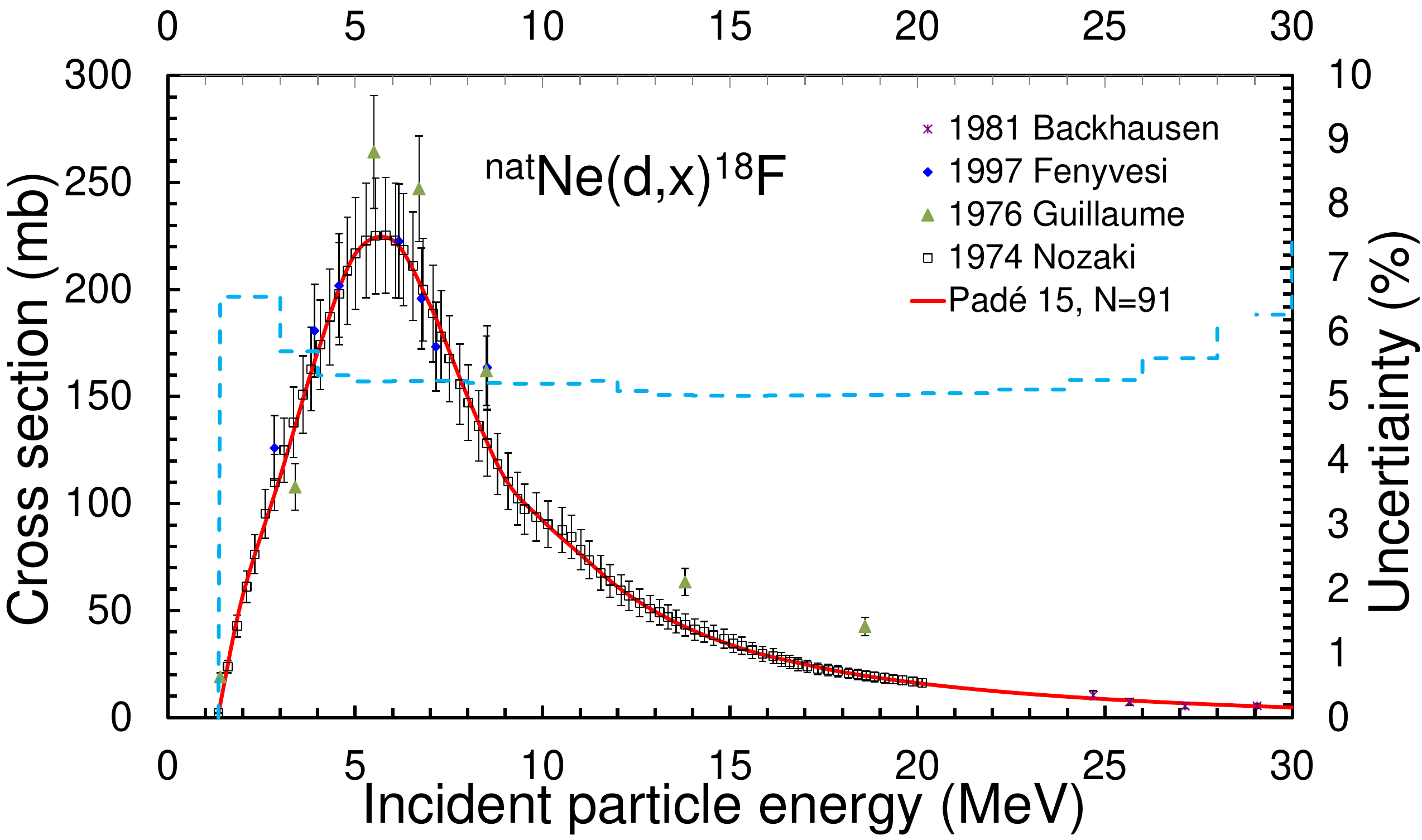}}
\centering
\vspace{-2mm}
\caption{(Color online) Evaluated Pad\'e fit and experimental data from
Refs.~\cite{Backhausen:1981, Fenyvesi:1997, Guillaume:1976, Morand:1970, Nozaki:1974, Takamatsu:1962} for the ${^\mathrm{nat}}$Ne(d,x)$^{18}$F reaction.}
\label{fig6:nenatdx-f18}
\vspace{-4mm}
\end{figure}

For production of $^{18}$F  through the nowadays rarely used ${^\mathrm{nat}}$Ne(d,x)$^{18}$F reaction, a grand total of six publications~\cite{Backhausen:1981, Fenyvesi:1997, Guillaume:1976,Morand:1970, Nozaki:1974, Takamatsu:1962} with experimental cross-section data were identified in the literature for incident particle energies up to 80~MeV  and are represented with uncertainties in Fig.~\ref{fig6:nenatdx-f18}(a). No new datasets were added after the previous update of the IAEA on-line charged particle database~\cite{database:2001} in 2003. The studies by Morand \etal(1970)~\cite{Morand:1970} and  Takamatsu (1962)~\cite{Takamatsu:1962} give partial cross sections for different excited states that were summed in a resulting value at one energy point in the review work of de Lassus St-Genies and  Tobailem (1972)~\cite{de Lassus:1972}.

After thorough critical study of the literature we conclude that the studies performed by Fenyvesi \etal (1997)~\cite{Fenyvesi:1997} and Nozaki et al.~\cite{Nozaki:1974}, on ${^\mathrm{nat}}$Ne targets reflect the cross sections for the ${^\mathrm{nat}}$Ne(d,x)$^{18}$F reaction even if in the manuscripts or in EXFOR they are identified as $^{20}$Ne(d,$\alpha$) $^{18}$F. The original data points of Guillaume (1976)~\cite{Guillaume:1976} show an energy shift and were back shifted 1.1 MeV by the compiler in order to have correspondence with the maximum of Refs.~\cite{Fenyvesi:1997,Nozaki:1974}. The results of 2 studies were rejected and not considered for further analysis, and the reasons for their removal are indicated: Morand \etal(1970) (only one data point from ~\cite{de Lassus:1972}, summation of partial cross section and no control on energy or consistency is possible)~\cite{Morand:1970}, Takamatsu \etal(1962) (only one data point from Ref.~\cite{de Lassus:1972}, summation of partial cross section and no control on energy or consistency is possible)~\cite{Takamatsu:1962}. The remaining four datasets~\cite{Backhausen:1981, Fenyvesi:1997, Guillaume:1976, Nozaki:1974} were considered  as input for a least-squares Pad\'e fit. The two highest energy points of Guillaume (1976) ~\cite{Guillaume:1976} were not taken into account and the energy range was limited to 30~MeV.

The Pad\'e functions with 15 parameters were fitted to 91 selected data points with a $\chi^2$=0.75 and covering the energy range up to 30~MeV as shown in Fig.~\ref{fig6:nenatdx-f18}(b). The uncertainties (including a 4\% systematic uncertainty) range between 7\% near the reaction threshold, decrease to below 6\% between 5 and 25 MeV and reach 7.5\% at the highest energy.

%-----------
\subsection{Production of $^{64}$Cu}\label{ssect-Cu64}

\noindent\textbf{Decay data:} \halflife=12.701(2)~h; \newline decay branches: $\epsilon$: 46.9\%, $\beta^+$: 17.6\%, $\beta^-$: 35.5\%.\newline
\noindent\textbf{Most abundant gammas:} \newline $E_{\gamma}=511$~keV, $I_{\gamma}=35.2$(4)\%; $E_{\gamma}=1345.77$(6)~keV, $I_{\gamma}=0.475$(11)\%.\newline
\T\noindent\textbf{Applications:} $^{\mathrm{64}}$Cu \textit{is used for PET imaging through labeling of both smaller molecules and larger, slower clearing proteins and nanoparticles. The biological activity of the hypoxia imaging agent,  }[$^{64}$Cu]--ATSM, \textit{has been described in great detail in animal models and in clinical PET studies. Also an emerging therapeutic (labelled peptides for tumor-receptor targeting, }[$^{64}$Cu]\textit{-labelled monoclonal antibodies for targeting tumor antigens, and }[$^{64}$Cu]\textit{-labelled nanoparticles for cancer targeting) because of $\beta^-$- and $\epsilon$-decay that produce low energy electrons.}\newline

Evaluation has been made of the $^{64}$Ni(p,n)$^{64}$Cu; $^{64}$Ni(d,2n)$^{64}$Cu, $^{68}$Zn(p,x)$^{64}$Cu and ${^\mathrm{nat}}$Zn(d,x)$^{64}$Cu charged-particle induced reactions previously evaluated in IAEA TRS-473~\cite{Betak:2011}.

\begin{figure}[!thb]
\vspace{-2mm}
\centering
\subfigure[~All experimental data are plotted with uncertainties and compared to TENDL-2017 evaluation \cite{TENDL}.]
{\includegraphics[width=0.98\columnwidth]{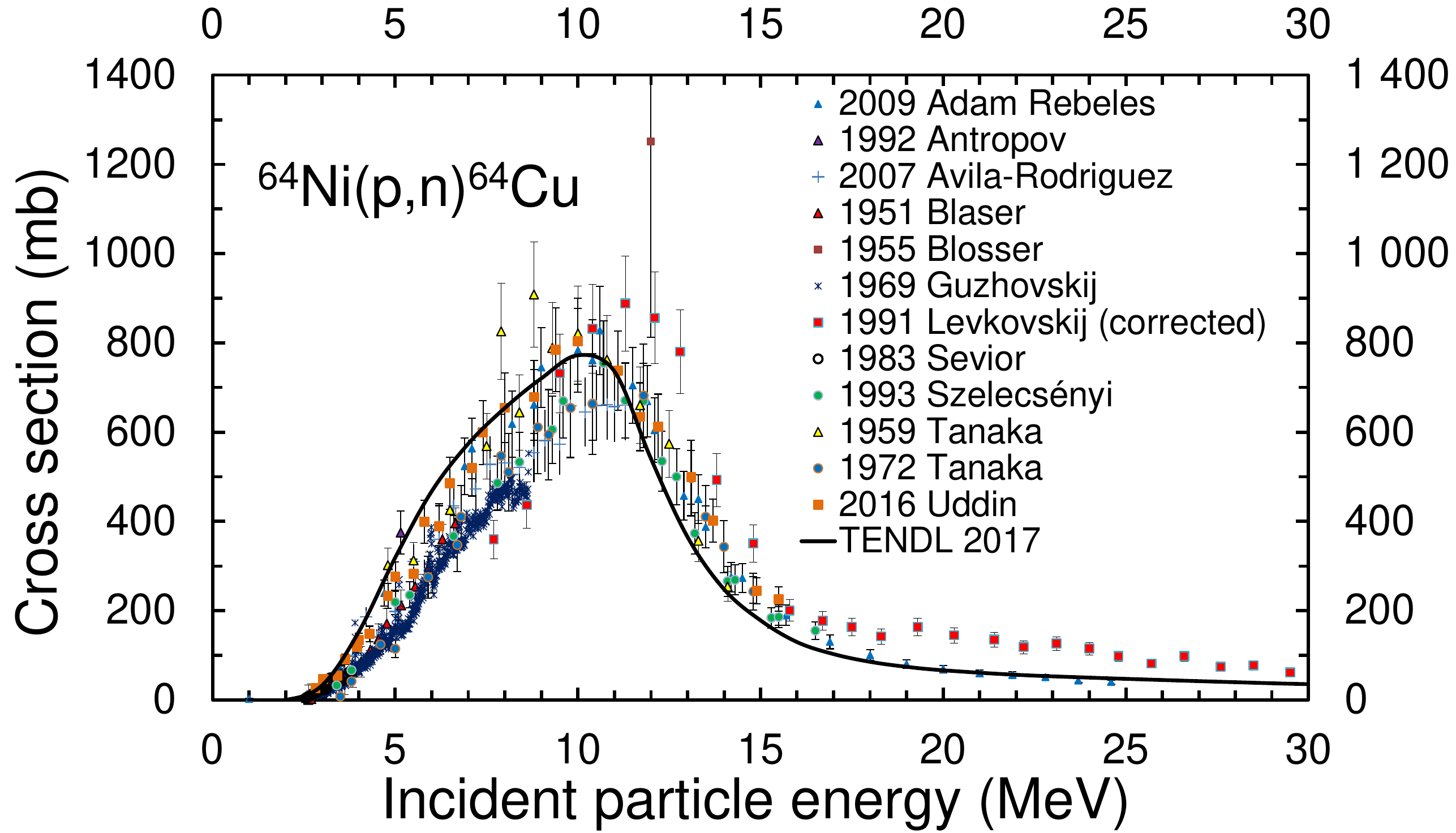}}
\subfigure[~Selected data compared with evaluated Pad\'e fit (L = 44, N = 576, $\chi^2$=2.24, solid line) and estimated total uncertainty in percentage including
a 4\% systematic uncertainty (dashed line, right-hand scale).]
{\includegraphics[width=0.98\columnwidth]{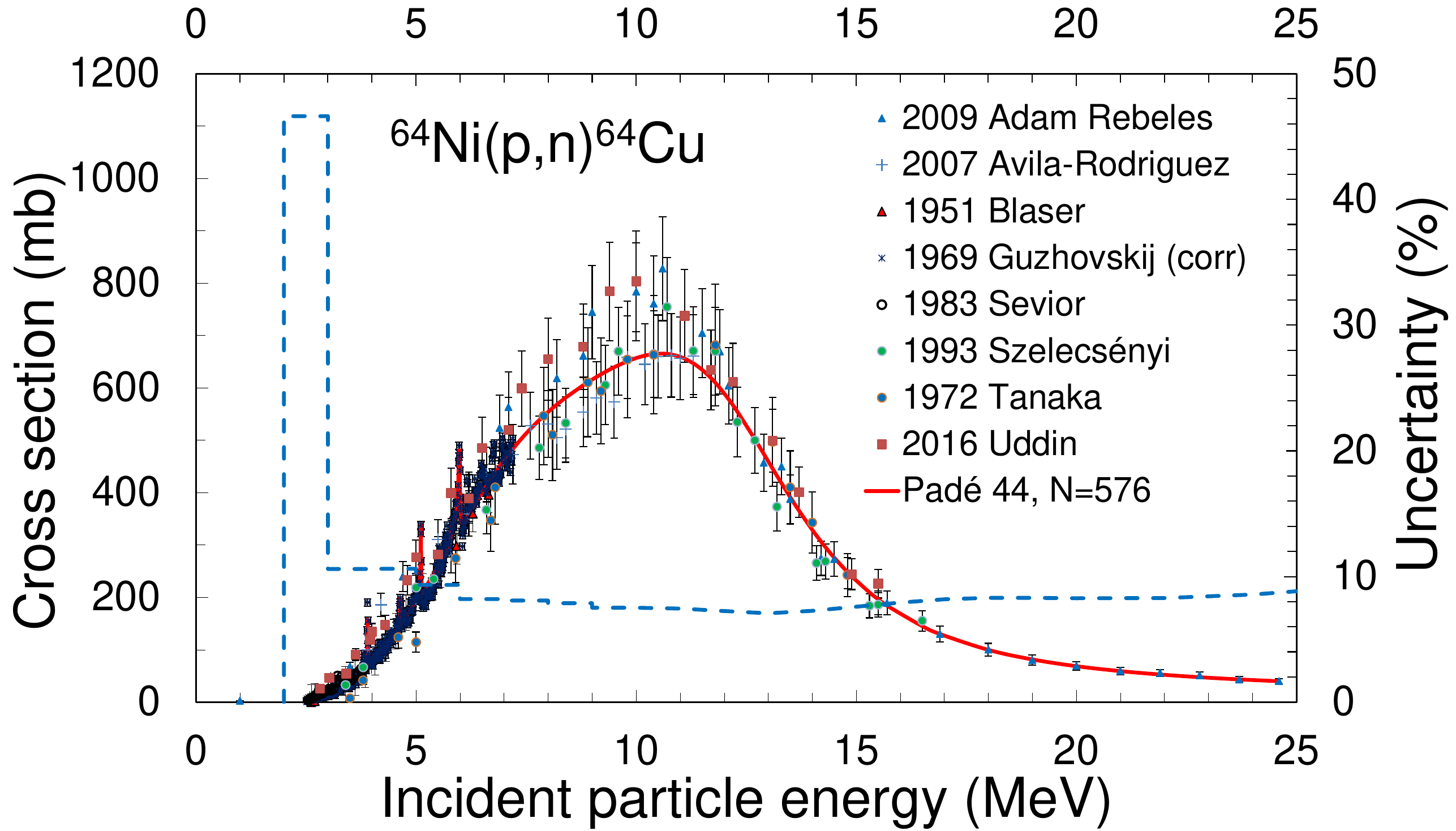}}
\centering
\vspace{-2mm}
\caption{(Color online) Evaluated Pad\'e fit and experimental data from
Refs.~\cite{Adam-Rebeles:2009,  Antropov:1992, Avila-Rodriguez:2007, Blaser:1951, Blosser:1955, Guzhovskij:1969, Levkovskij:1991, Sevior:1983, Szelecsenyi:1993, Tanaka:1959, Tanaka:1972, Uddin:2016} for the $^{64}$Ni(p,n)$^{64}$Cu reaction.}
\label{fig7:ni64pn-cu64}
\vspace{-2mm}
\end{figure}

\subsubsection{~~$^{\mathrm{64}}\mathrm{Ni(p,n)}^{\mathrm{64}}\mathrm{Cu}$ reaction}\label{sssect:Ni64pnCu64}
For formation of $^{64}$Cu through the commercially used $^{64}$Ni(p,n)$^{64}$Cu reaction, a grand total of twelve publications~\cite{Adam-Rebeles:2009,  Antropov:1992, Avila-Rodriguez:2007, Blaser:1951, Blosser:1955, Guzhovskij:1969, Levkovskij:1991, Sevior:1983, Szelecsenyi:1993, Tanaka:1959, Tanaka:1972, Uddin:2016} with experimental cross-section data were identified in the literature for incident particle energies up to 30 MeV  and are represented with uncertainties in Fig.~\ref{fig7:ni64pn-cu64}(a). Two additional sets by Nemashkalo \etal (1983)~\cite{Nemashkalo:1983} (partial cross sections near the threshold), and by Treytl \etal(1966)~\cite{Treytl:1966}, only including data points above 100~MeV, are not represented in the figure. Two datasets from Refs.~\cite{Adam-Rebeles:2009,Uddin:2016} were added after the discussion of this reaction in the IAEA-TRS 473 (2011)~\cite{Betak:2011}. The EXFOR data are used but were corrected for abundance of the total $\beta$ emission where necessary.
The original data of Levkovskij (1991)~\cite{Levkovskij:1991} were corrected by a factor 0.8 for the outdated value of the monitor reaction used as last discussed in section II.K of Ref.~\cite{Hermanne:2018} .

The results of four studies were rejected and not considered for further analysis, and the reasons for their removal are indicated: Antropov \etal(1992) (two data points, too high)~\cite{Antropov:1992}, Blosser  and Handley (1955) (single point, discrepant)~\cite{Blosser:1955}, Levkovskij (1991)~\cite {Levkovskij:1991} (values to high over the whole energy region and shifted to higher energy), Tanaka \etal (1959) (strange shape and scatter below the maximum)~\cite{Tanaka:1959}.
The remaining eight datasets~\cite{Adam-Rebeles:2009, Avila-Rodriguez:2007, Blaser:1951, Guzhovskij:1969, Sevior:1983, Szelecsenyi:1993, Tanaka:1972, Uddin:2016} were considered  as possible input for a least-squares Pad\'e fit. The data by Guzhovskij \etal\cite {Guzhovskij:1969}, obtained with high energy resolution, show several resonances between 2 and 6 MeV but are overall somewhat lower than the other selected datasets. An energy dependent correction factor, determined by comparing with a Pad\'e fit to the other selected data sets (about 20\%), was applied before fitting. The Pad\'e functions with 44 parameters were fitted to 576 selected data points with a $\chi^2=$2.24 and covering the energy range up to 25 MeV as shown in Fig.~\ref{fig7:ni64pn-cu64}(b).
The uncertainties (including a 4\% systematic uncertainty) range between 48\% near the reaction threshold, decrease to below 10\% at 5 MeV, remain around 8\% between 6 and 16 MeV and reach 10\% at the highest energy.

\begin{figure}[t]
\vspace{-1.4mm}
\centering
\subfigure[~All experimental data are plotted with uncertainties and compared to TENDL-2017 evaluation \cite{TENDL}.]
{\includegraphics[width=0.98\columnwidth]{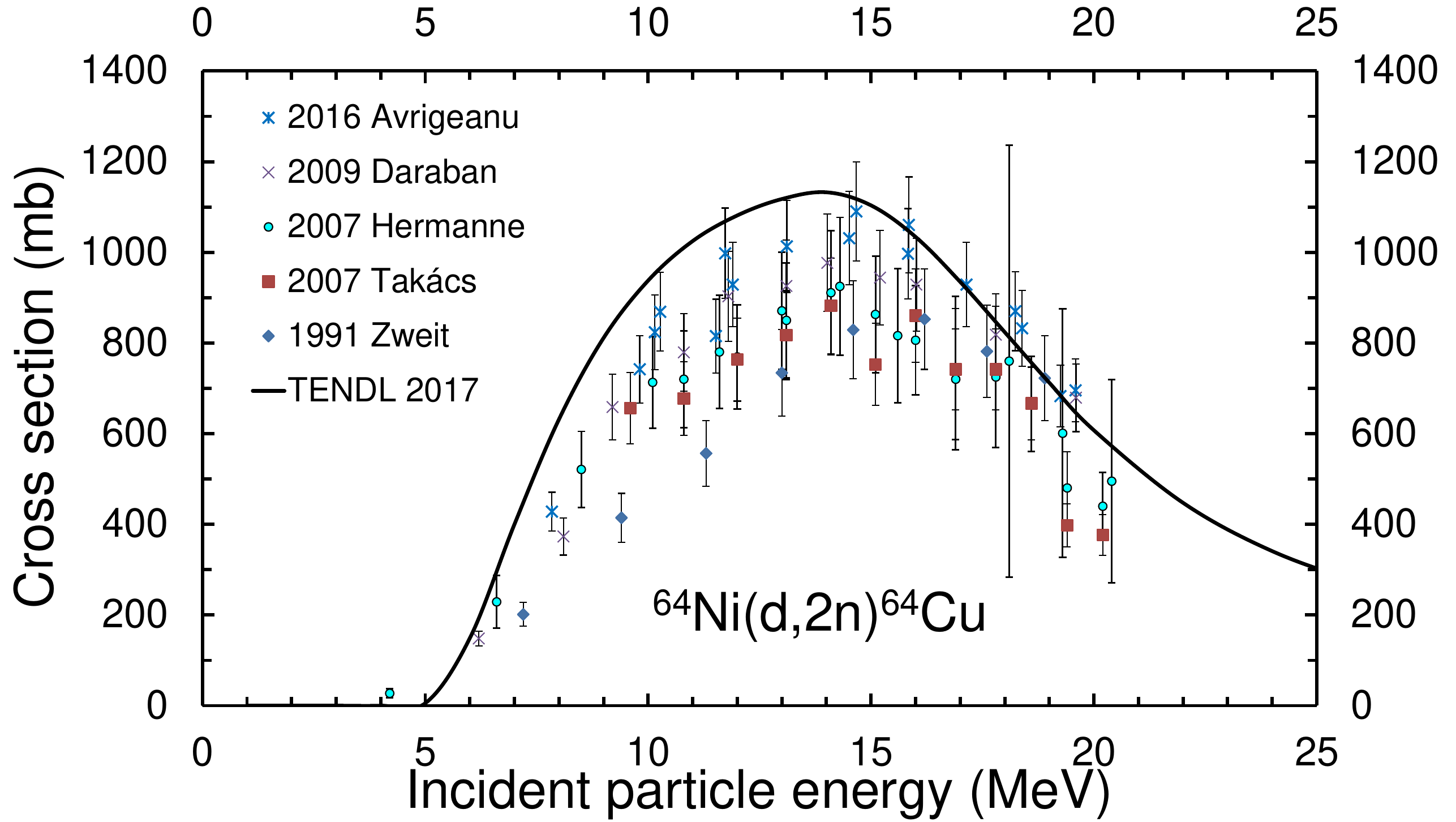}}
\subfigure[~Selected data compared with evaluated Pad\'e fit (L = 12, N = 77, $\chi^2$=1.11, solid line) and estimated total uncertainty in percentage including
a 4\% systematic uncertainty (dashed line, right-hand scale).]
{\includegraphics[width=0.98\columnwidth]{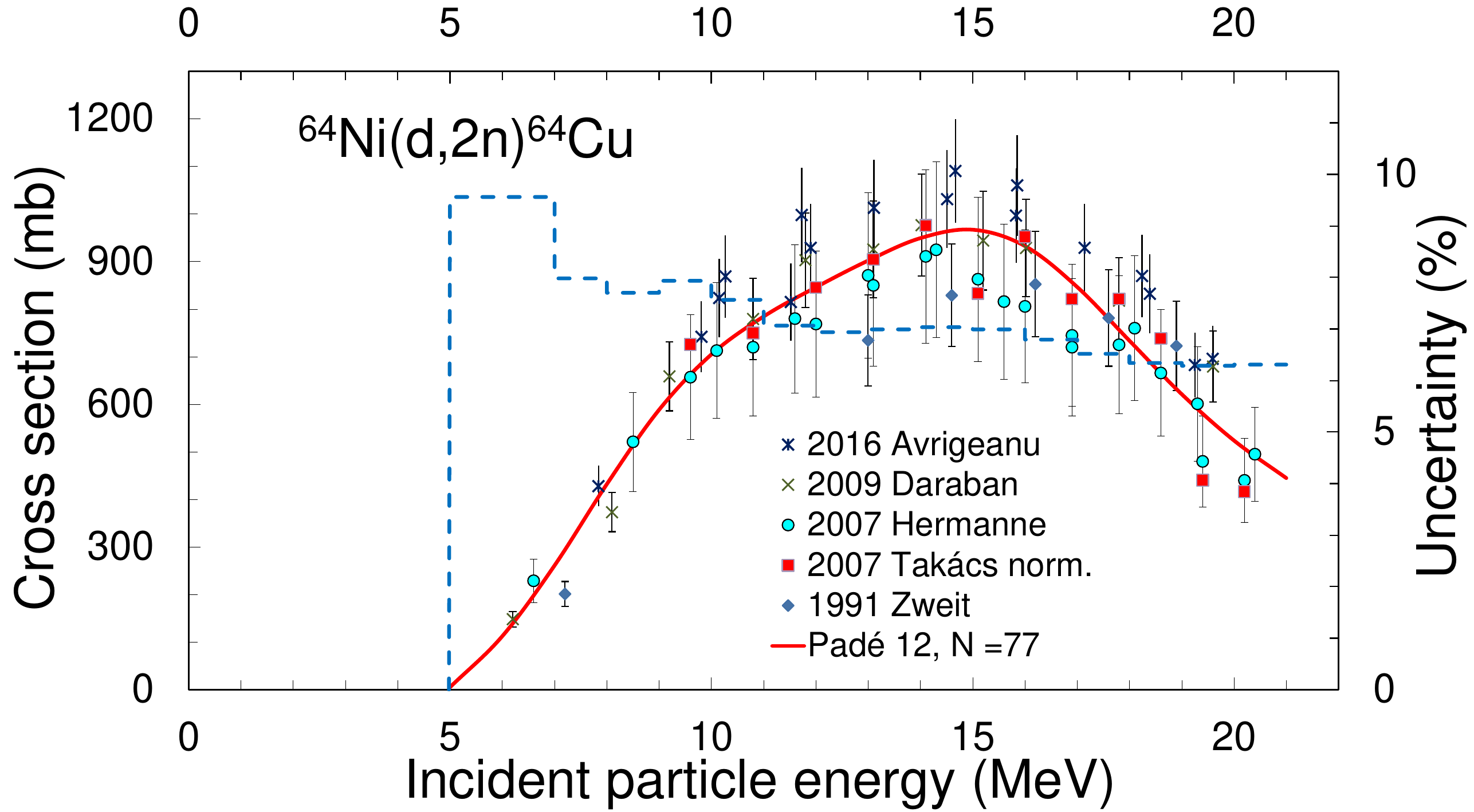}}
\centering
\vspace{-1mm}
\caption{(Color online) Evaluated Pad\'e fit and experimental data from Refs.~\cite{Avrigeanu:2016,Daraban:2009,Hermanne:2007,Takacs:2007,Zweit:1991} for the $^{64}$Ni(d,2n)$^{64}$Cu reaction.}
\label{fig8:ni64d2n-cu64}
%\vspace{-2mm}
\end{figure}

\subsubsection{~~$^{\mathrm{64}}\mathrm{Ni(d,2n)}^{\mathrm{64}}\mathrm{Cu}$ reaction}\label{sssect:Ni64d2nCu64}
For formation of $^{64}$Cu through the $^{64}$Ni(d,2n)$^{64}$Cu reaction, a grand total of five publications~\cite{Avrigeanu:2016, Daraban:2009, Hermanne:2007, Takacs:2007, Zweit:1991} with experimental cross-section data were identified in the literature for incident particle energies up to 25~MeV  and are represented with uncertainties in Fig.~\ref{fig8:ni64d2n-cu64}(a). Only the data set by Avrigeanu \etal(2016)~\cite{ Avrigeanu:2016} was added to the discussion of this reaction in  the IAEA TRS-473 (2011)~\cite{Betak:2011}. All sets were considered as possible input for a least-squares Pad\'e fit. In order to improve the statistical consistency of the sets the data points of Zweit \etal\cite{Zweit:1991} at 9.2 and 11~MeV were deleted , while the remaining points were corrected for $\beta^+$ abundance. To allow fitting above 20 MeV and near the threshold additional points derived from the TENDL-2017 prediction were added, while 5 low values of Hermanne \etal(2007)~\cite{Hermanne:2007} and Tak\'acs \etal(2007)~\cite{Takacs:2007} were discarded. The Pad\'e functions with 12 parameters were fitted to 77 selected data points with  a $\chi^2$=1.11 and covering the energy range  up to 21 MeV as shown in Fig.~\ref{fig8:ni64d2n-cu64}(b). The uncertainties (including a 4\% systematic uncertainty) range between 10\% near the reaction threshold, decrease to below 6\% between 6 and 23~MeV and rises slightly at higher energies.

\subsubsection{~~$^{\mathrm{68}}\mathrm{Zn(p,x)}^{\mathrm{64}}\mathrm{Cu}$ reaction}\label{sssect:Zn68pxCu64}

\begin{figure}[b]
%\vspace{-2mm}
\centering
\subfigure[~All experimental data are plotted with uncertainties and compared to TENDL-2017 evaluation \cite{TENDL}.]
{\includegraphics[width=\columnwidth]{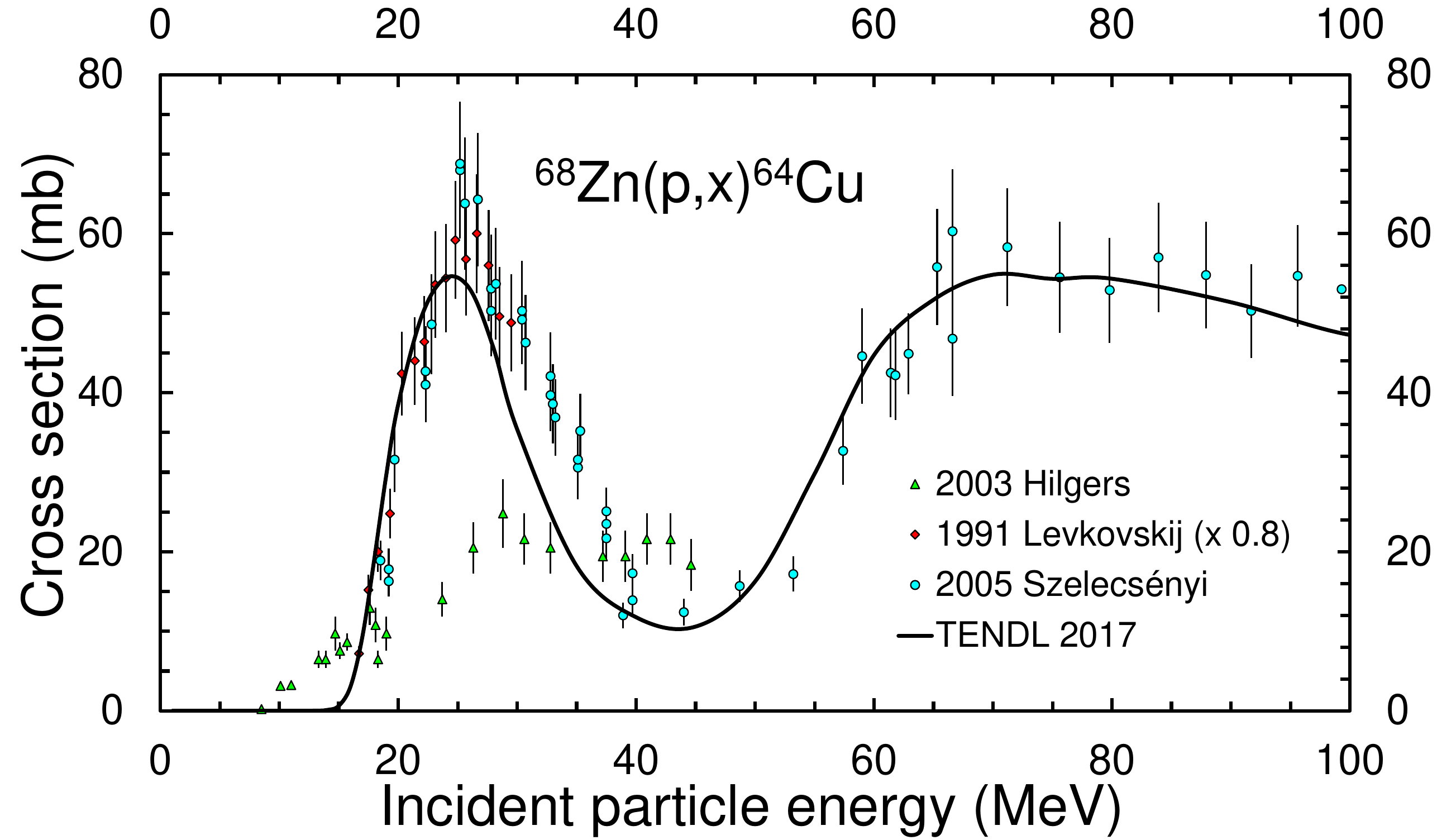}}
\subfigure[~Selected data compared with evaluated Pad\'e fit (L = 8, N = 68, $\chi^2$=1.13, solid line) and estimated total uncertainty in percentage including
a 4\% systematic uncertainty (dashed line, right-hand scale).]
{\includegraphics[width=\columnwidth]{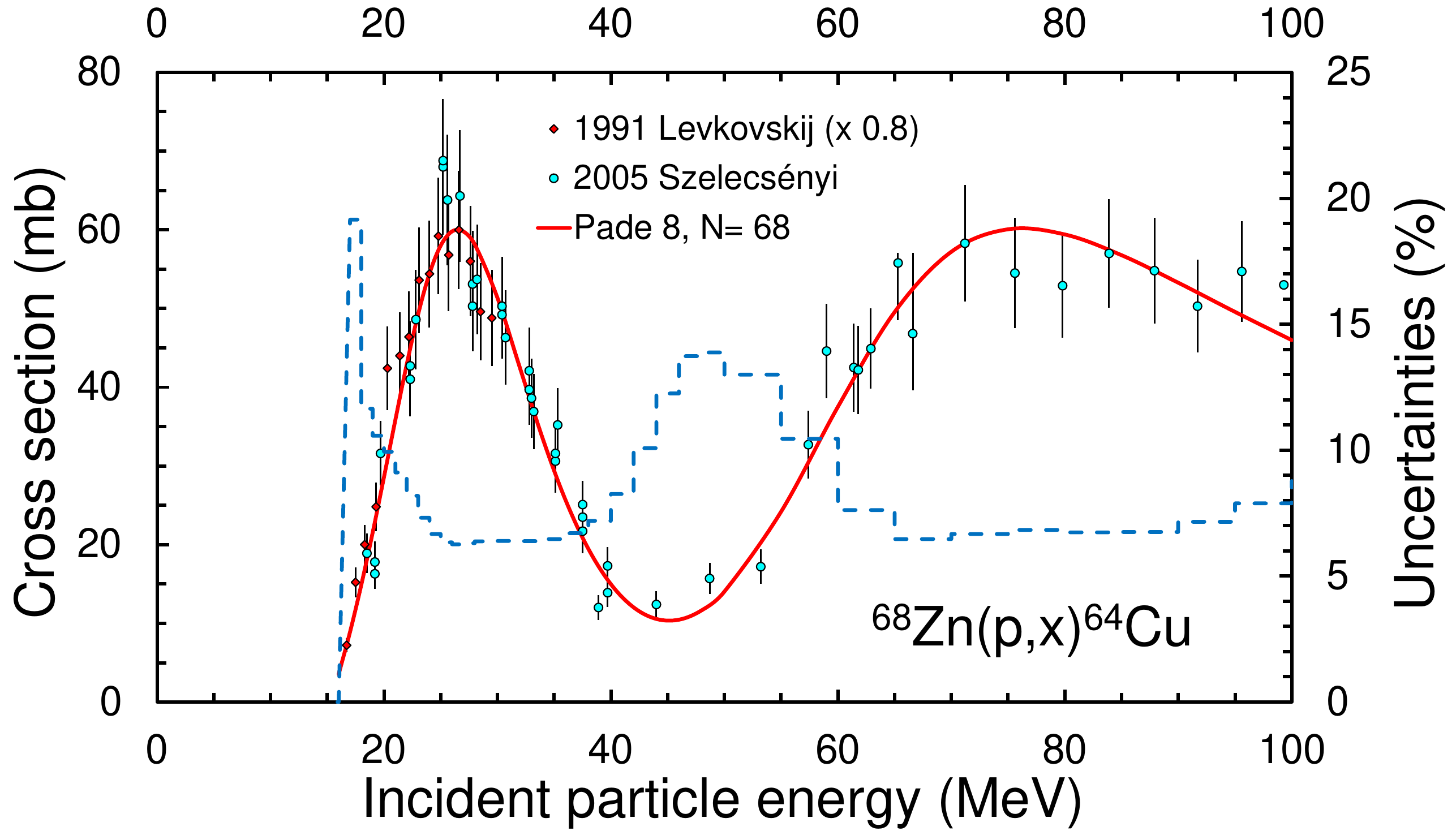}}
\centering
%\vspace{-2mm}
\caption{(Color online) Evaluated Pad\'e fit and experimental data from Refs.~\cite{Hilgers:2003,Levkovskij:1991,Szelecsenyi:2005} for the $^{68}$Zn(p,x)$^{64}$Cu reaction.}
\label{fig9:zn68px-cu64}
%\vspace{-4mm}
\end{figure}

For formation of $^{64}$Cu through the $^{68}$Zn(p,x)$^{64}$Cu pathway (including $^{68}$Zn(p,2p3n)$^{64}$Cu and different reactions with cluster emission) a grand total of three datasets~\cite{Hilgers:2003,Levkovskij:1991,Szelecsenyi:2005} with experimental cross-section data were identified in the literature for incident particle energies up to 100 MeV and are represented with uncertainties in Fig.~\ref{fig9:zn68px-cu64}(a). The main contribution at energies below 30 MeV comes from the $^{68}$Zn(p,$\alpha$n)$^{64}$Cu reaction while at higher energies reactions with nucleon emission take over.
The Levkovskij (1991)~\cite{Levkovskij:1991} data were corrected by a factor 0.8 for outdated monitor value. No new data were added to the discussion of this pathway in IAEA TRS-473 (2011)~\cite{Betak:2011}.
The results of Hilgers \etal(2003)~\cite{Hilgers:2003} were not considered for further analysis because the contribution of the $^{68}$Zn(p,$\alpha$n)$^{64}$Cu reaction is not well represented.
The remaining two datasets~\cite{Levkovskij:1991,Szelecsenyi:2005} were considered  as possible input for a least-squares Pad\'e fit.    		

The Pad\'e functions with 8 parameters were fitted to 68 selected data points with a $\chi^2$=1.13 and cover the energy range up to 100~MeV as shown in Fig.~\ref{fig9:zn68px-cu64}(b).
The uncertainties (including a 4\% systematic uncertainty) range between 20\% near the reaction threshold, decrease to below 7\% at higher energy, except in the range between 42 and 60 MeV (dip in cross sections) where a rise to 14\% is noted.

\begin{figure}[!htb]
%\vspace{-2mm}
\centering
\subfigure[~All experimental data are plotted with uncertainties and compared to TENDL-2017 evaluation \cite{TENDL}.]
{\includegraphics[width=\columnwidth]{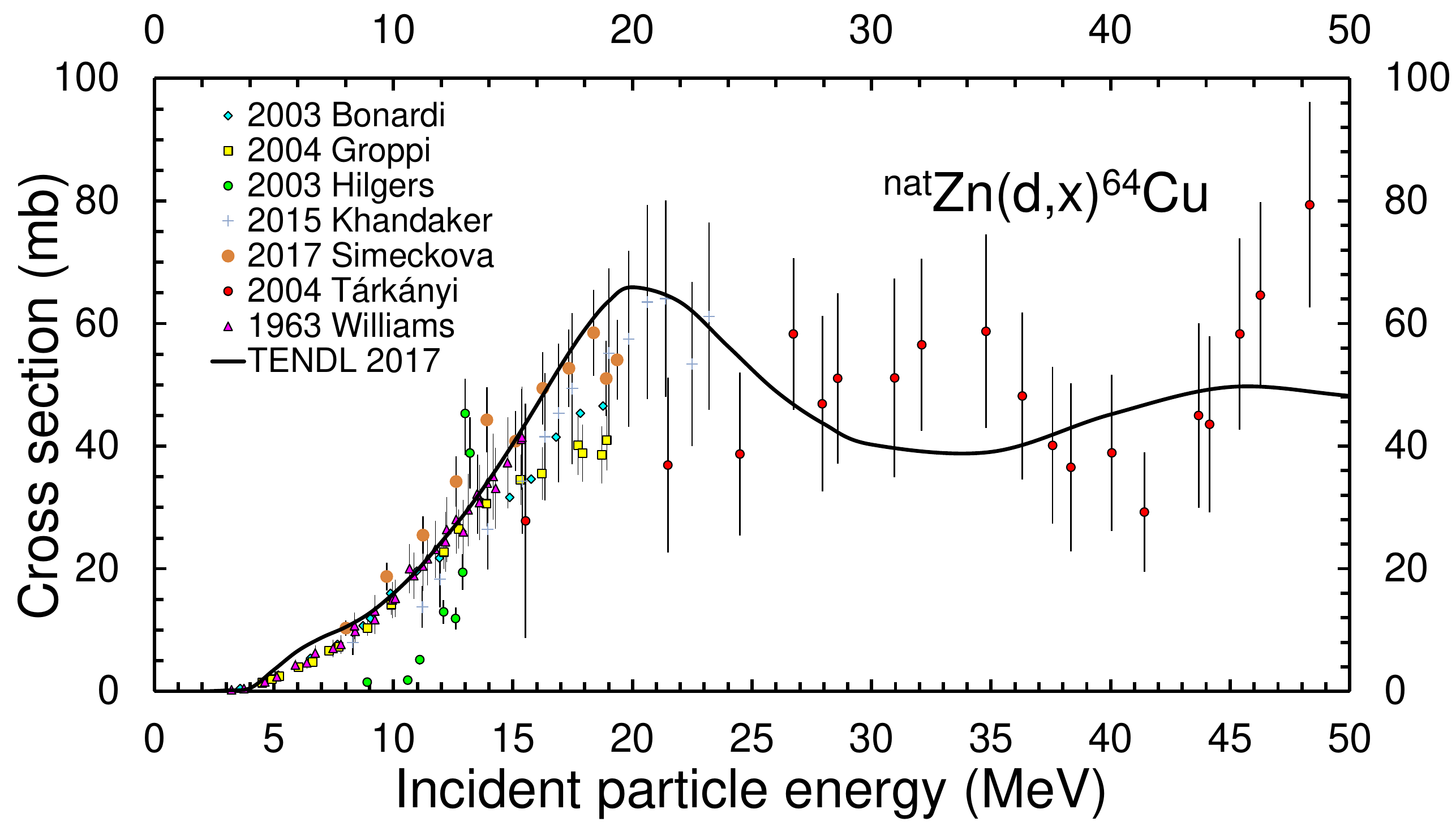}}
\subfigure[~Selected data compared with evaluated Pad\'e fit (L = 19, N = 105, $\chi^2$=0.66, solid line) and estimated total uncertainty in percentage including a 4\% systematic uncertainty (dashed line, right-hand scale).]
{\includegraphics[width=\columnwidth]{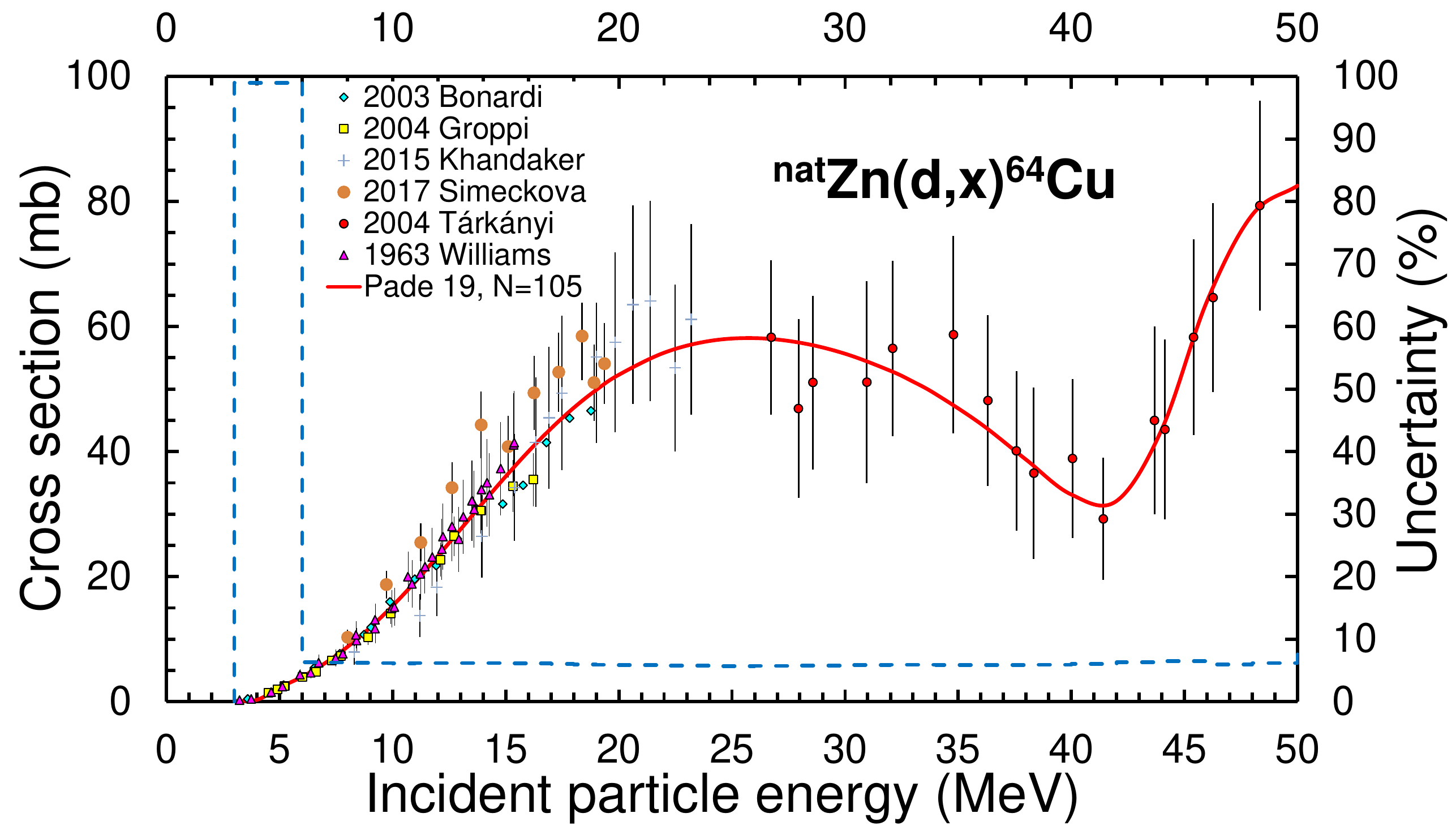}}
\centering
%\vspace{-3mm}
\caption{(Color online) Evaluated Pad\'e fit and experimental data from Refs.~\cite{Bonardi:2003,Groppi:2004,Hilgers:2003,Khandaker:2015,Simeckova:2017,Tarkanyi:2004,Williams:1963} for the ${^\mathrm{nat}}$Zn(d,x)$^{64}$Cu reaction.}
\label{fig10:znnatdx-cu64}
\vspace{-4mm}
\end{figure}

\subsubsection{~~$^{\mathrm{nat}}\mathrm{Zn(d,x)}^{\mathrm{64}}\mathrm{Cu}$ reaction}\label{sssect:ZndxCu64}
For formation of $^{64}$Cu through the ${^\mathrm{nat}}$Zn(d,x)$^{64}$Cu pathway (including essentially $^{64}$Zn(d,2p)$^{64}$Cu, $^{66}$Zn(d,$\alpha$)$^{64}$Cu,  $^{67}$Zn(d,$\alpha$n)$^{64}$Cu, and $^{68}$Zn(d,$\alpha$2n)$^{64}$Cu at lower energies and different reactions with nucleon emission above 25~MeV) a grand total of seven datasets~\cite{Bonardi:2003,Groppi:2004,Hilgers:2003,Khandaker:2015,Simeckova:2017,Tarkanyi:2004,Williams:1963} with experimental cross-section data were identified in the literature for incident particle energies up to 100~MeV and are represented with uncertainties in Fig.~\ref{fig10:znnatdx-cu64}(a).
Two new data sets  were added to the discussion of this pathway in the IAEA TRS-473 (2011)~\cite{Betak:2011}: Khandaker \etal(2015)~\cite{Khandaker:2015} and \^Sime\^ckov\'a \etal(2017)~\cite{Simeckova:2017}.  The results of Hilgers \etal(2003)~\cite{Hilgers:2003} were not considered for further analysis because the shape of the excitation function is not well represented.

The remaining six datasets~\cite{Bonardi:2003,Groppi:2004,Khandaker:2015,Simeckova:2017,Tarkanyi:2004,Williams:1963} were considered  as possible input for a least-squares Pad\'e fit.
The Pad\'e functions with 19 parameters were fitted to 105 selected data points with a $\chi^2$=0.66 and covering the energy range up to 100~MeV as shown in Fig.~\ref{fig10:znnatdx-cu64}(b). The uncertainties (including a 4\% systematic uncertainty) are about 100\% near the reaction threshold, but quickly decrease to below 6.5\% and stay almost constant over the whole energy range.

%-----------
\subsection{Production of $^{124}$I}\label{ssect-I124}

\noindent\textbf{Decay data:} \halflife=4.1760(3)~d; \newline decay branches: $\epsilon$: 6.9\%, $\beta^+$: 22.7\%, $\beta^-$: 93.1\%.\newline
\noindent\textbf{Most abundant gammas:} \newline $E_{\gamma}=602.73$(8)~keV, $I_{\gamma}=62.9$(7)\%; $E_{\gamma}=722.78$(8)~keV, $I_{\gamma}=10.36$(12)\%; $E_{\gamma}=1690.96$(8)~keV, $I_{\gamma}=11.15$(17)\%; $E_{\gamma}=511$~keV, $I_{\gamma}=45$(3)\%.\newline
\T\noindent\textbf{Applications:} \textit{Diagnosis and treatment of differentiated thyroid carcinoma. Labelling of various antibodies for diagnosis of different cancers. Cardiovascular imaging through labelled MIBG.}\newline

Evaluation has been made of the cross sections of $^{124}$Te(p,n)$^{124}$ I, $^{124}$Te(d,2n)$^{124}$I and $^{125}$Te(p,2n)$^{124}$I production reactions.

\subsubsection{~~$^{\mathrm{124}}\mathrm{Te(p,n)}^{\mathrm{124}}\mathrm{I}$ reaction}\label{sssect:Te124pnI124}
\begin{figure}[!thb]
%\vspace{-2mm}
\centering
\subfigure[~All experimental data are plotted with uncertainties and compared to TENDL-2017 evaluation \cite{TENDL}.]
{\includegraphics[width=\columnwidth]{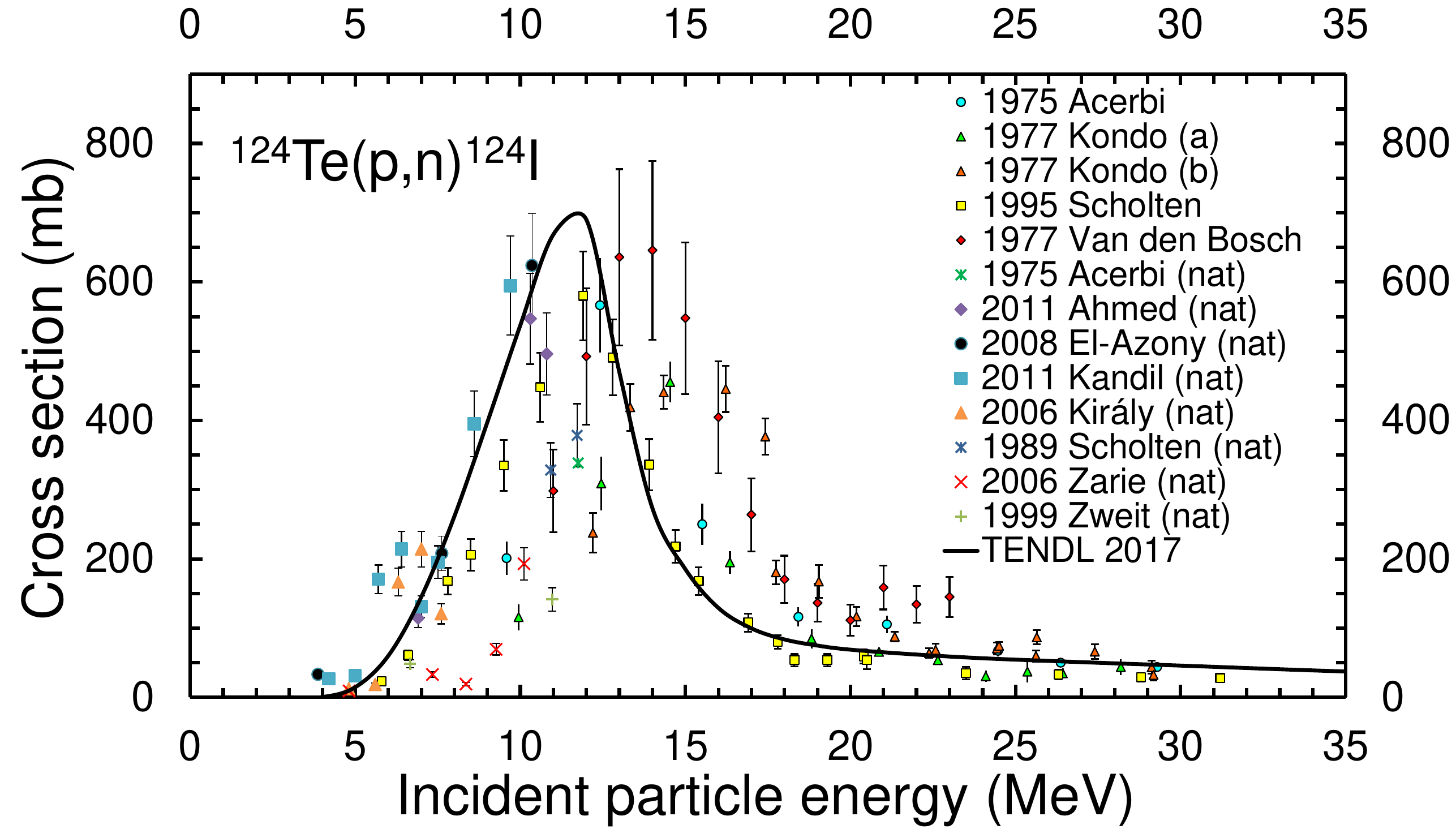}}
\subfigure[~Selected data compared with evaluated Pad\'e fit (L = 11, N = 60, $\chi^2$=3.21, solid line) and estimated total uncertainty in percentage including a 4\% systematic uncertainty (dashed line, right-hand scale).]
{\includegraphics[width=\columnwidth]{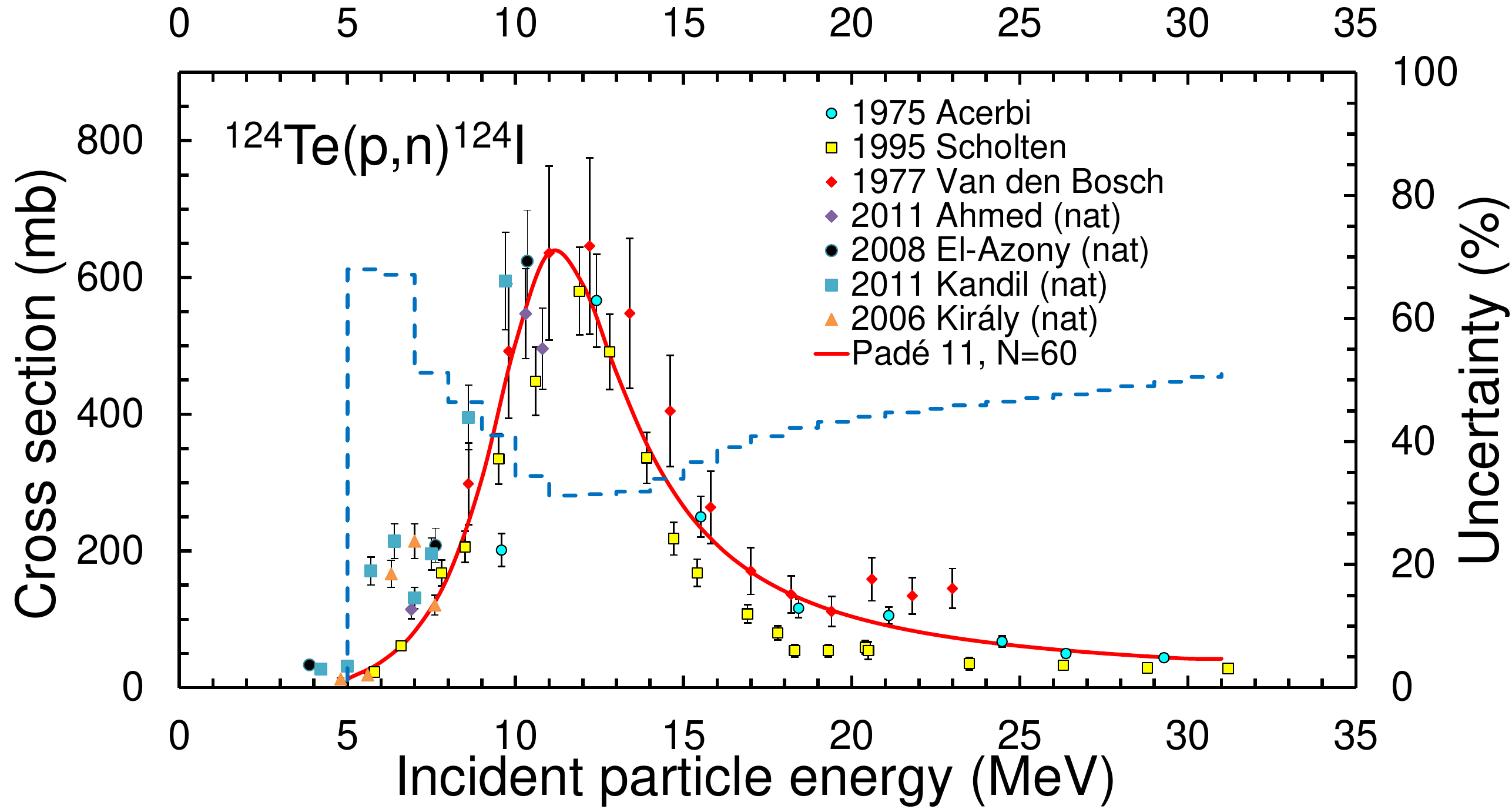}}
\centering
%\vspace{-2mm}
\caption{(Color online) Evaluated Pad\'e fit and experimental data from Refs.~\cite{Acerbi:1975,Kondo:1977,Scholten:1995,Vandenbosch:1977} on enriched $^{124}$Te targets and Refs.~\cite{Zweit:1991a,Acerbi:1975,Ahmed:2011,El Azony:2008,Kandil:2013,Kiraly:2006,Scholten:1989,Zarie:2006} on ${^\mathrm{nat}}$Te targets  for the $^{124}$Te(p,n)$^{124}$I reaction.}
\label{fig11:te124pn-i124}
%\vspace{-2mm}
\end{figure}

For formation of $^{124}$I through its only practical production pathway $^{124}$Te(p,n)$^{124}$I  (natural abundance of $^{124}$Te: 4.82\%), only 5 datasets (in 4 publications) obtained on enriched or partially enriched $^{124}$Te targets~\cite{Acerbi:1975,Kondo:1977,Scholten:1995,Vandenbosch:1977} with experimental cross-section data were identified in the literature for incident particle energies up to 32~MeV  and are represented with uncertainties in Fig.~\ref{fig11:te124pn-i124}(a). The article by Kondo \etal(1977)~\cite{Kondo:1977} contains two series of data obtained on targets with different enrichment and are indicated as (a) and (b) in the figure. The represented data of Ref.~\cite{Vandenbosch:1977} are yield values converted to cross sections.

Additional data obtained on ${^\mathrm{nat}}$Te targets can also be considered for energy points below 10.6~MeV, the threshold of the $^{125}$Te(p,2n)$^{124}$I reaction. In the present evaluation we included, apart from the data of Zweit \etal(1999)~\cite{Zweit:1991a}, already used in the IAEA TRS-473 (2011)~\cite{Betak:2011}, seven datasets obtained by normalisation of data in experiments with ${^\mathrm{nat}}$Te~\cite{Acerbi:1975,Ahmed:2011,El Azony:2008,Kandil:2013,Kiraly:2006,Scholten:1989,Zarie:2006}. These datasets are identified by (nat) in the figure. The EXFOR data are used but were corrected for abundance of the total $\beta$-emission where necessary.

The data of 5 sets were rejected and not considered for further analysis, because the published values are too low or too high: Acerbi \etal(1975) (natural targets)~\cite{Acerbi:1975}, Kondo \etal(1977) (series a and b)~\cite{Kondo:1977},  Scholten \etal(1989) (${^\mathrm{nat}}$Te)~\cite{Scholten:1989}, Zarie \etal(2006)~\cite {Zarie:2006}, and Zweit \etal(1991)~\cite{Zweit:1991a}.
The remaining seven datasets~\cite{Acerbi:1975,Scholten:1995,Vandenbosch:1977,Ahmed:2011,El Azony:2008,Kandil:2013,Kiraly:2006} were considered  as possible input for a least-squares Pad\'e fit. In order to improve the agreement with the maximum of other sets the data of Van den Bosch \etal\cite{Vandenbosch:1977} were energy shifted (-1.4 MeV at the maximal cross section).

The Pad\'e functions with 11 parameters were fitted to 60 selected data points with a $\chi^2$=3.21 and covering the energy range up to 32~MeV as shown in Fig.~\ref{fig11:te124pn-i124}(b). Due to the large scatter of experimental results the uncertainties (including a 4\% systematic uncertainty) are large and are below 45\% only between 8 and 22~MeV.

\subsubsection{~~$^{\mathrm{124}}\mathrm{Te(d,2n)}^{\mathrm{124}}\mathrm{I}$ reaction}\label{sssect:Te124d2nI124}
\begin{figure}[!thb]
%\vspace{-2mm}
\centering
\includegraphics[width=\columnwidth]{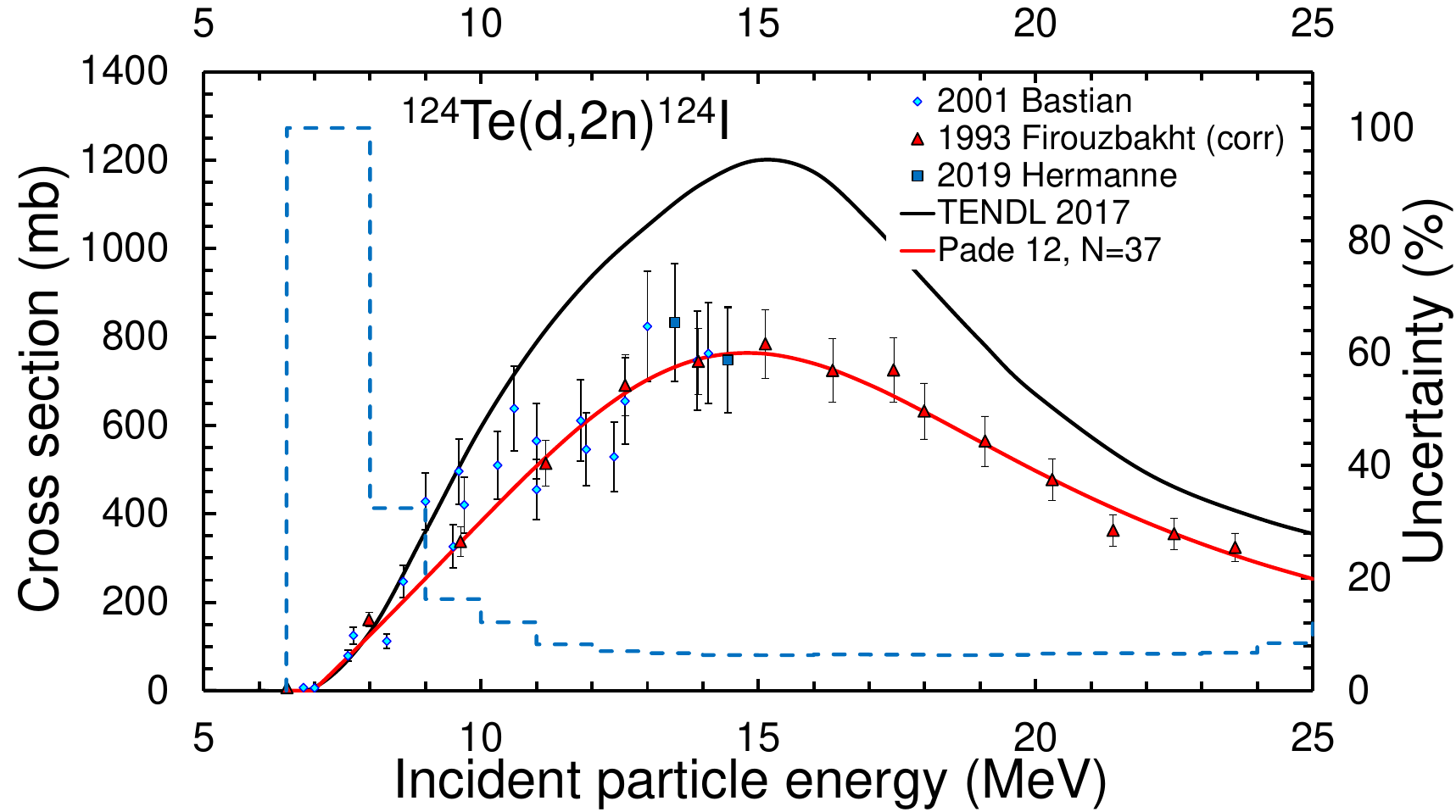}
%\vspace{-2mm}
\caption{(Color online) All experimental (and selected) data from Refs.~\cite{Bastian:2001,Firouzbakht:1993,Hermanne:2019} are compared with TENDL-2017 and evaluated Pad\'e fit (L =12, N = 37, $\chi^2$=1.86, solid line) and estimated total uncertainties in percentages, including 4\% systematic uncertainty (dashed line, right-hand scale) for the $^{124}$Te(d,2n)$^{124}$I reaction.}
\label{fig12:te124d2n-i124}
%\vspace{-2mm}
\end{figure}

For formation of $^{124}$I through the less used $^{124}$Te(d,2n)$^{124}$I reaction only 3 datasets~\cite{Bastian:2001,Firouzbakht:1993,Hermanne:2019} with experimental cross-section data were identified in the literature for incident particle energies up to 25~MeV  and are represented with uncertainties in Fig.~\ref{fig12:te124d2n-i124}.
The cross section values reported in Firouzbakht \etal(1993)~\cite{Firouzbakht:1993} seem to be wrong and therefore new cross sections were deduced from the integral yield data, published in the same article, which are in reasonable agreement with  Bastian \etal\cite{ Bastian:2001} data, as was already stated in the IAEA TRS-473~\cite{Betak:2011}. From the recent study by Hermanne \etal\cite{Hermanne:2019} using ${^\mathrm{nat}}$Te targets, two data points data below 15 MeV (threshold for the $^{125}$Te(d,3n)$^{124}$I reaction) are used. The three datasets sets were considered as possible input for a least-squares Pad\'e fit.  In order to reduce the energy scatter below 13~MeV, the Firouzbakht \etal\cite{Firouzbakht:1993} data points were progressively shifted in energy as represented in Fig.~\ref{fig12:te124d2n-i124}. The Pad\'e functions with 12 parameters were fitted to 37 selected data points with a $\chi^2$=1.86 covering the deuteron energy range up to 24~MeV as shown in Fig.~\ref{fig12:te124d2n-i124}. Due to the large energy scatter and reduced number of data points the uncertainties (including a 4\% systematic uncertainty) are large at low energies and are below 10\% only from 11~MeV on.

\subsubsection{~~$^{\mathrm{125}}\mathrm{Te(p,2n)}^{\mathrm{124}}\mathrm{I}$ reaction}\label{sssect:Te125p2nI124}
\begin{figure}[!thb]
%\vspace{-2mm}
\centering
\includegraphics[width=\columnwidth]{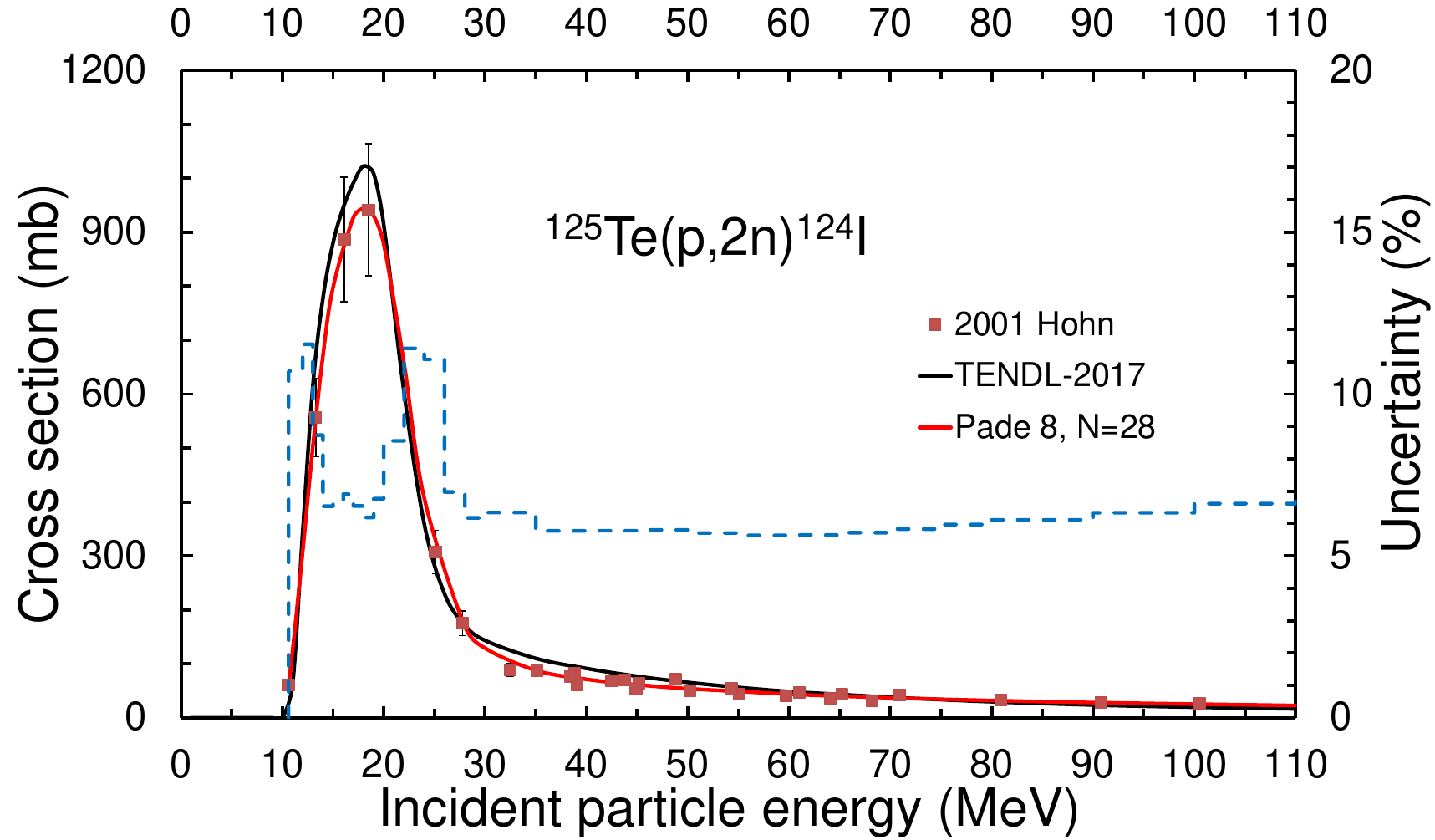}
%\vspace{-2mm}
\caption{(Color online) Single available dataset from Ref.~\cite{Hohn:2001} is compared with TENDL-2017 and evaluated Pad\'e fit (L =8, N = 28, $\chi^2$=0.94, solid line) and estimated total uncertainties in percentages, including 4\% systematic uncertainty (dashed line, right-hand scale) for the $^{125}$Te(p,2n)$^{124}$I reaction.}
\label{fig13:te125p2n-i124}
\vspace{+2mm}
\end{figure}

For formation of $^{124}$I through $^{125}$Te(p,2n)$^{124}$I reaction on somewhat more abundant $^{125}$Te (7.14\% abundance) only 1 dataset by Hohn \etal(2001)~\cite{Hohn:2001} with experimental cross-section data was identified in the literature for incident particle energies up to 110~MeV  and is represented with uncertainties in Fig.~\ref{fig13:te125p2n-i124}. This dataset was, after correction for the abundance of the 602.73~keV $\gamma$-line, considered as input for a least-squares Pad\'e fit. The Pad\'e functions with 8 parameters were fitted to 28 selected data points with a $\chi^2$=0.94 and cover the energy range up to 110~MeV as shown in Fig.~\ref{fig13:te125p2n-i124}. The uncertainties (including a 4\% systematic uncertainty) are below 8\% over nearly the whole energy region.

%======================================================================
\section{RADIONUCLIDES USED IN SINGLE PHOTON EMISSION COMPUTER TOMOGRAPHY (SPECT)} %-- $\gamma$--EMITTERS}
%======================================================================
\label{sect-SPECT}
%-----------
\subsection{Production of $^{81}$Rb}\label{ssect-Rb81}
\noindent\textbf{Decay data:} \halflife=4.572(4)~h; \newline decay branches: $\epsilon$: 72.8\%, $\beta^+$: 27.2\%.\newline
\noindent\textbf{Most abundant gammas:} \newline $E_{\gamma}=190.46$(16)~keV, $I_{\gamma}=64.9$(22)\%; $E_{\gamma}=446.15$(3)~keV, $I_{\gamma}=23.5$(9)\%; $E_{\gamma}=511$~keV, $I_{\gamma}=54.4$(20)\%.\newline
\T\noindent\textbf{Applications:} $^{81}$Rb \textit{is mainly used as mother in a generator of very short-lived }$^{\mathrm{81m}}$Kr \textit{(\halflife=13.1~s) (gas) for study of lung functions (ventilation and perfusion).}\newline

Evaluation has been made of the $^{82}$Kr(p,2n)$^{81}$Rb and ${^\mathrm{nat}}$Kr(p,x)$^{81}$Rb reactions.

\subsubsection{~~$^{\mathrm{82}}\mathrm{Kr(p,2n)}^{\mathrm{81}}\mathrm{Rb}$ reaction}\label{sssect:Kr82p2nRb81}
\begin{figure}[!thb]
%\vspace{-2mm}
\centering
\subfigure[~All experimental data are plotted with uncertainties and compared to TENDL-2017 evaluation \cite{TENDL}.]
{\includegraphics[width=0.99\columnwidth]{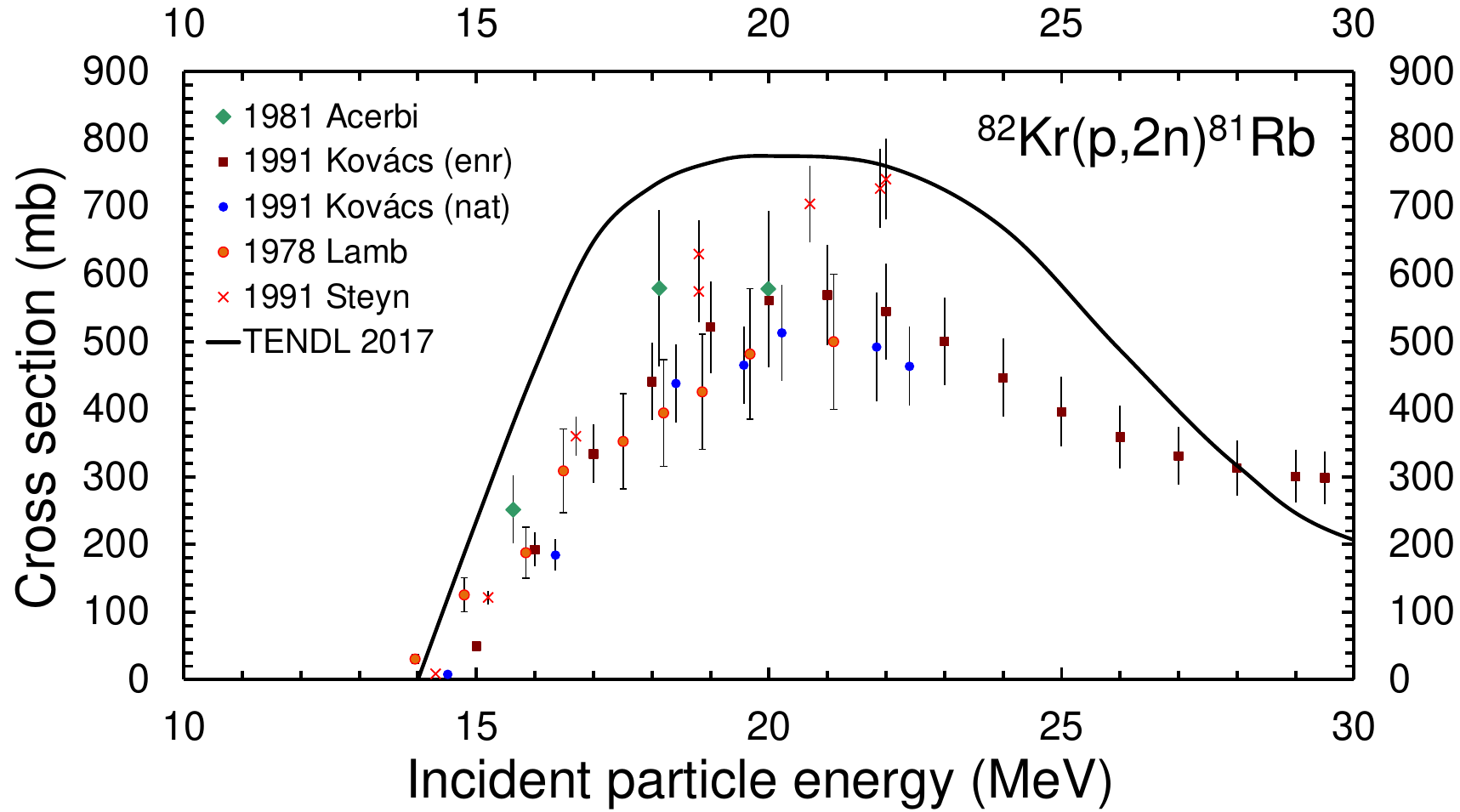}}
\subfigure[~Selected data compared with evaluated Pad\'e fit (L = 17, N = 30, $\chi^2$=1.37, solid line) and estimated total uncertainty in percentage including
a 4\% systematic uncertainty (dashed line, right-hand scale).]
{\includegraphics[width=0.99\columnwidth]{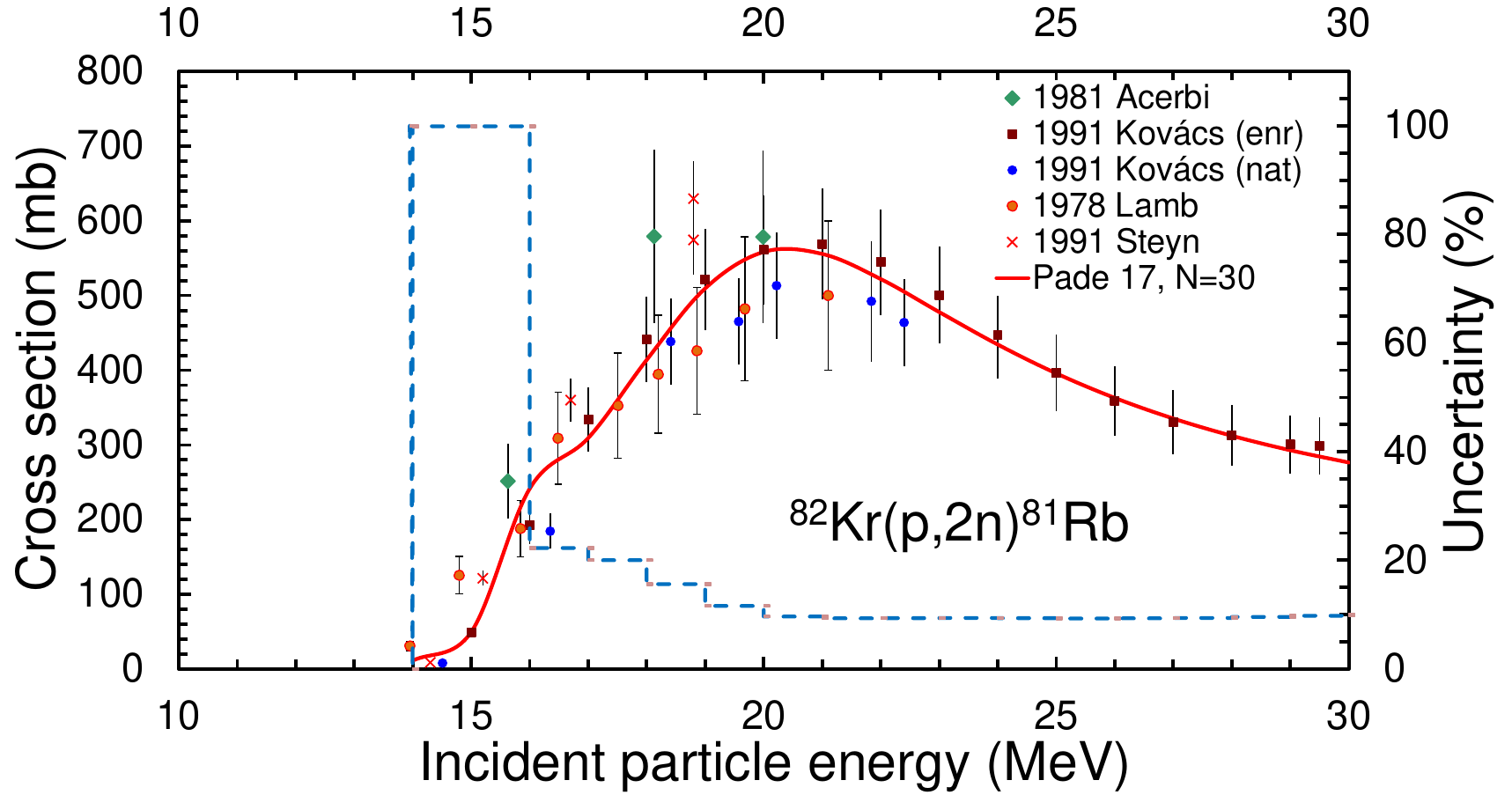}}
\centering
\vspace{-2mm}
\caption{(Color online) Evaluated Pad\'e fit and experimental data from Refs.~\cite{Acerbi:1981,Kovacs:1991,Lamb:1978,Steyn:1991} for the $^{82}$Kr(p,2n)$^{81}$Rb reaction.}
\label{fig14:kr82p2n-rb81}
\vspace{-2mm}
\end{figure}

For formation of $^{81}$Rb through the $^{82}$Kr(p,2n)$^{81}$Rb reaction, a total of four useful publications~\cite{Acerbi:1981,Kovacs:1991,Lamb:1978,Steyn:1991} with experimental cross-section data were identified in the literature for incident particle energies up to 30~MeV and are represented with uncertainties in Fig.~\ref{fig14:kr82p2n-rb81}(a). No new datasets were added after the update of the gamma-emitter section of the IAEA on-line charged particle database~\cite{database:2001} in 2004, also documented in Tak\'acs \etal(2005)~\cite{Takacs:2005}. The article by Kov\'acs \etal(1991)~\cite{Kovacs:1991} contained two data sets on natural and enriched targets, respectively, and are represented as (nat) and (enr) in the figure.  The data of Lamb \etal(1978) were obtained on enriched $^{82}$Kr targets. The data originally published  by Acerbi \etal(1981), Kov\'acs \etal(1991)(nat) and Steyn \etal(1991) were obtained on ${^\mathrm{nat}}$Kr targets and were normalized to the abundance of $^{82}$Kr and are limited to an energy of  21.719~MeV, the threshold of the $^{83}$Kr(p,3n)$^{81}$Rb reaction.

The five sets obtained in that way from Refs.~\cite{Acerbi:1981,Kovacs:1991,Lamb:1978,Steyn:1991} were considered  as possible input for a least-squares Pad\'e fit.  To improve the statistical coherence near the maximum of the excitation curve, the four highest energy points of Steyn \etal(1991)~\cite{Steyn:1991} were deselected and not used in the fitting procedure. The Pad\'e functions with 17 parameters were fitted to 30 selected data points with a $\chi^2$=1.37 covering the energy range up to 30~MeV as shown in Fig.~\ref{fig14:kr82p2n-rb81}(b). The uncertainties (including a 4\% systematic uncertainty) are 100\% near the reaction threshold, drop to 20\% at 16 MeV and decrease steadily to below 10\% from 20 MeV on.

\subsubsection{~~$^{\mathrm{nat}}\mathrm{Kr(p,xn)}^{\mathrm{81}}\mathrm{Rb}$ reaction}\label{sssect:KrpxRb81}
\begin{figure}[!thb]
%\vspace{-2mm}
\centering
\subfigure[~All experimental data are plotted with uncertainties and compared to TENDL-2017 evaluation \cite{TENDL}.]
{\includegraphics[width=\columnwidth]{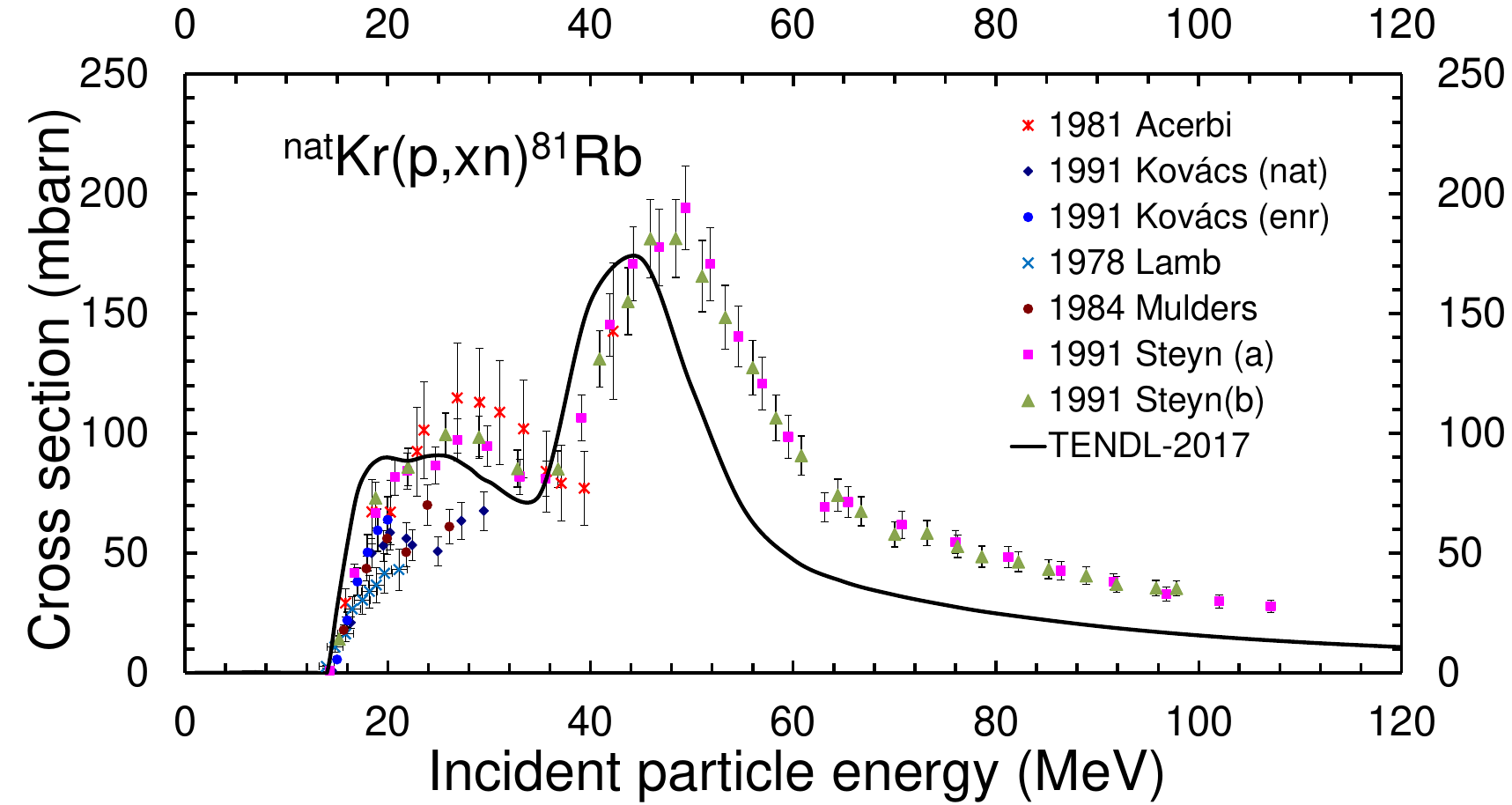}}
\subfigure[~Selected data compared with evaluated Pad\'e fit (L = 14, N = 78, $\chi^2$=0.99, solid line) and estimated total uncertainty in percentage including
a 4\% systematic uncertainty (dashed line, right-hand scale).]
{\includegraphics[width=\columnwidth]{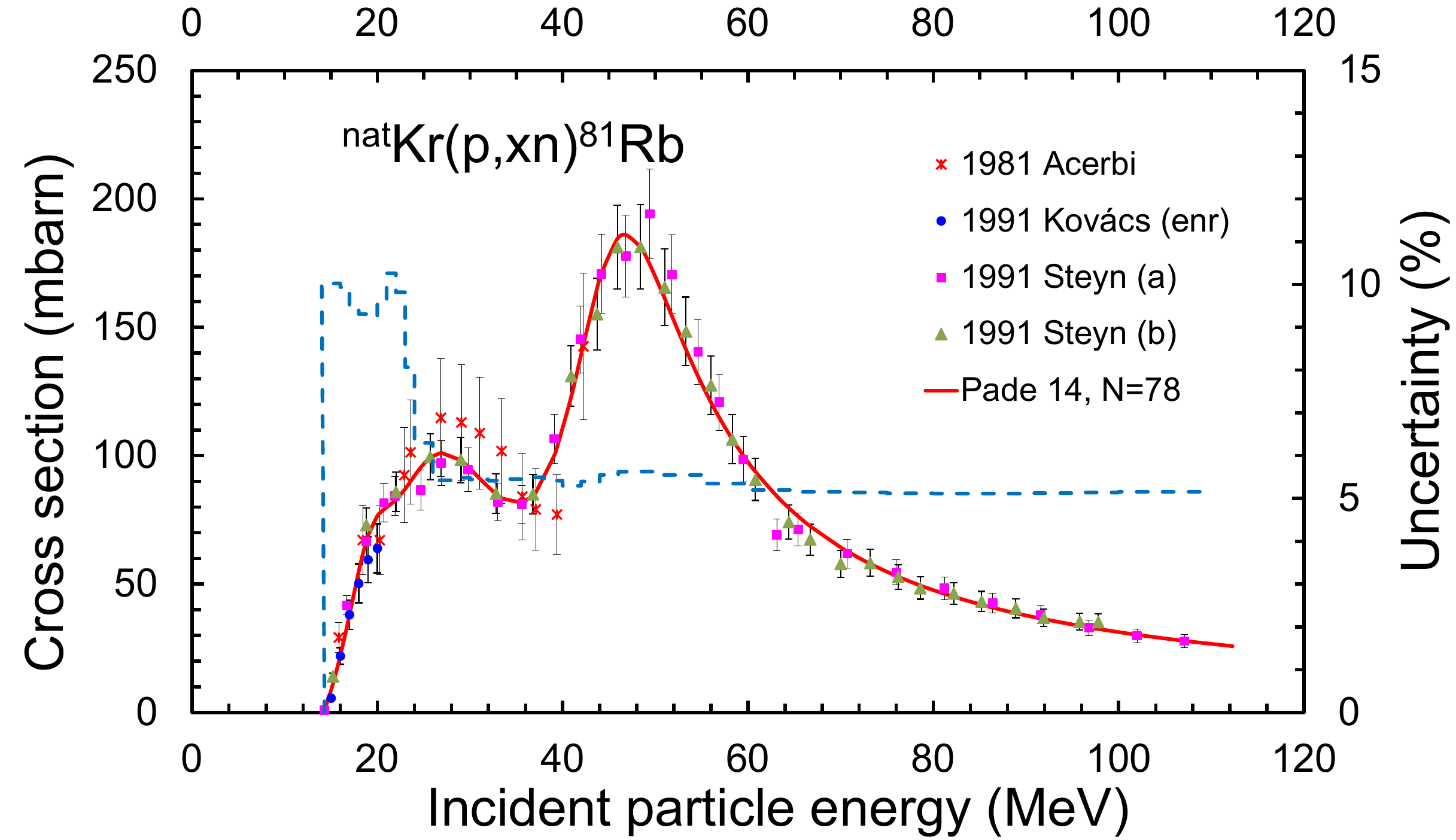}}
\centering
%\vspace{-2mm}
\caption{(Color online) Evaluated Pad\'e fit and experimental data from Refs.~\cite{Acerbi:1981,Kovacs:1991,Lamb:1978,Mulders:1984,Steyn:1991} for the ${^\mathrm{nat}}$Kr(p,x)$^{81}$Rb reaction.}
\label{fig15:krnatpx-rb81}
%\vspace{-2mm}
\end{figure}

For formation of $^{81}$Rb through the ${^\mathrm{nat}}$Kr(p,xn)$^{81}$Rb reaction, a total of five useful publications~\cite{Acerbi:1981,Kovacs:1991,Lamb:1978,Mulders:1984,Steyn:1991} with experimental cross-section data were identified in the literature for incident particle energies up to 120~MeV and are represented with uncertainties in Fig.~\ref{fig15:krnatpx-rb81}(a). The article by Steyn \etal(1991)~\cite{Steyn:1991} contained two data sets represented as (a) and (b) in the figure. The article by Kov\'acs \etal(1991)~\cite{Kovacs:1991} contained two data sets, respectively measured on natural Kr targets and on enriched $^{82}$Kr targets represented as (nat) and (enr) in the figure. The data measured on enriched targets from Lamb \etal(1978)~\cite{Lamb:1978} and Kov\'acs \etal(1991)(enr)~\cite{Kovacs:1991} were normalised and included in the evaluation up to 20 MeV. No new datasets were added since the update of the gamma-emitter section of the IAEA on-line charged particle database~\cite{database:2001} in 2004, also documented in Tak\'acs \etal(2005)~\cite{Takacs:2005}. The data of 3 sets were rejected and not considered for further analysis, and the reasons for their removal are indicated: Kov\'acs \etal(1991)(nat)~\cite {Kovacs:1991} (too low values), Lamb \etal(1978)~\cite {Lamb:1978} (too low values), Mulders (1984)~\cite{Mulders:1984} (too low values).

The remaining four datasets of Refs.~\cite{Acerbi:1981,Kovacs:1991,Steyn:1991} were considered  as possible input for a least-squares Pad\'e fit.  The Pad\'e functions with 14 parameters were fitted to 78 selected data points with a $\chi^2$=0.99 covering the energy range up to 120~MeV as shown in Fig.~\ref{fig15:krnatpx-rb81}(b). The uncertainties (including a 4\% systematic uncertainty), are 100\% near the reaction threshold, but decrease steadily to below 10\% from 20~MeV on.

%-----------
\subsection{Production of $^{123}$I}\label{ssect-I123}

\noindent\textbf{Decay data:} \halflife=13.2235(19)~h; \newline decay branch: $\epsilon$: 100\%.\newline
\noindent\textbf{Most abundant gammas:} \newline $E_{\gamma}=158.97$(5)~keV, $I_{\gamma}=83.3$(4)\%.\newline
\T\noindent\textbf{Applications:} \textit{Most used halogen for wide variety of labelled products.}\newline

Evaluation has been made of the older production pathways: direct $^{123}$Te(p,n)$^{123}$I, $^{124}$Te(p,2n)$^{123}$I reactions and the indirect $^{127}$I(p,3n)$^{125}$Xe$\rightarrow$$^{125}$I (contaminant) and  $^{127}$I(p,5n)$^{123}$Xe$\rightarrow$$^{123}$I reactions. Present commercial pathways through $^{124}$Xe(p,x)$^{123}$Cs$\rightarrow$$^{123}$Xe$\rightarrow$$^{123}$I and $^{124}$Xe(p,x)$^{123}$Xe$\rightarrow$$^{123}$I were already evaluated  in the previous project and published by T\'ark\'anyi \etal(2019)~\cite{Tarkanyi:2019}.

\subsubsection{~~$^{\mathrm{123}}\mathrm{Te(p,n)}^{\mathrm{123}}\mathrm{I}$ reaction}\label{sssect:Te123pnI123}
\begin{figure}[!thb]
%\vspace{-2mm}
\centering
\subfigure[~All experimental data are plotted with uncertainties and compared to TENDL-2017 evaluation \cite{TENDL}.]
{\includegraphics[width=\columnwidth]{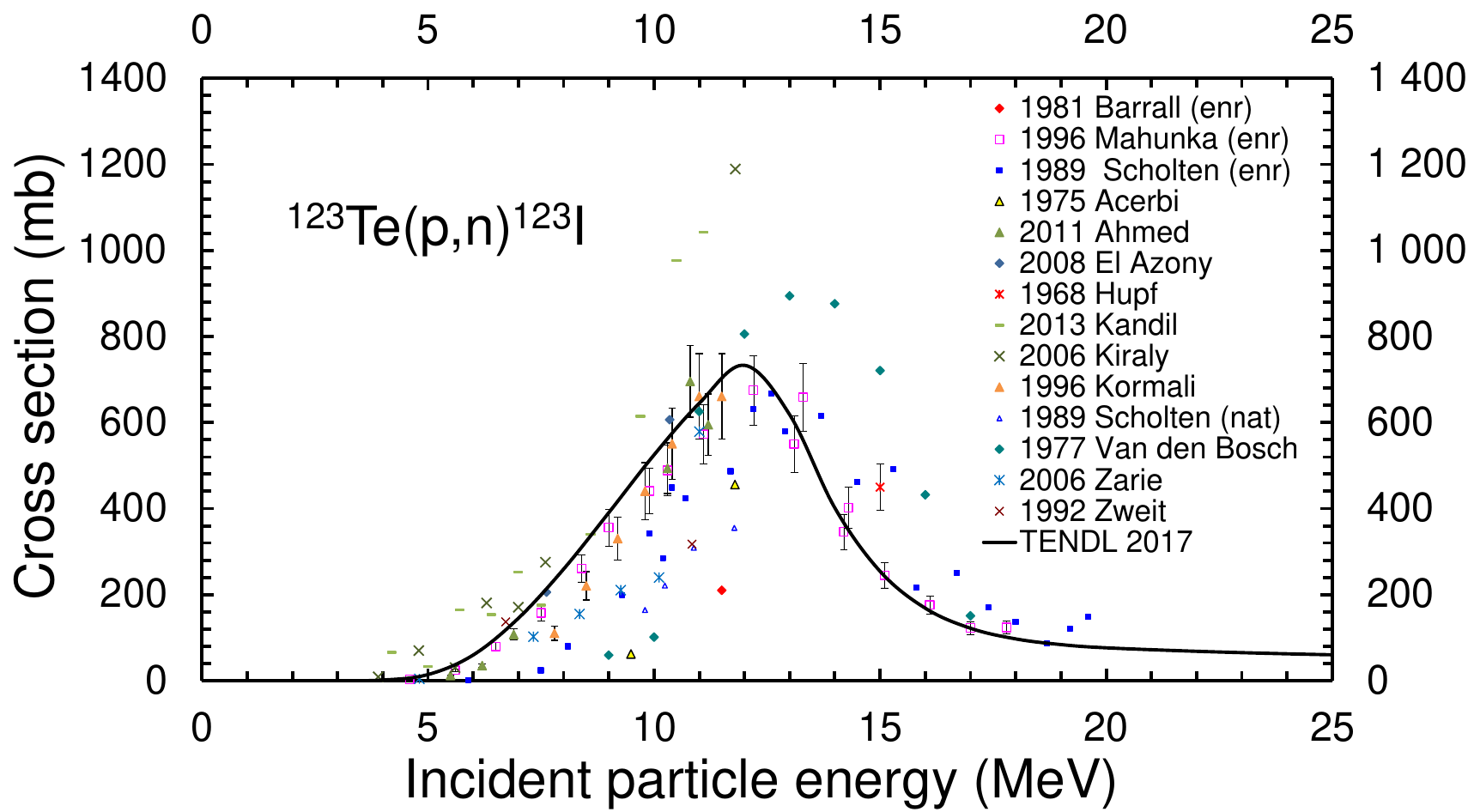}}
\subfigure[~Selected data compared with evaluated Pad\'e fit (L = 10, N = 31, $\chi^2$=1.63, solid line) and estimated total uncertainty in percentage including
a 4\% systematic uncertainty (dashed line, right-hand scale).]
{\includegraphics[width=\columnwidth]{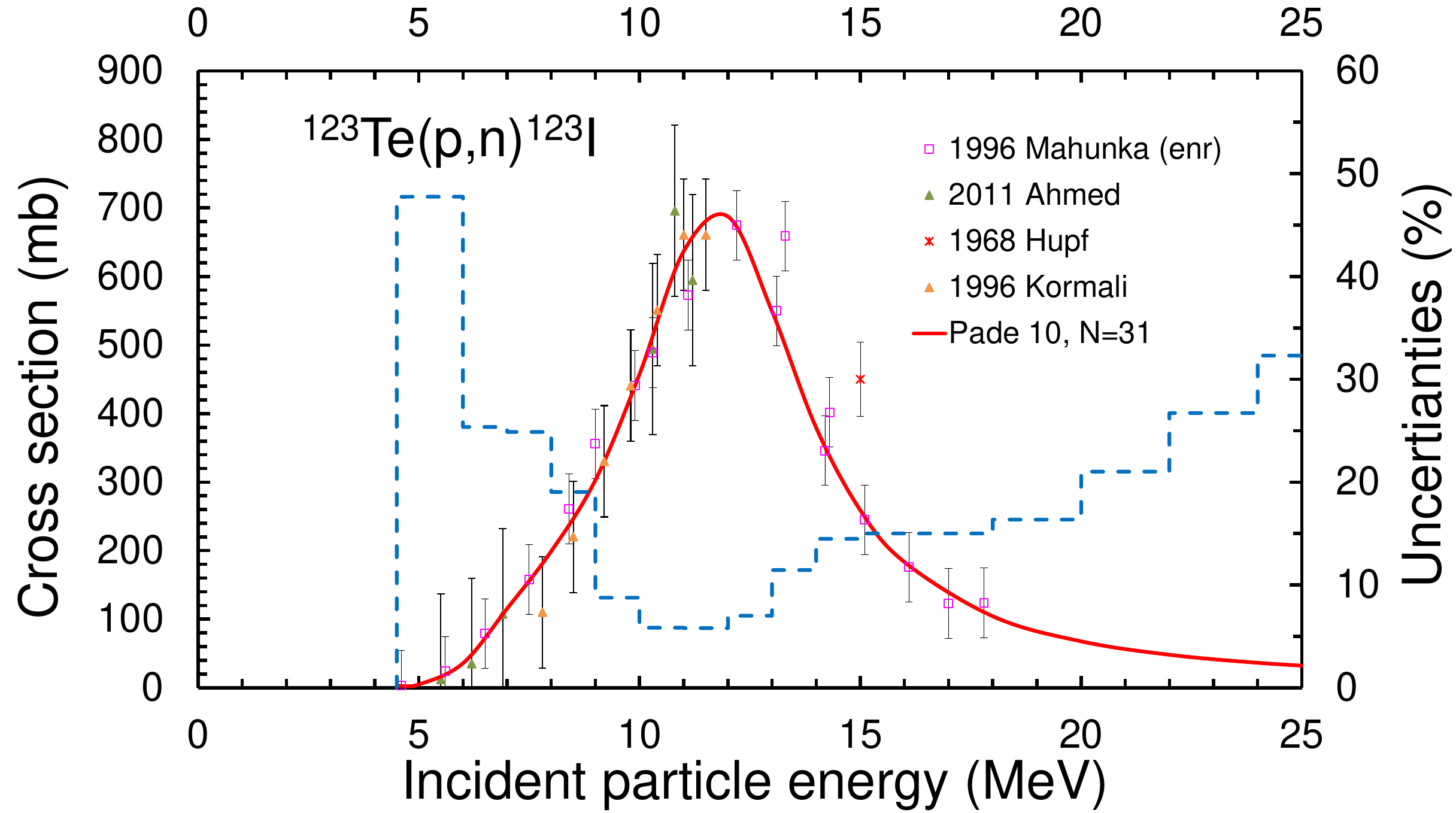}}
\centering
\vspace{-2mm}
\caption{(Color online) Evaluated Pad\'e fit and experimental data from Refs.~\cite{Acerbi:1975,Ahmed:2011,Barall:1981,El Azony:2008,Hupf:1968,Kandil:2013,Kiraly:2006,Kormali:1976,Mahunka:1996,Scholten:1989,Vandenbosch:1977,Zarie:2006,Zweit:1991a} for the $^{123}$Te(p,n)$^{123}$I reaction.}
\label{fig16:te123pn-i123}
%\vspace{-2mm}
\end{figure}

For formation of $^{123}$I through the $^{123}$Te(p,n)$^{123}$I reaction on low abundance enriched $^{123}$Te targets (0.908\% in ${^\mathrm{nat}}$Te) a grand total of 13 works~\cite{Acerbi:1975,Ahmed:2011,Barall:1981,El Azony:2008, Hupf:1968,Kandil:2013,Kiraly:2006,Kormali:1976,Mahunka:1996,Scholten:1989,Vandenbosch:1977,Zarie:2006,Zweit:1991a} with experimental cross-section data were identified in the literature for incident particle energies up to 25~MeV and are represented with uncertainties in Fig.~\ref{fig16:te123pn-i123}(a). An extremely high cross section value (1700 mb at 11 MeV) of El-Azony \etal(2008)~\cite{El Azony:2008} is not represented in the figure.  The cross sections were mostly obtained from experiments on ${^\mathrm{nat}}$Te targets~\cite{Acerbi:1975,Ahmed:2011,El Azony:2008,Hupf:1968,Kandil:2013,Kiraly:2006,Kormali:1976,Scholten:1989,Vandenbosch:1977,Zarie:2006,Zweit:1991a} and the results of these studies up to 11.5~MeV, the threshold of the $^{124}$Te(p,2n)$^{123}I$ reaction, were normalized and added to the 3 datasets obtained on enriched $^{124}$Te targets~\cite{Barall:1981,Mahunka:1996,Scholten:1989}. The two sets of data in Scholten \etal(1989)~\cite{Scholten:1989} are indicated as (nat) and (enr). All datasets published after 1999 and Kormali \etal(1976)~\cite{Kormali:1976} have to be considered as new with respect to the 2004 update of the gamma-emitter section of the IAEA on-line charged particle database~\cite{database:2001}, also documented in Tak\'acs \etal(2005)~\cite{Takacs:2005}. The results of Refs.~\cite{Acerbi:1975,Barall:1981,El Azony:2008,Kandil:2013,Kiraly:2006,Scholten:1989,Vandenbosch:1977,Zarie:2006,Zweit:1991a} were not considered for further analysis as the presented results are either too high or too low or show discrepant data points.

The remaining four datasets~\cite{Ahmed:2011,Hupf:1968,Kormali:1976,Mahunka:1996} were considered  as possible input for a least-squares Pad\'e fit.   		
The Pad\'e functions with 10 parameters were fitted to 31 selected data points with a $\chi^2$=1.63 and covering the energy range up to 100~MeV as shown in Fig.~\ref{fig16:te123pn-i123}(b).
The uncertainties (including a 4\% systematic uncertainty), are 50\% near the reaction threshold, decrease steadily to below 10\% from 10 MeV to 14 MeV and increase again to 35\% for higher energies.

\subsubsection{~~$^{\mathrm{124}}\mathrm{Te(p,2n)}^{\mathrm{123}}\mathrm{I}$ reaction}\label{sssect:Te124p2nI123}
\begin{figure}[!thb]
%\vspace{-2mm}
\centering
\subfigure[~All experimental data are plotted with uncertainties and compared to TENDL-2017 evaluation \cite{TENDL}.]
{\includegraphics[width=\columnwidth]{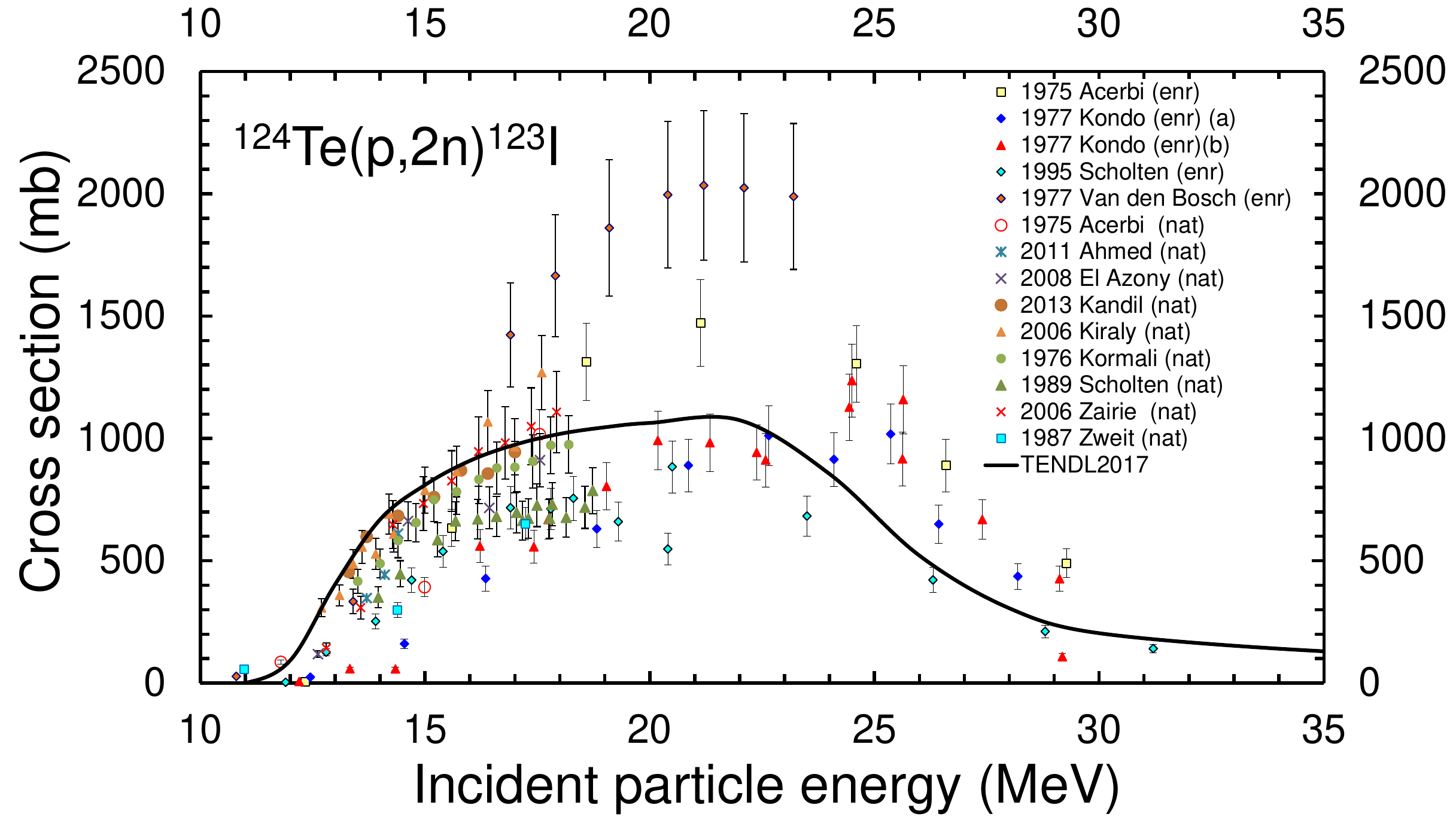}}
\subfigure[~Selected data compared with evaluated Pad\'e fit (L = 8, N = 87, $\chi^2$=1.29, solid line) and estimated total uncertainty in percentage including
a 4\% systematic uncertainty (dashed line, right-hand scale).]
{\includegraphics[width=\columnwidth]{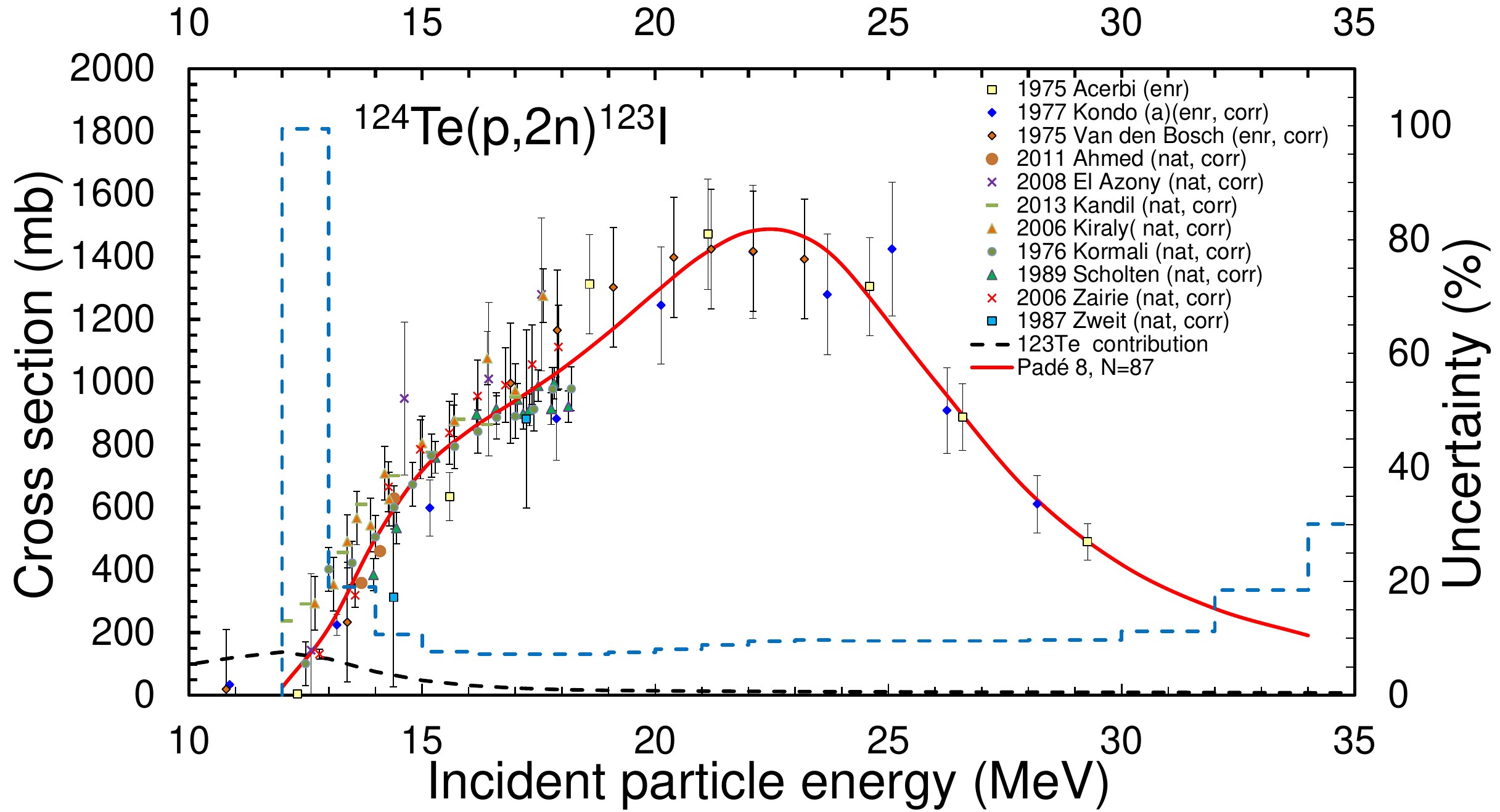}}
\centering
%\vspace{-2mm}
\caption{(Color online) Evaluated Pad\'e fit and experimental data from Refs.~\cite{Acerbi:1975,Ahmed:2011, El Azony:2008,Kandil:2013,Kiraly:2006,Kondo:1977,Kormali:1976,Scholten:1989,Scholten:1995,Vandenbosch:1977,Zarie:2006,Zweit:1991a} for the $^{124}$Te(p,2n)$^{123}$I reaction.}
\label{fig17:te124p2n-i123}
%\vspace{-4mm}
\end{figure}

For formation of $^{123}$I through the $^{124}$Te(p,2n)$^{123}$I reaction on higher abundance enriched $^{124}$Te targets (4.816\% in ${^\mathrm{nat}}$Te) a grand total of 12 works~\cite{Acerbi:1975,Ahmed:2011, El Azony:2008,Kandil:2013,Kiraly:2006,Kondo:1977,Kormali:1976,Scholten:1989,Scholten:1995,Vandenbosch:1977,Zarie:2006,Zweit:1991a} with experimental cross-section data were identified in the literature for incident particle energies up to 25~MeV  and are represented with uncertainties in Fig.~\ref{fig17:te124p2n-i123}(a). Only 5 datasets in 4 publications were obtained on enriched $^{124}$Te targets~\cite{Acerbi:1975,Kondo:1977,Scholten:1995,Vandenbosch:1977}. The publication by Kondo \etal(1977)~\cite{Kondo:1977}, contained two sets (of different degree of enrichment), labelled (a) and (b) in the figure while two sets of data are available in  Acerbi (1975)~\cite{Acerbi:1975}, obtained on enriched $^{124}$Te (labelled (enr)) and on ${^\mathrm{nat}}$Te (labelled nat). The cross sections obtained from experiments on ${^\mathrm{nat}}$Te targets in Refs.~\cite{Acerbi:1975,Ahmed:2011,El Azony:2008,Kandil:2013,Kiraly:2006,Kormali:1976,Scholten:1989,Zarie:2006,Zweit:1991a} up to 18.15~MeV, the threshold of the $^{125}$Te(p,3n) reaction, were normalized and corrected for the contribution of  the $^{123}$Te(p,n) reaction, relevant between 11 and 17 MeV (by using the recommended data of 2004 update of the gamma-emitter section of the IAEA on-line charged particle database~\cite{database:2001}).

All datasets on ${^\mathrm{nat}}$Te have to be considered as new with respect to the 2004 update of the gamma-emitter section of the IAEA on-line charged particle database~\cite{database:2001}, also documented in Tak\'acs \etal(2005)~\cite{Takacs:2005} as only the data on enriched $^{124}$Te were used at that time. A large scatter is observed between the 5 sets on enriched $^{124}$Te targets from the different authors while also the nine normalized datasets are dispersed and show a high and low group. This fact made a selection and fit similar to what was done in the IAEA database 2004 update~\cite{database:2001} not relevant.

In order to obtain a more coherent set, it was decided to rely essentially on the values of cross sections measured on enriched targets of Acerbi \etal(1975)(enr)~\cite{Acerbi:1975} and the studies on natural targets by Kandil \etal(2013)~\cite{Kandil:2013} and Kir\'aly \etal(2006)~\cite{Kiraly:2006} and to make corrections to other sets. For enriched targets the two sets of Kondo \etal(1977)~\cite{Kondo:1977} were energy shifted and multiplied by a factor of 1.4, while the very high values of Van den Bosch \etal(1977)~\cite{Vandenbosch:1977} were multiplied by 0.7. For natural targets the values of El-Azony \etal(2008)~\cite{El Azony:2008}, Scholten \etal(1989)~\cite{Scholten:1989} and Zweit \etal(1991)~\cite{Zweit:1991a}  (all low data group) were multiplied by 1.4.	The results of  Acerbi \etal(1975)(nat)~\cite{Acerbi:1975}, Kondo \etal(1977)(b)(enr)~\cite{Kondo:1977}  and  Scholten \etal(1995)(enr)~\cite{Scholten:1995} were not considered for further analysis as the presented results are either too high or too low or show discrepant data points. The remaining 11 datasets from Refs.~\cite{Acerbi:1975,Ahmed:2011,El Azony:2008,Kandil:2013,Kiraly:2006,Kondo:1977,Kormali:1976,Scholten:1989,Vandenbosch:1977,Zarie:2006,Zweit:1991a} were considered  as possible input for a least-squares Pad\'e fit.   		

The Pad\'e functions with 8 parameters were fitted to 87 selected data points with a $\chi^2$=1.29 and covering the energy range up to 28~MeV as shown in Fig.~\ref{fig17:te124p2n-i123}(b).
The uncertainties (including a 4\% systematic uncertainty), are above 50\% near the reaction threshold, decrease steadily to below 10\% from 13 MeV to 25 MeV and increase again to 40\% for higher energies.
Although this reaction is in principle more efficient, because of the abundance of the target material, than the route $^{123}$Te(p,n)$^{123}$I discussed before, practical use is limited because of the unavoidable formation of longer-lived contaminant $^{124}$I through the $^{124}$Te(p,n)$^{124}$I reaction evaluated in Sect.~\ref{sssect:Te124pnI124}.

\begin{figure}[!thb]
%\vspace{-2mm}
\centering
\subfigure[~All experimental data are plotted with uncertainties and compared to TENDL-2017 evaluation \cite{TENDL}.]
{\includegraphics[width=\columnwidth]{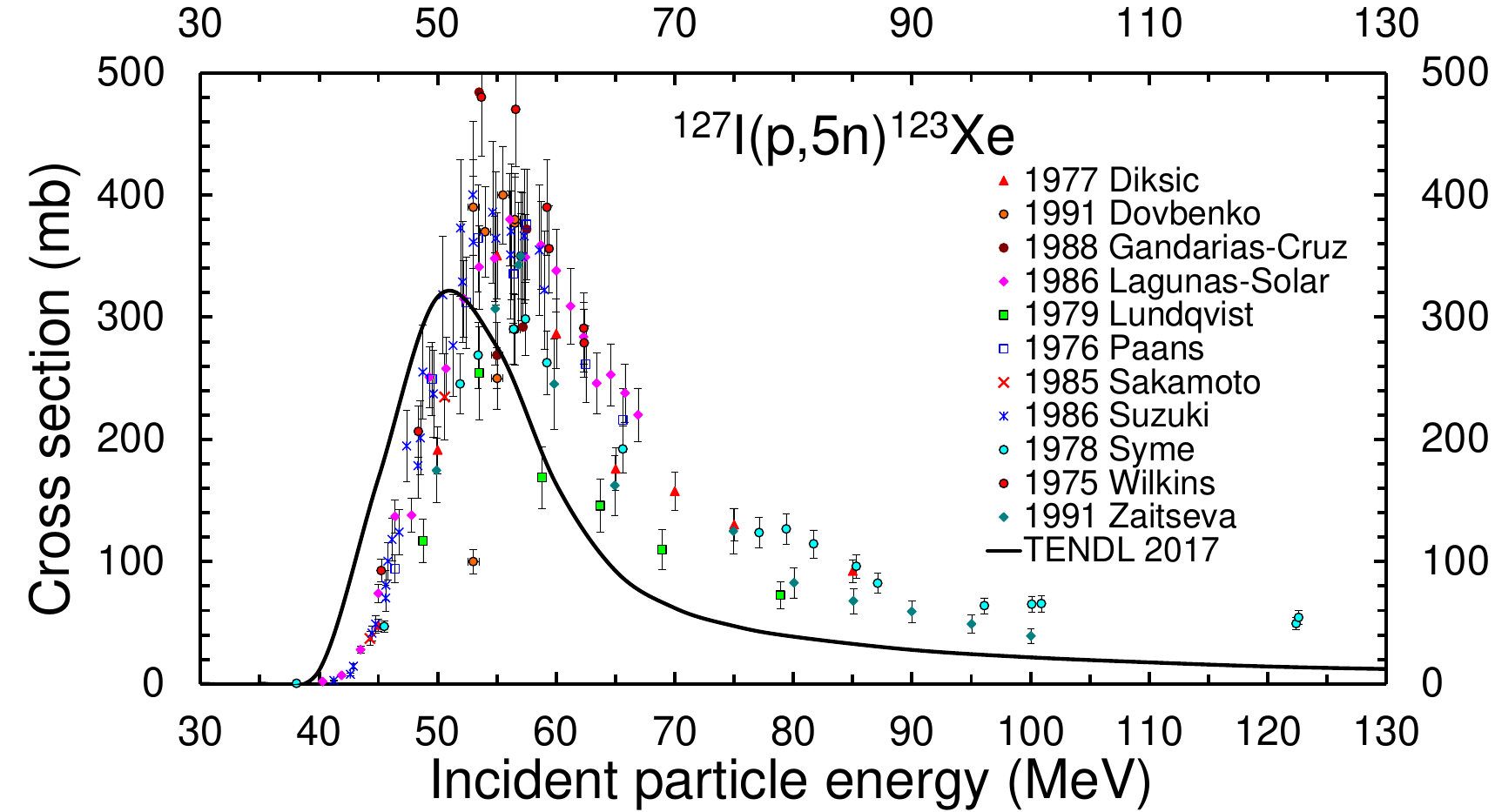}}
\subfigure[~Selected data compared with evaluated Pad\'e fit (L = 15, N = 108, $\chi^2$=1.87, solid line) and estimated total uncertainty in percentage including
a 4\% systematic uncertainty (dashed line, right-hand scale).]
{\includegraphics[width=\columnwidth]{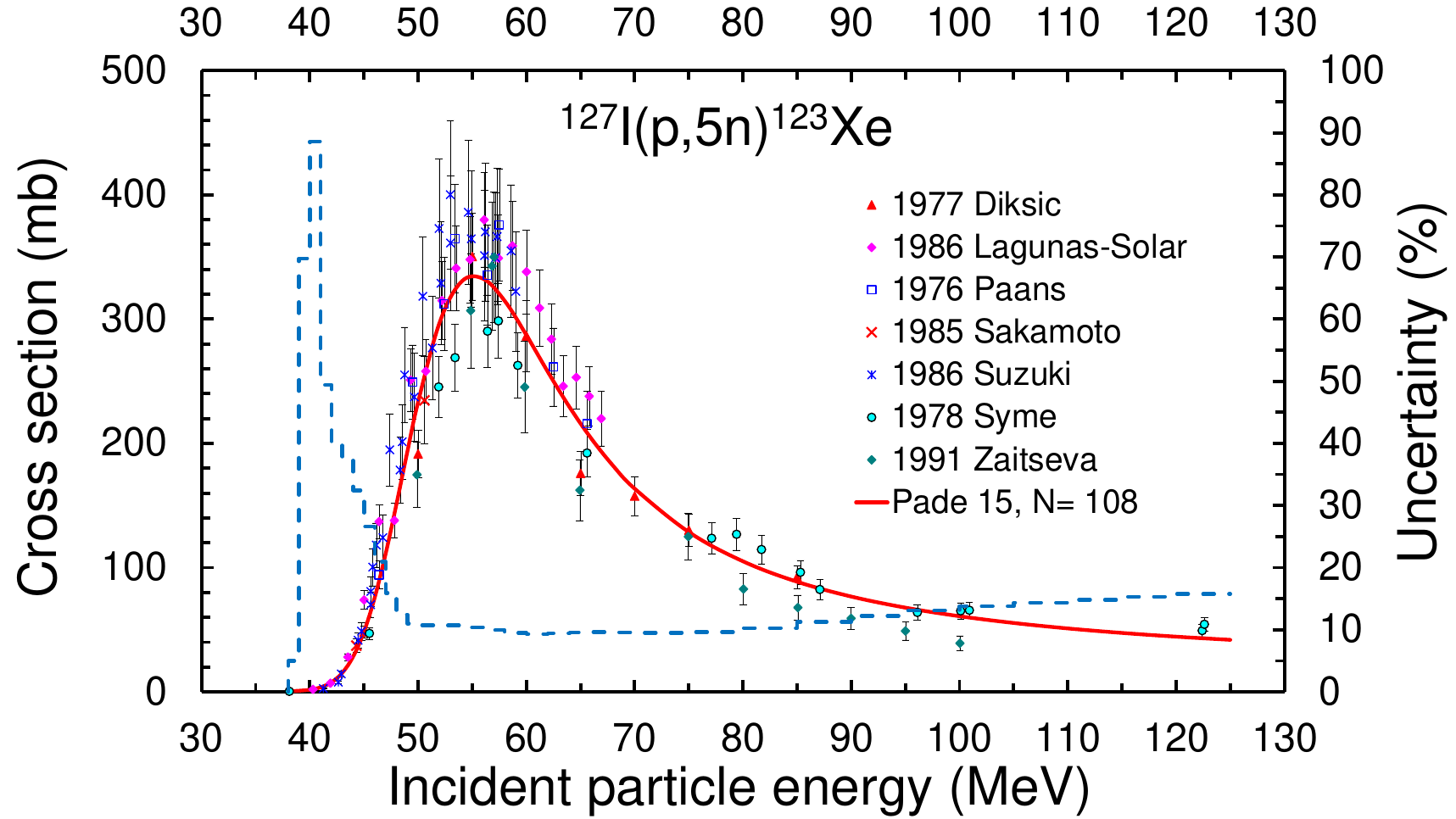}}
\centering
\vspace{-2mm}
\caption{(Color online) Evaluated Pad\'e fit and experimental data from Refs.~\cite{Diksic:1977,Lagunas-Solar:1986,Lundqvist:1979,Paans:1976,Sakomoto:1985,Suzuki:1986,Syme:1978,Wilkins:1975,Zaitseva:1991} for the $^{127}$I(p,5n)$^{123}$Xe$\rightarrow$$^{123}$I reaction.}
\label{fig18:i127p5n-xe123}
%\vspace{-4mm}
\end{figure}

\subsubsection{~~$^{\mathrm{127}}\mathrm{I(p,5n)}^{\mathrm{123}}\mathrm{Xe}\rightarrow$$^{\mathrm{123}}\mathrm{I}$ reaction}\label{sssect:I127p5nI123}
The reference pathway formerly used in commercial production of $^{123}$I was a high energy proton bombardment (at least 37.06~MeV incident energy) of natural iodine (monoisotopic $^{127}$I) relying on the indirect route $^{127}$I(p,5n)$^{123}$Xe$\rightarrow$$^{123}$I. For this reaction a grand total of 9 works~\cite{Diksic:1977,Lagunas-Solar:1986,Lundqvist:1979,Paans:1976,Sakomoto:1985,Suzuki:1986,Syme:1978,Wilkins:1975,Zaitseva:1991} with experimental cross-section data were identified in the literature for incident particle energies up to 125~MeV and are represented with uncertainties in Fig.~\ref{fig18:i127p5n-xe123}(a). Additionally two sets based on theoretical calculations published by Dovbenko \etal(1991)~\cite{Dovbenko:1991} and Gandarias-Cruz (1988)~\cite{Gandarias:1988} are represented. The values of Ref.~\cite{Diksic:1977} were multiplied by 1.3 to correct for the very old $\gamma$-line abundance used. No new datasets are added with respect to the 2004 update of the gamma-emitter section of the database~\cite{database:2001}, also documented in Tak\'acs \etal(2005)~\cite{Takacs:2005}. The results of the two experimental sets by Lundqvist \etal(1979) (too low values)~\cite{Lundqvist:1979} and Wilkins \etal(1975) (very high near maximum)~\cite{Wilkins:1975} were not considered for further analysis. Also the values of the two theoretical calculations published in Refs.~\cite{Dovbenko:1991,Gandarias:1988} were not selected.

The remaining seven datasets~\cite{Diksic:1977,Lagunas-Solar:1986,Paans:1976,Sakomoto:1985,Suzuki:1986,Syme:1978,Zaitseva:1991} were considered  as possible input for a least-squares Pad\'e fit.   		
The Pad\'e functions with 15 parameters were fitted to 108 selected data points with a $\chi^2$=1.87 and covering the energy range up to 125 MeV as shown in Fig.~\ref{fig18:i127p5n-xe123}(b).
The uncertainties (including 4\% systematic uncertainty), are 90\% near the reaction threshold, decrease steadily to below 10\% from 48 MeV to 80 MeV and increase again to 15\% for higher energies.	

\subsubsection{~~$^{\mathrm{127}}\mathrm{I(p,3n)}^{\mathrm{125}}\mathrm{Xe}\rightarrow$$^{\mathrm{125}}\mathrm{I}$ reaction, an impurity for $^{123}\mathrm{I}$ production}\label{sssect:I127p3nI125}
\begin{figure}[!thb]
%\vspace{-2mm}
\centering
\subfigure[~All experimental data are plotted with uncertainties and compared to TENDL-2017 evaluation \cite{TENDL}.]
{\includegraphics[width=\columnwidth]{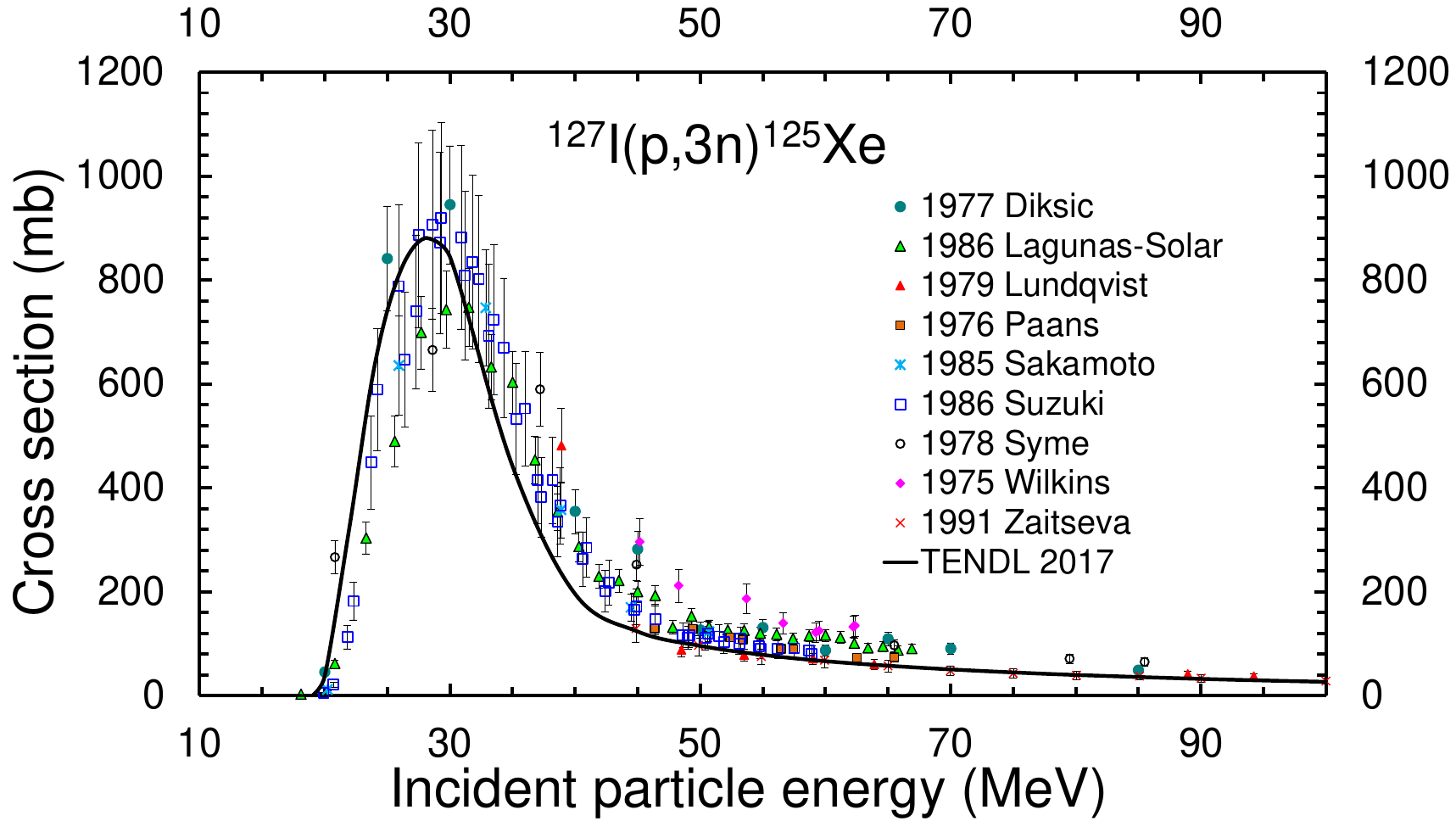}}
\subfigure[~Selected data compared with evaluated Pad\'e fit (L = 11, N = 125, $\chi^2$=1.43, solid line) and estimated total uncertainty in percentage including
a 4\% systematic uncertainty (dashed line, right-hand scale).]
{\includegraphics[width=\columnwidth]{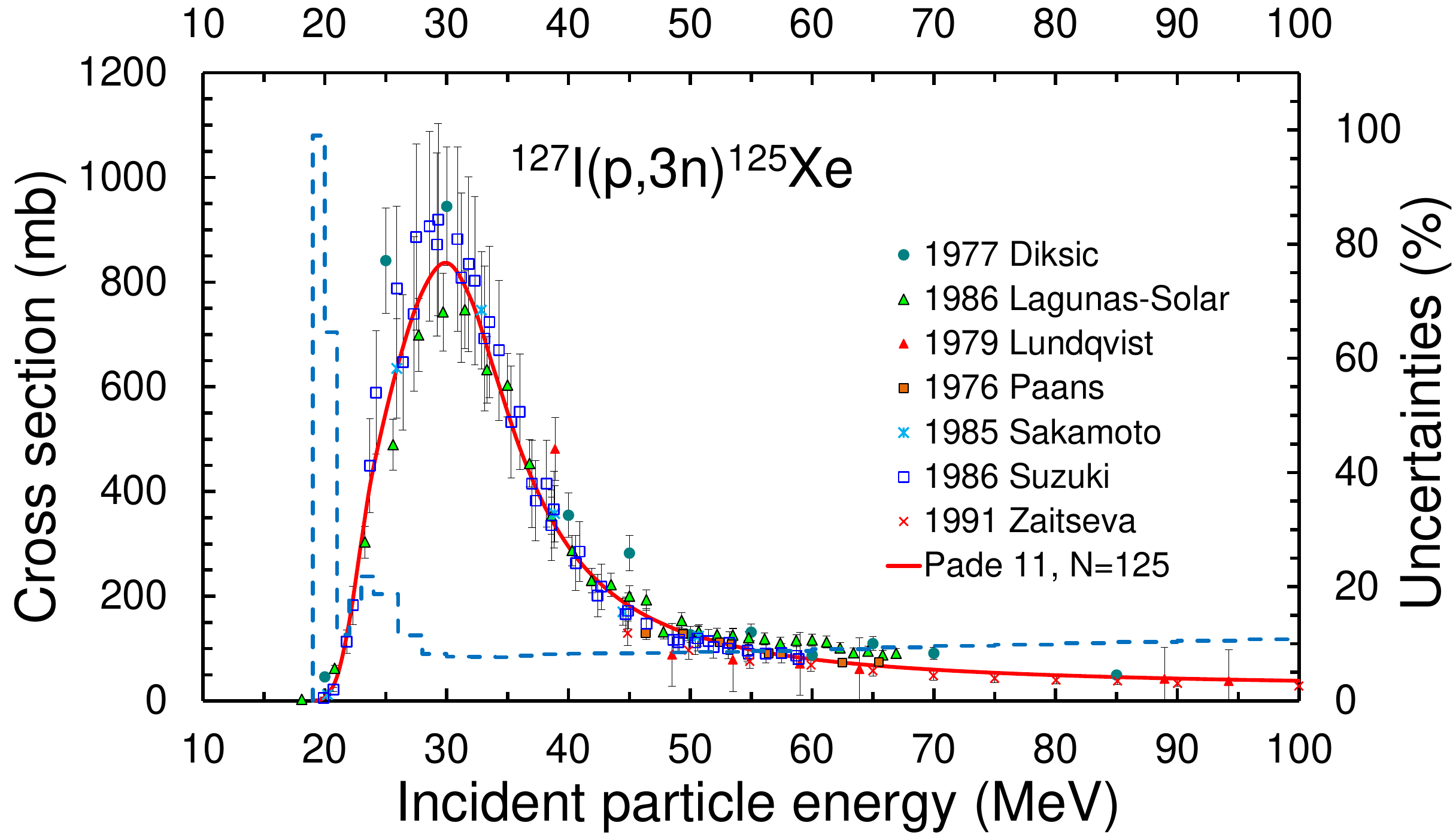}}
\centering
\vspace{-2mm}
\caption{(Color online) Evaluated Pad\'e fit and experimental data from Refs.~\cite{Diksic:1977,Lagunas-Solar:1986,Lundqvist:1979,Paans:1976,Sakomoto:1985,Suzuki:1986,Syme:1978,Wilkins:1975,Zaitseva:1991} for the $^{127}$I(p,3n)$^{125}$Xe$\rightarrow$$^{125}$I reaction.}
\label{fig19:i127p3n-xe125}
%\vspace{-4mm}
\end{figure}

When using the route for production of $^{123}$I discussed above in Sect.~\ref{sssect:I127p5nI123} the contamination with long-lived $^{125}$I impurity (\halflife=59.41~d), formed in the $^{127}$I(p,3n)$^{125}$Xe$\rightarrow$$^{125}$I reaction with 18.864 MeV threshold, is almost unavoidable. Therefore, we have evaluated the production cross section of this impurity for a proper quantification. For this reaction a grand total of nine works~\cite{Diksic:1977,Lagunas-Solar:1986,Lundqvist:1979,Paans:1976,Sakomoto:1985,Suzuki:1986,Syme:1978,Wilkins:1975,Zaitseva:1991} with experimental cross-section data were identified in the literature for incident particle energies up to 100~MeV  and are represented with uncertainties in Fig.~\ref{fig19:i127p3n-xe125}(a). No new datasets are added with respect to the 2004 update of the gamma-emitter section of the database~\cite{database:2001}, also documented in Tak\'acs \etal(2005)~\cite{Takacs:2005}.

The results of two experimental sets -- Syme \etal(1978) (too high values)~\cite{Syme:1978} and Wilkins \etal(1975) (too high values)~\cite{ Wilkins:1975} -- were not considered for further analysis.
The remaining seven datasets from Refs.~\cite{Diksic:1977,Lagunas-Solar:1986,Lundqvist:1979,Paans:1976,Sakomoto:1985,Suzuki:1986,Zaitseva:1991} were considered as possible input for a least-squares Pad\'e fit.   		

The Pad\'e functions with 11 parameters were fitted to 125 selected data points with a $\chi^2$=1.43 and covering the energy range up to 100~MeV as shown in Fig.~\ref{fig19:i127p3n-xe125}(b).
The uncertainties (including a 4\% systematic uncertainty), are 90\% near the reaction threshold, decrease steadily to below 10\% from 26 MeV to 80 MeV and increase slightly for higher energies.
This reaction will also be included in the upcoming update for therapeutic isotopes.
%
%
%======================================================================
\section{SUMMARY AND CONCLUSIONS}
%======================================================================
Although during the 2012--2017 IAEA Coordinated Research Project on Nuclear Data for Charged-particle Monitor Reactions and Medical Isotope Production more than 100 reaction cross sections have been evaluated and added  to  the IAEA-NDS recommended cross-section database, it appeared during preparation of the publications~\cite{Hermanne:2018,Tarkanyi:2019,Tarkanyi:2019b,Engle:2019} that about 20 reactions for SPECT and PET imaging and 20 reactions for therapeutic radionuclides, which were evaluated in the TECDOC-1211~\cite{Gul:2001} or in TRS-473~\cite{Betak:2011} were not included in this CRP. To remediate this deficiency an additional effort was set up and we report here on the newly evaluated cross sections with uncertainties for diagnostic (PET and SPECT) isotopes. A similar upgrade for therapeutic isotopes is in progress.

Evaluations, including data compilation and selection for Pad\'e fitting resulting in recommended values with uncertainties, were performed for 19 reactions leading to direct, indirect or generator production of $^{11}$C, $^{13}$N, $^{15}$O, $^{18}$F, $^{64}$Cu, $^{124}$I, $^{81}$Kr, and $^{123}$I. Although for the short lived PET isotopes $^{11}$C, $^{13}$N, $^{15}$O, and $^{18}$F only a few new datasets became available since the 2003 update of the IAEA website~\cite{database:2001}, the quality of the database has been strengthened by the inclusion of many more accurately digitized cross section values from earlier experimental studies in EXFOR, making data retrieval easier. In most of the present evaluations these newer EXFOR values were used but no or only minor influence on the results of the fits was observed. On the other hand particular attention has been paid to selection of relevant experimental data (with sufficient energy resolution) in the resonance region for the proton induced reactions.

The positron emitters $^{64}$Cu and $^{124}$I (evaluated earlier as therapeutic isotopes in IAEA TRS 473~\cite{Betak:2011}) were included in this report because their clinical use nowadays is predominantly in imaging. For both isotopes all charged-particle induced reactions studied in IAEA TRS-473 have been re-evaluated, but not the neutron-induced ones. For reactions leading to $^{64}$Cu one or two new publications are available and were included in the selected sets. For the $^{64}$Ni(p,n)$^{64}$Cu reaction the data with very high resolution of Guzhovskij \etal(1969)~\cite{{Guzhovskij:1969}} were fitted separately to show the resonances. Overall no important changes in the fits for the four evaluated production reactions are observed compared to previous work~\cite{Betak:2011}.

For formation of $^{124}$I by bombardment of enriched Te targets only for the $^{124}$Te(p,n)$^{124}$I reaction important changes compared to the evaluation in Ref.~\cite{Betak:2011} have been made. In addition to the experimental results obtained on $^{124}$Te targets also some cross sections measured on ${^\mathrm{nat}}$Te (limited to the threshold of the $^{125}$Te(p,2n) reaction and normalized to the abundance of $^{124}$Te) were included in the compilation. This resulted in a quite different selection compared to previous work~\cite{Betak:2011} and a drastic change in the fit and recommended values.

For the two reactions leading to formation of the SPECT mother isotope $^{81}$Rb (protons on ${^\mathrm{nat}}$Kr and $^{82}$Kr) no new datasets were found since the update of the database in 2001~\cite{database:2001} but a somewhat different selection results in modified fits and recommended values. For formation of $^{123}$I, two reactions on enriched Te and two reactions on $^{127}$I (including the formation of contaminant $^{125}$I) present in IAEA TECDOC-1211~\cite{Gul:2001} were evaluated. For the two direct reactions on $^{123,124}$Te isotopes results from experimental studies on ${^\mathrm{nat}}$Te were included (in limited energy region and normalized). This completion and extension of the database at lower energies resulted in quite different selection and fits. Nearly no changes for the indirect reactions on $^{127}$I leading to Xe precursors of $^{123,125}$I occur.

Where possible the compiled experimental data were compared to the theoretical predictions by the TALYS 1.6 code as available in the on-line library TENDL-2017. Although the overall shape of the excitation functions is mostly well described, still sometimes significant disagreements in the magnitude and shifts of calculated excitation functions at higher energies are noted, reflecting the current status of reaction modelling and model parameters at those energies.

All recommended cross-section data with their corresponding uncertainties and deduced integral thick target yields are available on-line at the IAEA-NDS medical portal \href{https://www-nds.iaea.org/medportal/}{\textit{www-nds.iaea.org/medportal/}}
and also at the IAEA-NDS web page \href{https://www-nds.iaea.org/medical/}{\textit{www-nds.iaea.org/medical/}}.

\section*{Acknowledgments}
The IAEA is grateful to all participant laboratories for their assistance in the work and for support of the activities. The work described in this paper would not have been possible without IAEA
Member State contributions.

\end{document}